\documentclass[
aps, 
prb, 
10pt,  
amssymb, 
amsmath, 
superscriptaddress, 
tightenlines, 
twocolumn, 
notitlepage, 
] {revtex4-2}
\usepackage{graphicx}
\usepackage[colorlinks, linkcolor = blue, citecolor = blue, urlcolor=blue, pdfborder={0 0 0 [0 0]},bookmarks=false]{hyperref}
\usepackage{bm}
\usepackage{physics}
\usepackage{amssymb}
\usepackage{amsmath}
\usepackage[dvipsnames]{xcolor}
\usepackage{soul}

\usepackage{lipsum}

\newcommand{\be}{\begin{equation}}
\newcommand{\ee}{\end{equation}}

\begin{abstract} 
We introduce the notion of nonreciprocal superconductors where inversion and time-reversal symmetries are broken, giving rise to an asymmetric energy dispersion. 
We demonstrate that nonreciprocal superconductivity can be detected by Andreev reflection. 
In particular, a transparent junction between a normal metal and a nonreciprocal superconductor generally exhibits an asymmetric current-voltage characteristic, which serves as a defining feature of nonreciprocal superconductivity. 
Unlike the superconducting diode effects, our detection scheme has the advantage of avoiding large critical currents that turn the superconducting state to normal. 
Finally, we discuss candidates for nonreciprocal superconductivity, including graphene, UTe$_2$, as well as engineered platforms.
\end{abstract}

\begin{document}
\author{Margarita Davydova}
\affiliation{Department of Physics, Massachusetts Institute of Technology, Cambridge, MA 02139, USA }
\author{Max Geier}
\affiliation{Department of Physics, Massachusetts Institute of Technology, Cambridge, MA 02139, USA }
\author{Liang Fu}
\affiliation{Department of Physics, Massachusetts Institute of Technology, Cambridge, MA 02139, USA }

\title{Nonreciprocal superconductivity}

\maketitle


\section{Introduction}

The best-known example of nonreciprocal transport is realized in the  $p$-$n$ semiconductor junction, one of the building blocks for modern electronics. There, the nonreciprocity is manifested as the dependence of the resistance on the direction of the current and is governed by an inversion-breaking depletion region with a built-in electric field.

Recently, there has been a surge of interest in nonreciprocal phenomena in superconducting materials and devices. A large body of theoretical and experimental efforts has been dedicated to studying diode effects in superconductors~\cite{jiang2022superconducting,
Nadeem2023Oct,ando2020observation,yuan2021supercurrent,PhysRevLett.128.037001,PhysRevLett.128.177001,
hou2022ubiquitous,bauriedl2022supercurrent,kochan2023phenomenological,yuan2023edelsteineffectsupercurrentdiode,hasan2024supercurrentdiodeeffecthelical,shaffer2024superconductingdiodeeffectmultiphase,PHFu2024May} and Josephson junctions~\cite{PhysRevLett.99.067004,Baumgartner2022Jan,Costa2023Nov,wu2022field,davydova2022universal,pal2022josephson,zhang2022general,PhysRevB.109.024504,zazunov2024approachingidealrectificationsuperconducting}, wherein the critical supercurrent differs in two opposite directions. These phenomena can occur when inversion ($P$) and time-reversal ($T$) symmetries are broken,  
which can be achieved, for example, by the polar crystal structure of the superconducting material~\cite{PismaZhETF.41.365,Edelstein1996Jan} 
and by the applied magnetic field, respectively \cite{He_2022,kochan2023phenomenological}. 
While theoretical studies have mostly focused on intrinsic superconducting diode effects, the interpretation of experimental data is often complicated by extrinsic factors related to Joule heating, vortex dynamics~\cite{Gaggioli2024May} and device geometry~\cite{hou2022ubiquitous}, which can all affect the asymmetry in the critical current that drives the superconductor into a resistive state.

In this work, we define the concept of nonreciprocal superconductors as the superconducting states that break both time-reversal and inversion symmetries.  
We study the essential physics of nonreciprocal superconductors and identify their universal features through Andreev reflection spectroscopy. Importantly, our proposed method for detecting nonreciprocal superconductivity does not invoke the critical current, thus avoiding the complication of heating and vortex dynamics. 
We also propose a simple setup for engineering nonreciprocal superconducting states with conventional superconductors. Finally, we discuss the promising prospect of intrinsic nonreciprocal superconductivity in graphene systems and UTe$_2$.

\begin{figure}[!t] 
	\includegraphics[width= 1\columnwidth]{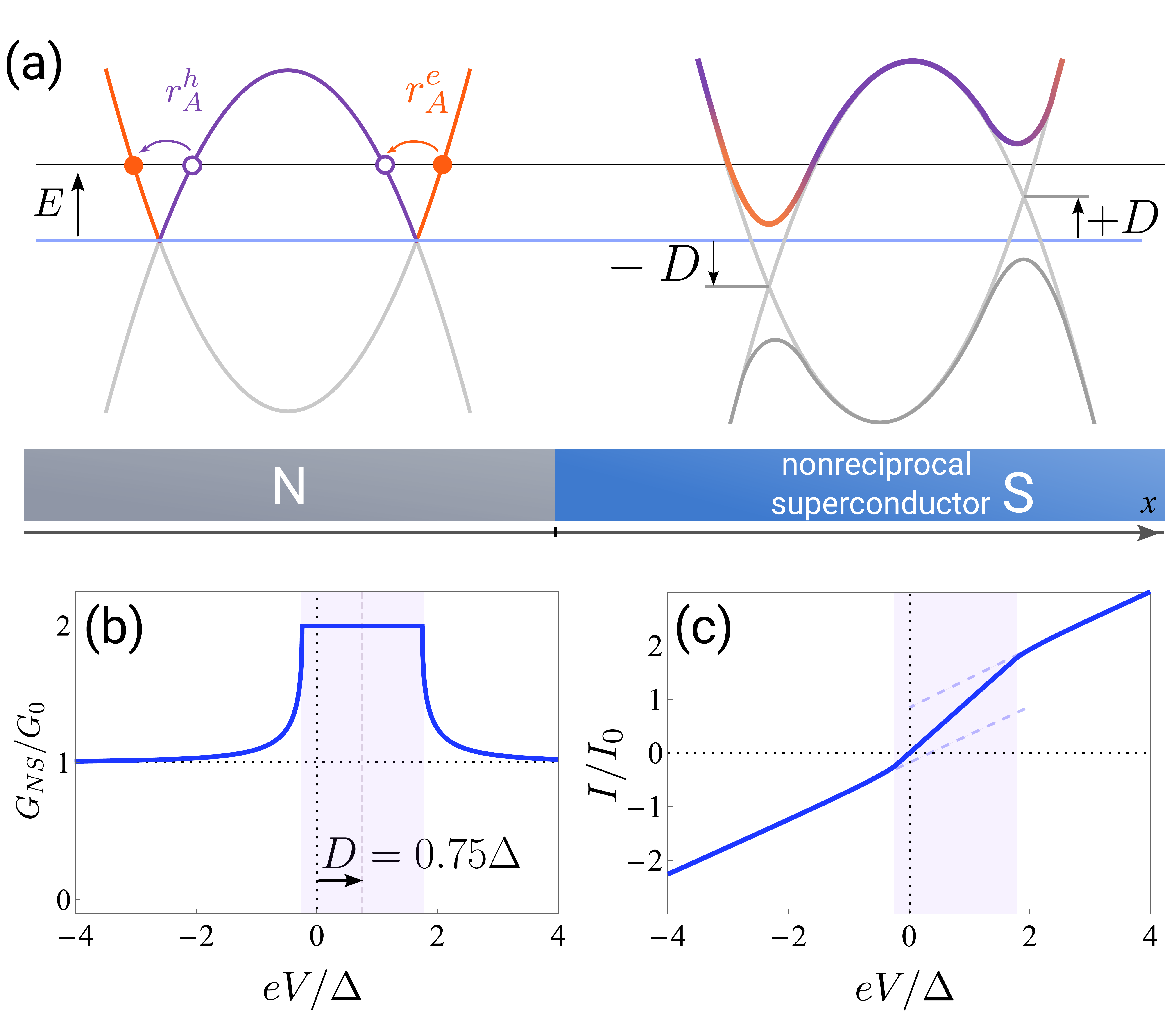}
	\caption{(a) A normal metal-nonreciprocal superconductor junction, with the quasiparticle dispersion is shown schematically in each region. The quasiparticle dispersion in the nonreciprocal superconductor is asymmetric (see eq.~\eqref{eq:asymm1}) with the superconducting gaps shifted in energy by $\pm D$.
    The Andreev reflection process for the electrons incoming from the normal region is sensitive to this shift $r^e_A (E) = r_A( E - D)$, and the situation is similar but opposite in sign for the incoming holes. 
    (b) Nonreciprocal conductance $G_{NS}$ and (c) the $I$-$V$ characteristic of a transparent one-dimensional junction. $G_0$ denotes the conductance in the normal state.  
	} \label{fig:schematic}
\end{figure}

The key idea behind sensing this type of superconductivity in transport is simple. Nonreciprocal superconductors breaking time-reversal and inversion symmetries generally have an asymmetric quasiparticle energy spectrum $E({\bm k})\neq E(-\bm k)$. This asymmetry directly gives rise to nonreciprocal Andreev reflection that can be easily detected in a junction between a normal metal and a nonreciprocal superconductor,  as shown in Fig.~\ref{fig:schematic}. 
In transparent junctions, only one direction of the asymmetric quasiparticle dispersion is probed in transport, because Andreev reflection relates incoming and reflected quasiparticles on the same branch of the normal state Fermi surface (shown on the left in Fig.~\ref{fig:schematic}(a)). 
As a consequence, the conductance in the forward direction can be twice that in the opposite direction, see Fig.~\ref{fig:schematic}(b), (c). 
This nonreciprocal behavior can be measured in a point-contact normal metal-superconductor (NS) junction, and, unlike superconducting diode effects, avoids the need of a large current driving the superconductor far from equilibrium.

The nonreciprocal Andreev reflection in NS junctions can be used as a smoking-gun feature for identifying nonreciprocal superconductors and probing their gap structure. 
Candidates for nonreciprocal superconductivity include finite-momentum Fulde-Ferrel states~\cite{PhysRev.135.A550} and superconductors with mixed order parameters~\cite{M_ckli_2022,Hu_2021,PhysRevB.93.134512,PhysRevLett.119.187003,Hu_2021,shaffer2024superconductingdiodeeffectmultiphase}. Recently, spontaneous supercurrent diode effect has been observed in twisted graphene multilayers at zero magnetic field~\cite{diez2021magnetic,lin2022zero} as well as twisted cuprates~\cite{zhao2021emergent},  
which could give evidence for the nonrecirpocal superconductivity with unconventional order parameters.  %

Importantly, nonreciprocal superconductivity can also be achieved using conventional $s$-wave superconductors.  For example, nonreciprocal superconductivity can be inherited from a time-reversal and inversion symmetry-breaking normal system that is proximitized by an ordinary $s$-wave parent superconductor, resulting in an asymmetric quasiparticle dispersion.  

\begin{figure}[!t] 
	\includegraphics[width= 0.7\columnwidth]{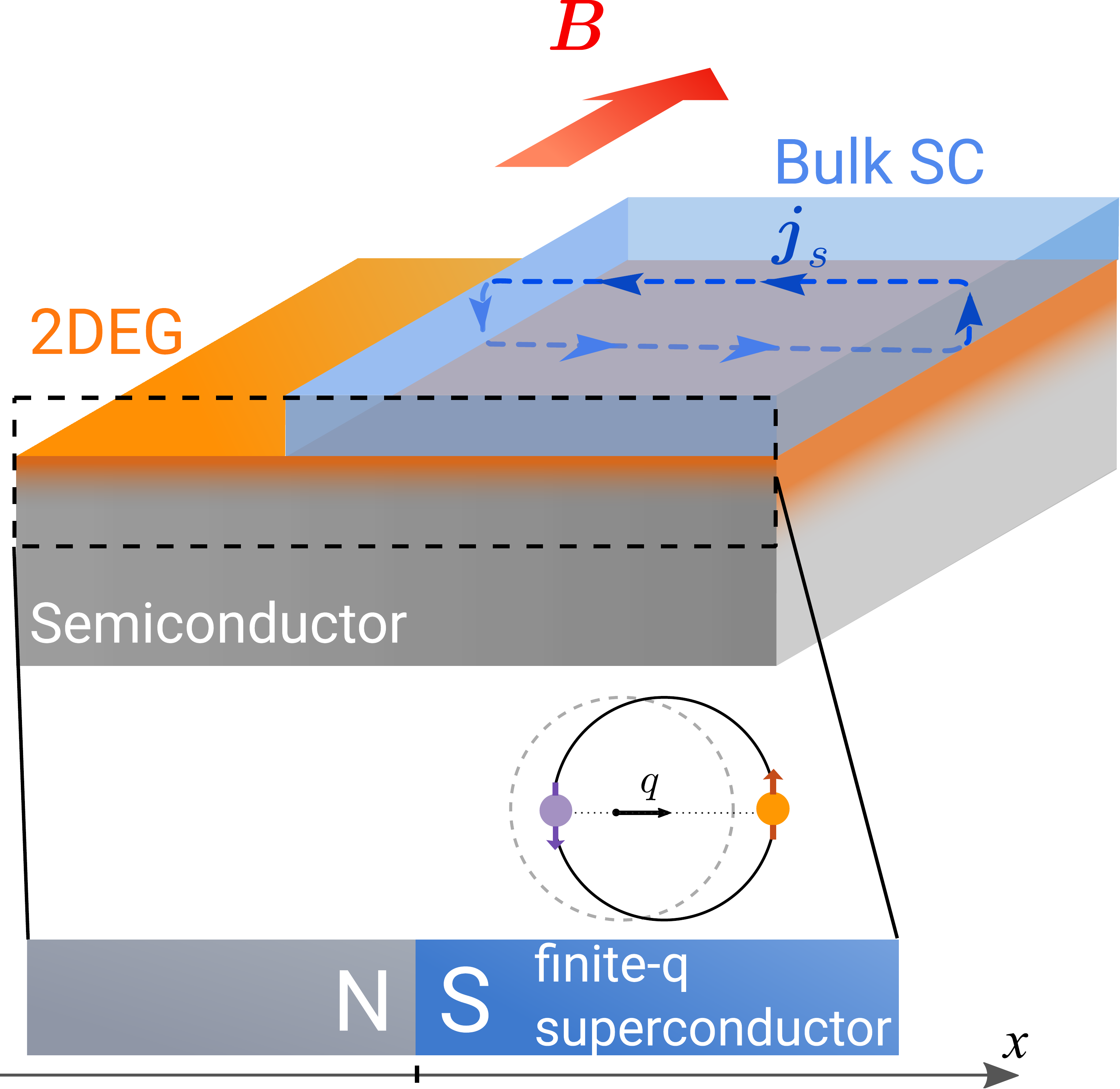}
	\caption{A junction between a normal metal N and a finite-momentum superconductor S. In the simplest setup, the 2D electron gas  (2DEG, orange) is proximitized by a slab of a parent superconductor (blue) on the right ($x>0$) and is in the normal state on the left ($x<0$). A small magnetic field parallel to the junction induces Meissner current in the parent superconductor and induces finite-momentum pairing perpendicular to the junction.  One-dimensional setup can also be realized, by gating to collimate electrons or by using semiconducting nanowires instead.
 \vspace{-15 pt}
	} \label{fig:schematic2}
\end{figure}

In this work, we also consider another simple realization of nonreciprocal superconductivity from $s$-wave pairing, shown in Fig.~\ref{fig:schematic2}. It involves a thin normal layer, for example, 2D electron gas proximitized by  $s$-wave superconductor that carries Meissner screening current due to a small in-plane magnetic field. This induces finite Cooper pair momentum near the surface which is inherited by the proximitized layer~\cite{zhu2020discovery, davydova2022universal}. This universal method of inducing finite-momentum pairing through proximity effect and Meissner screening has been demonstrated in topological insulator thin films~\cite{zhu2020discovery,chen2018finite}  and quantum wells in semiconductors~\cite{PhysRevB.80.220507,banerjee2023phase}, requiring magnetic fields as low as  20~mT~\cite{zhu2020discovery}. 
Finally, nonreciprocal superconducting states originating from $s$-wave pairing may be also achieved in proximitized Rashba systems under Zeeman field~\cite{hart2017controlled,baumgartner2021josephson,yuan2021supercurrent}. %

As another salient feature of nonreciprocal superconductivity, we demonstrate that
the differential conductance measured using N-S junctions and scanning tunneling microscopy near the edge of a nonreciprocal superconductor will produce a result that \emph{significantly differs} from the bulk quasiparticle density of states in a wide range of energies. This can have wide implications for interpreting the results of scanning tunneling microscopy in unconventional superconductors.

As a remark, the nonreciprocal Andreev reflection discussed in this paper is drastically different from the asymmetry frequently seen in scanning tunneling microscopy experiments in the presence of localized states at magnetic impurities in the superconductor~\cite{Yazdani1997Mar,Heinrich2018Feb}. As we discuss later in the paper, the latter effect relies on particle-hole asymmetry introduced by the impurities and on the presence of inelastic relaxation processes. The effect considered here is due to the asymmetry in quasiparticle dispersion, involves large changes in conductance, and occurs already in the transparent junction regime with elastic scattering.

\section{N-S junction with nonreciprocal superconductor}

\subsection{Symmetry-breaking normal state}

There are several simple ways to realize nonreciprocal superconductivity from $s$-wave zero-momentum pairing. As the first example, consider a thin layer where the normal state is already time-reversal and inversion-breaking and its normal-state Hamiltonian is 
\begin{equation} \label{eq:H1}
    \mathcal{H}_0 = \sum_{\bm k, \sigma} \left ( \varepsilon( \bm k) - \mu \right)c^\dagger_{\sigma, \bm k}c_{\sigma, \bm k}  
\end{equation}
where we assumed a single spin-degenerate band for simplicity, and $\mu$ is the Fermi energy. The symmetry breaking allows the normal-state dispersion to be asymmetric, i.e. $\varepsilon(\bm k) \neq \varepsilon(- \bm k)$.  In proximity to a parent $s$-wave superconductor, the pairing term is induced which has the form  
$
\mathcal{H}_{int}  =\sum_{\bm k} \Delta c^\dagger_{\uparrow,\bm k} c^\dagger_{\downarrow, -\bm k}+  h.c.$
Diagonalizing the full Bogoliubov-de-Gennes (BdG) Hamiltonian 
 \be
 \mathcal{H} = \mathcal{H}_{0} + \mathcal{H}_{int}
 \ee
 leads to quasiparticle dispersion with an asymmetric gap:
\be
\begin{split} \label{eq:disp1}
E_{\pm}(\bm k) &= \left ( \frac{\varepsilon(\bm k) - \varepsilon(- \bm k) }{2}\right) 
\\
&\pm \sqrt{\left (  \frac{\varepsilon(\bm k) + \varepsilon(-\bm k) - 2 \mu }{2}\right )^2 + |\Delta|^2 }.
\end{split}
\ee
where the $\pm$ index labels the two BdG solutions.

To show the asymmetry more explicitly, we consider the one-dimensional case and expand the dispersion near \emph{the gap-opening momenta} $\pm k_0$ defined by the equation $\varepsilon (k_0) + \varepsilon ( - k_0)  -2 \mu =0$.    The asymmetry in dispersion becomes characterized by a single quantity $D = \frac{1}{2} \left ( \varepsilon (k_0) - \varepsilon ( - k_0)\right ) $.  Assuming that $\delta k = k_0 - k_F \ll k_F$ and that $v(k_0) \approx v(-k_0) \approx v_F$~\footnote{
Generically, for a non-reciprocal metal, the Fermi velocities $v_{F+},\, v_{F-}$ around the Fermi momenta $k_{F+}$ and $k_{F-}$ are distinct.
Assuming that $\Delta, v_{F\pm} (\pm k_0 - k_{F\pm}) \ll E_F$, the quasiparticle dispersion, Eq.~\eqref{eq:asymm1}, for this case can be expressed as
$$
E(k) = D(k)\pm\sqrt{\left(\frac{v_{F+}\delta k_{+}+v_{F-}\delta k_{-}}{2}\right)^{2}+\Delta^{2}}
$$
with $\delta k_\pm = \pm k-k_{F\pm}$ and $D(k) = \frac{v_{F+}\delta k_{+}-v_{F-}\delta k_{-}}{2}$, which is valid around both $k_{F\pm}$.
In this limit, the gap opening occurs around $k = \pm k_0$ with $k_{0}=\frac{k_{F+}v_{F+}+k_{F-}v_{F-}}{v_{F+}-v_{F-}}$.
}, we can write
$\varepsilon (k) + \varepsilon ( - k)  = 2 (\mu +  v_F \delta k )$, and we assume $\hbar = 1$ throughout.  
The quasiparticle dispersion for the left- and right-movers labeled by $\alpha = \pm$ is then
\be
\begin{split} \label{eq:asymm1}
(1D) \quad E^\alpha_{\pm}(k) \approx \alpha D 
\pm \sqrt{\left ( v_F \delta k \right )^2 + |\Delta|^2 },
\end{split}
\ee
The resulting dispersion is shifted by $\pm D$ for right(left)-movers and is identical to the one shown in Fig.~\ref{fig:schematic}(a) on the right.  In the rest of this paper, we will study nonreciprocal transport in one- and two-dimensional setups, but the effect manifests itself fully already in one dimension, which we will frequently use for illustration.  

This mechanism for non-reciprocal superconductivitiy may be realized in
time-reversal and inversion-breaking materials when zero-momentum superconducting pairing is introduced.  Potential candidates include time-reversal-breaking (for example, valley-polarized) states in multilayer graphene~\cite{diez2021magnetic,lin2022zero,chiles2023nonreciprocal}.
 
\subsection{Finite-momentum pairing}

Another simple example of nonreciprocal superconductivity emerging purely in the presence of $s$-wave pairing is proximity-induced finite-momentum superconductivity in the presence of a screening current~\cite{davydova2022universal}. This scenario is shown in Fig.~\ref{fig:schematic2} for proximitized two-dimensional electron gas (2DEG), which is a favorable setup for realizing exceptionally transparent N-S junctions.  In the presence of an in-plane magnetic field,  the screening supercurrent $\bm j_s$ will flow near the surface of the parent superconductor (as well as in the proximitized layer) producing finite-momentum pairing.  In the geometry shown in Fig.~\ref{fig:schematic2}, the Cooper pair momentum  is related to the screening supercurrent near the surface of the parent superconductor
as   $\bm q =  \bm v_s k_{F}/v_{F}$, where $v_s$ is the supercurrent velocity associated with the screening current and $k_F, v_F$ are the Fermi momentum and velocity in the proximitized layer.  

This setup is readily realizable in experiment, for example in InAs/Al heterostructures \cite{baumgartner2021josephson}. In these systems Zeeman field and spin-orbit coupling render the InAs two-dimensional electron gas a nonreciprocal metal while the Al remains an ordinary s-wave superconductor. In this platform, the contribution from the nonreciprocal metal can combine with that from an induced Cooper pair momentum due to orbital coupling to an applied in-plane magnetic field \cite{banerjee2023phase}. We estimate that the magnetic field on the order of 100 mT or less would be sufficient for inducing significant finite-momentum Cooper pairing (see Appendix~\ref{sec:appendix_estimates} for the details).

The induced finite-momentum interaction term is spatially inhomogeneous and can be written as 
\begin{equation} \label{eq:H_int}
\begin{split}
    \mathcal{H}_{int, \bm q}  &= \sum_{\bm r} \Delta e^{2 i \bm q \bm r} \psi^\dagger_{\uparrow} (\bm r) \psi^\dagger_{\downarrow}(\bm r) + h.c.
    \\
    &=\sum_{\bm k} \Delta c^\dagger_{\uparrow,\bm k} c^\dagger_{\downarrow, -(\bm k- 2 \bm q)}+  h.c.
\end{split}
\end{equation}
where $2 \bm q$ is the momentum of the Cooper pairs. 
In the Bogoliubov-de-Gennes (BdG) form in the single-band approximation, the effective description is that of an $s$-wave superconductor where the electrons with dispersion $ \varepsilon( \bm k) - \mu $ are coupled with the holes with momentum-shifted dispersion $-\left ( \varepsilon( -\bm k + 2 \bm q) - \mu \right)$, where we now assume symmetric normal-state dispersion. The BdG Hamiltonian is
\begin{equation} \label{eq:BdG_matrix}
     H_{BdG} =\begin{pmatrix}
\left ( \varepsilon( \bm k) - \mu \right)  &  \Delta \\
\Delta^* & - \left ( \varepsilon( -\bm k + 2 \bm q) - \mu \right) \\
\end{pmatrix}
\end{equation}
which acts bilinearly on the spinor $(\psi_{\uparrow, \bm k},\psi^\dagger_{\downarrow, -(\bm k-2 \bm q)})^T$. Upon linearizing in momentum we obtain the quasiparticle dispersion 
\be \label{eq:E_dop}
E_{\pm}(\bm k) \approx \bm q  \cdot \bm v_F \pm \sqrt{(\bm v_F \delta \bm k)^2 + |\Delta|^2 },
\ee 
which is a Doppler-shifted dispersion of a finite-momentum superconductor. Here, $\bm v_F $ is the Fermi velocity in the direction of the wavevector, i.e. $\bm v_F  = v_F \bm k/k$. We also define $\delta \bm k = \bm k - \bm k_0$, where $\bm k_0 = k_F + \bm q  \bm v_F$ is gap-opening momentum which is shifted from the Fermi momentum due to the finite-momentum pairing. It is again determined by the condition $\varepsilon(\bm k_0) + \varepsilon(-\bm k_0 + 2 \bm q) - 2 \mu = 0$.
The one-dimensional version of this dispersion reads 
\be \label{eq:E_dop}
(1D) \quad E^\alpha_{\pm}(k) \approx \alpha  q   v_F \pm \sqrt{( v_F \delta k )^2 + |\Delta|^2 } .
\ee 
which is completely analogous to \eqref{eq:asymm1} if we associate $D \leftrightarrow  q v_F$.  In fact, both of these simple realizations of nonreciprocal superconductivity exhibit the same physics in the superconducting state. In what follows, we will use finite-momentum superconductors as a default example of nonreciprocal superconductivity in our considerations, however, the conclusions of this work apply broadly to nonreciprocal superconductors.

\subsection{Origin of the nonreciprocal $I-V$ characteristic in N-S junction}

While this paper discusses several related ways of probing nonreciprocal superconductivity in DC transport, the most drastic signature is the asymmetry of the conductance through a transparent N-S junction. 
In this section, we discuss the physical origin of the nonreciprocity of the $I-V$ characteristic (and consequently, the conductance). Let us use the explicit example of the transparent juntion between a normal metal and a nonreciprocal superconductor with gap asymmetry $D$ in one-dimensional geometry.  Nevertheless, our arguments can be naturally adapted for higher dimensions and for transparency $0<T<1$. 

Recall first the expression for the differential conductance of a fully transparent single-channel N-S junction with a conventional $s$-wave superconductor is~\cite{RevModPhys.69.731} at zero temperature:
\begin{equation} \label{eq:1D-jNS}
    G_{NS}^{T = 1}(\bm q=0, V) \propto 1 + |r_A (eV)|^2,
\end{equation}
where the superscript $T = 1$ specifies the junction transparency.  The Andreev reflection coefficient for incoming electrons and holes is $r_A(E) = e^{- i \arccos {\frac{E}{|\Delta|}}}$ when $|E|<|\Delta|$, and $r_A(E) = e^{-  \mathrm{arcosh} {\frac{|E|}{|\Delta|}}}$ otherwise. 
%
From conservation of quasiparticle currents upon Andreev reflection, a standard argument~\cite{RevModPhys.69.731} shows that the differential conductance of a transparent N-S interface must be symmetric under voltage reversal~\footnote{In more detail, the symmetry under the voltage reversal of the differential conductance of an N-S interface can be derived from the elastic scattering theory by employing unitarity of the reflection matrix (i.e. the incoming quasiparticle current is entirely reflected into the same lead) and the particle-hole conjugation constraint inherited from Nambu formalism of the Bogoliubov-de Gennes description of the superconducting state.}\footnote{In addition, above the spectral gap of the superconductor, the incoming electrons and holes can enter the superconductor as Bogoliubov quasiparticles. Above this voltage, the reflection matrix is not unitary, and the local conductance is generically not symmetric in voltage.}. 
 

When the quasiparticle dispersion and the superconducting gap are asymmetric, the conductance is generically nonreciprocal, as long as the current flows in an asymmetric direction. 
Crucially, for an N-S interface with a nonreciprocal superconductor, the window for the total Andreev reflection 
is shifted. Because the incoming and the Andreev reflected quasiparticles are on the same side of the Fermi surface, depending on whether the incoming particle in the normal metal is an electron or a hole, the window for the total Andreev reflection is shifted by $+D$ or $-D$.  
The result can be captured by the simple expression
\begin{equation} \label{eq:GNS1}
\begin{split}
    G_{NS}^{T = 1}(q, V)  &= \ G_{NS}^{T = 1}( q=0, V -  D/e) \\
    &\propto \ 1 + |r_A (eV -  D)|^2.
    \end{split}
\end{equation}
And the differential conductance is symmetric with respect to the voltage $V =  D/e$ as opposed to $V = 0$:
\begin{equation}
G_{NS}^{T = 1}( q, V) = G_{NS}^{T = 1}( q,  2  D/e - V ).
\end{equation}
The window for the total Andreev reflection is still realized inside the true gap of the system, i.e. for $|eV| < |\Delta| - D$. 
Inside this window, the differential conductance is symmetric with respect to voltage reversal. In contrast, for $|\Delta| - D < |eV| < |\Delta| + D$, total Andreev reflection occurs for one type of incident quasiparticles (an electron or a hole) but not the other. In this voltage range and above it, the $I-V$ characteristic is asymmetric. This effect has been missed in the previous work~\cite{PhysRevB.109.024504}.

At large bias voltage, the non-reciprocity of the conductance also implies a non-reciprocity of the excess current~\cite{PhysRevB.25.4515} (see Fig.~\ref{fig:schematic}(c)). At high junction transparency $T \approx 1$, the excess current $I_{\rm exc, \pm} \approx \frac{2 e^2}{\hbar} (|\Delta| \pm D)$ for $0 < D < |\Delta|$. The asymmetry is especially drastic when  $D > \Delta$, where we have $I_{\rm exc, +} \approx \frac{4 e^2}{\hbar} |\Delta|$, $I_{\rm exc, -} = 0$. The excess current is a signature of the nonreciprocal Andreev reflection that is observable even at large voltage bias $eV \gg \Delta$.

In the following sections, we elaborate on these results using Blonder-Klapwijk-Tinkham formalism~\cite{PhysRevB.25.4515} (see Sec.~\ref{sec:GNS}) and confirm them independently using {\it kwant} software~\cite{Groth2014Jun} simulations (see App.~\ref{sec:appendix_current}). The shift in the conductance has been previously remarked in the literature for the Fulde-Ferrel superconductors~\cite{Partyka_2010,Kaczmarczyk_2011}. This shift also appears in the proximitized 3D topological insulator surface states \cite{PhysRevB.92.205424} in the vicinity of a ferromagnet (which could be replaced by either Zeeman field or screening supercurrent).

Finally, we remark that for the nonreciprocal transport, both inversion and time-reversal symmetries have to be broken. Any of these symmetries require $E(\bm k) = E(- \bm k)$, which is incompatible with the asymmetry required by nonreciprocity. 
In addition, the current must have a component parallel to the polar vector (i.e. the vector determining the direction of the asymmetry) of the nonreciprocal superconductor (in example of the finite-momentum superconductivity, it is the Cooper pair momentum $\bm q$).

\subsection{Bogoliubov Fermi surface}

\begin{figure}[b] 
	\includegraphics[width= 1\columnwidth]{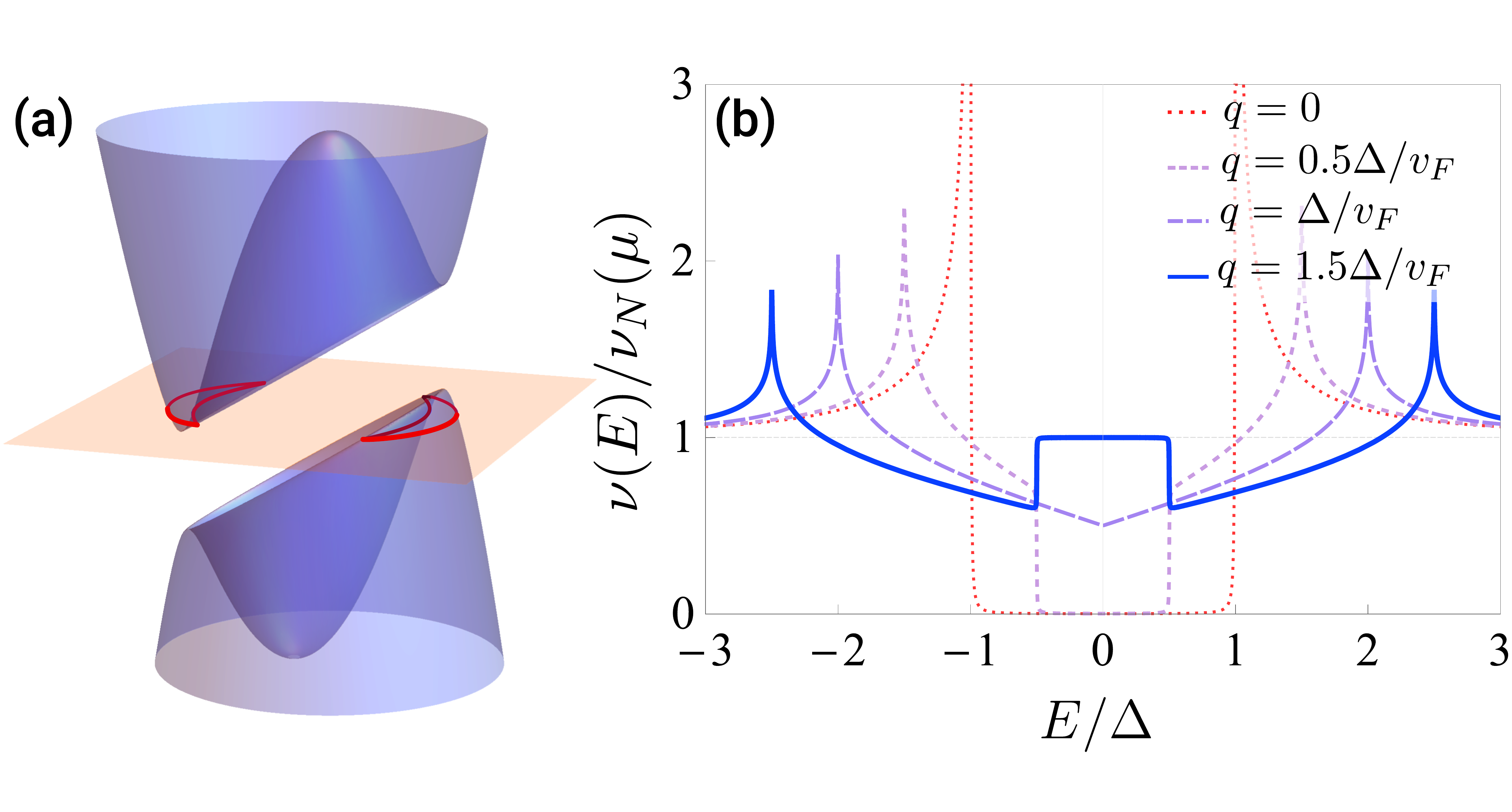}
	\caption{ (a) Schematic of the Bogoliubov Fermi surface (red contour) emerging from asymmetric quasiparticle dispersion in a two-dimensional nonreciprocal superconductor. The orange surface shows the Fermi level. We assume normal state parabolic dispersion here and  finite Cooper pair momentum $q  > |\Delta| / v_F$. (b) The density of states of a two-dimensional finite-momentum superconductor. The dashed lines indicate the positions of band extrema $\pm \left (|\Delta| \pm q v_F \right)$.  We included the broadening factor $\Gamma = 0.001 \Delta$ and set $\Delta = 0.01 \mu$.
	} \label{fig:dos}
\end{figure}

The regime where the gap asymmetry $D$ is close to or larger than $|\Delta|$ 
is the most interesting for nonreciprocal trunsport through an N-S junction because it gives rise to a large difference between the current in the forward and reverse direction, see Fig.~\ref{fig:schematic}(b,c). Moreover, at $D > |\Delta|$, the region of the doubled conductance (the window of perfect Andreev reflection) is shifted to strictly positive/negative voltages, as seen in Fig.~\ref{fig:schematic}(a).
When $D > |\Delta|$, the branch of the dispersion $E_+(\bm k)$ (see eq.~\eqref{eq:asymm1}) can become negative near the Fermi surface and the spectrum  becomes gapless, see Fig.~\ref{fig:dos}(a). The proximitized layer still remains superconducting if the gap in the parent superconductor is not closed. 
As a result, {\em Bogoliubov Fermi surface} \cite{yuan2018,papaj2021creating,PhysRevLett.128.107701,PhysRevB.92.205424} emerges around a segment of the normal-state Fermi surface in the proximitzed layer, as illustrated in Fig. \ref{fig:dos}(a). 
Bogoliubov Fermi surface in the surface of proximitized topological insulator occurring due to the Meissner current has been recently observed experimentally~\cite{zhu2020discovery}.

The quasiparticle density of states is also significantly changed in this regime and sharply reveals the signatures of the nonreciprocal superconductivity and the Bogoliubov Fermi surface. For illustration, we assume that the nonreciprocal superconductivity is caused by finite-momentum pairing such that $D = q  v_F$ and the direction of asymmetry is specified by the vector $\bm q$. The density of states in this case is shown in Fig.~\ref{fig:dos}(b) for two-dimensional geometry and is featured in the case of one-dimensional geometry in Fig.~\ref{fig:32}. Let us focus on the 2D geometry:
\be \label{eq:2Ddos}
\begin{split}
    \nu(E)&= - \frac{1}{\pi} \int \frac{dk}{2 \pi} \Im G_{\bm q}^R(E, \bm k)  \\
&=\nu_{N}(\mu) \int \frac{d 
    \theta}{2 \pi} \frac{ |E - q v_F \cos \theta|}{\sqrt{(E - q v_F \cos \theta)^2 - \Delta^2}}  \\
    &\times \Theta|(E -  q v_F \cos \theta| - \Delta),
\end{split}
\ee
where $\Theta(x)$ is the Heavyside step function, and $\nu_N (\mu)$ is the normal 2D density of states (DOS)  at the Fermi level. At finite Cooper pair momentum, the edges of the gap, producing a characteristic square-root divergence in the density of states, split into two extrema. In 2D, these correspond to the band minima at the edges of the true gap $E = \pm ||\Delta| - \bm q \bm v_F  |$ and the saddle points at $E = \pm ||\Delta| + \bm q \bm v_F  |$. The square-root divergence is replaced with a weaker logarithmic divergence at the saddle points and approaches a constant at band minima. When $q v_F > |\Delta|$, the density of states near zero energy approaches that of the normal state thanks to both quasiparticle bands overlapping in energy in this region. This produces behavior resembling an in-gap peak.

In Sec.~\ref{sec:tunneling-stm}, we confirm that scanning tunneling microscopy measurements in a bulk region of a nonreciprocal superconductor can reveal the quasiparticle density of states. At the same time, we demonstrate an exceptional sensitivity of such measurements to the presence of edges in such superconductors, in which case the measurement outcome will be drastically modified.

\subsection{Conductance of a normal - nonreciprocal superconductor junction}%
\label{sec:GNS}

Let us now discuss the results for the differential conductance for junctions between a normal metal and a nonreciprocal superconductor superconductor in one- and two-dimensional geometries. We use an example of nonreciprocity caused by the finite-momentum pairing ($D = q  v_F$ and the polar vector determined by $\bm q$), though the conclusions apply to the general case of a nonreciprocal superconductor, are as we discussed earlier.  From symmetry considerations, only the component of the finite Cooper pair momentum $\bm q$  perpendicular to the junction (and parallel to the current direction) contributes to the nonreciprocity. To focus on this effect, we assume that  $\bm q$ is always perpendicular to the junction.  The effect of finite-momentum Cooper pairing with momentum parallel to the junction, in which case there is no nonreciprocity, is discussed in Appendix~\ref{sec:appendix_current}. It has also been addressed in the literature in the context of supercurrent-carrying~\cite{PhysRevB.70.172508,PhysRevB.76.144508} and finite-momentum~\cite{PhysRevB.73.214514} superconductors.

\begin{figure}[!t] 
	\includegraphics[width= 1\columnwidth]{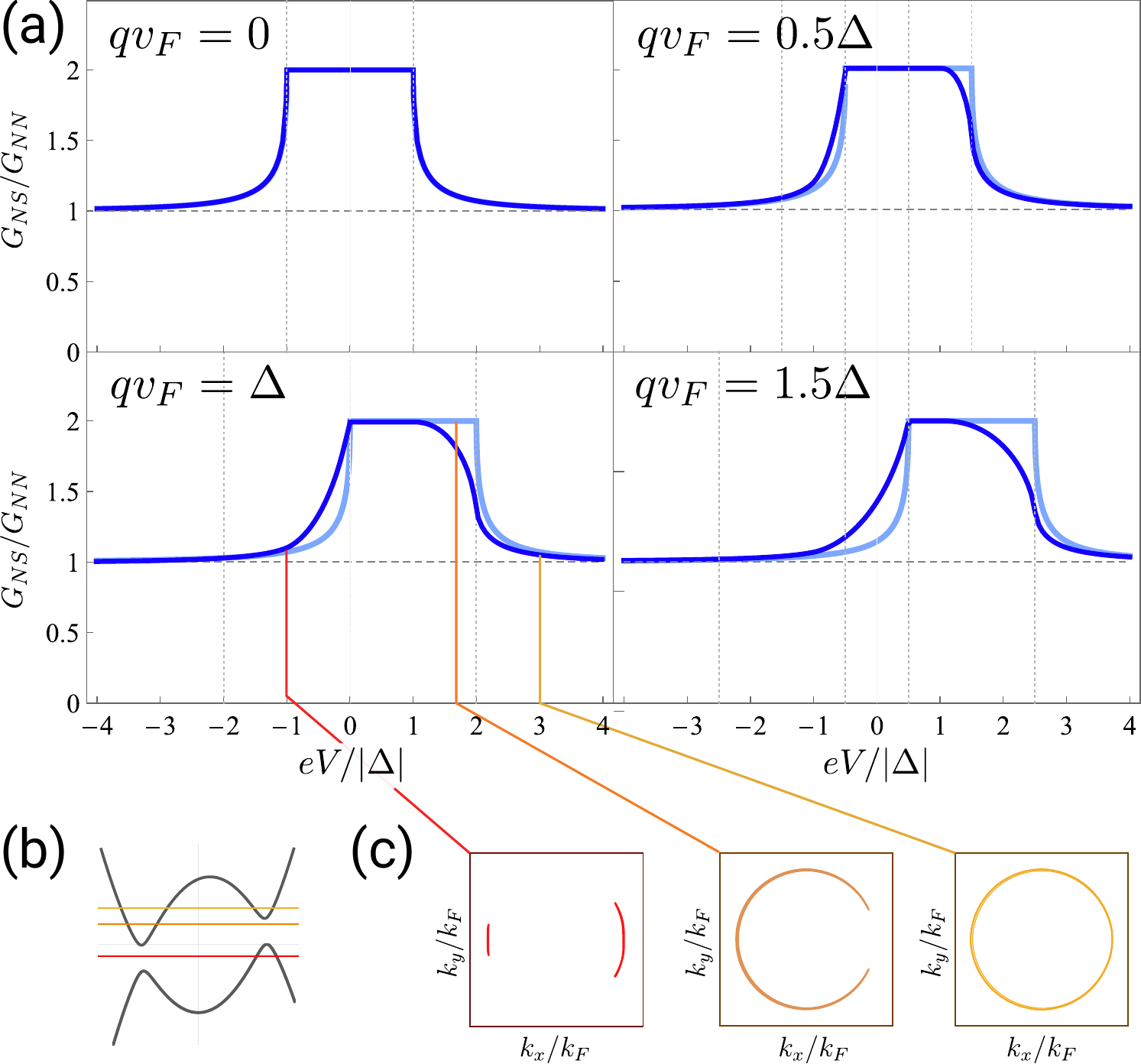}
	\caption{(a) Differential conductance of a junction between a normal metal and a  nonreciprocal superconductor in one and two dimensions is shown by light blue and blue lines, respectively. The nonreciprocity is quantified the Cooper pair momentum $q$ in the superconductor, directed perpendicularly to the junction. The vertical dashed lines show the energies $\pm \Delta \pm q v_F$. Cuts of the (b) one-dimensional dispersion and (c)  two-dimensional quasiparticle dispersion at several values $E = eV$ and fixed $q v_F = \Delta$. We assumed $\Delta = 0.01 \mu$.
	} \label{fig:3}
\end{figure}

\begin{figure}[!t] 
	\includegraphics[width= 1\columnwidth]{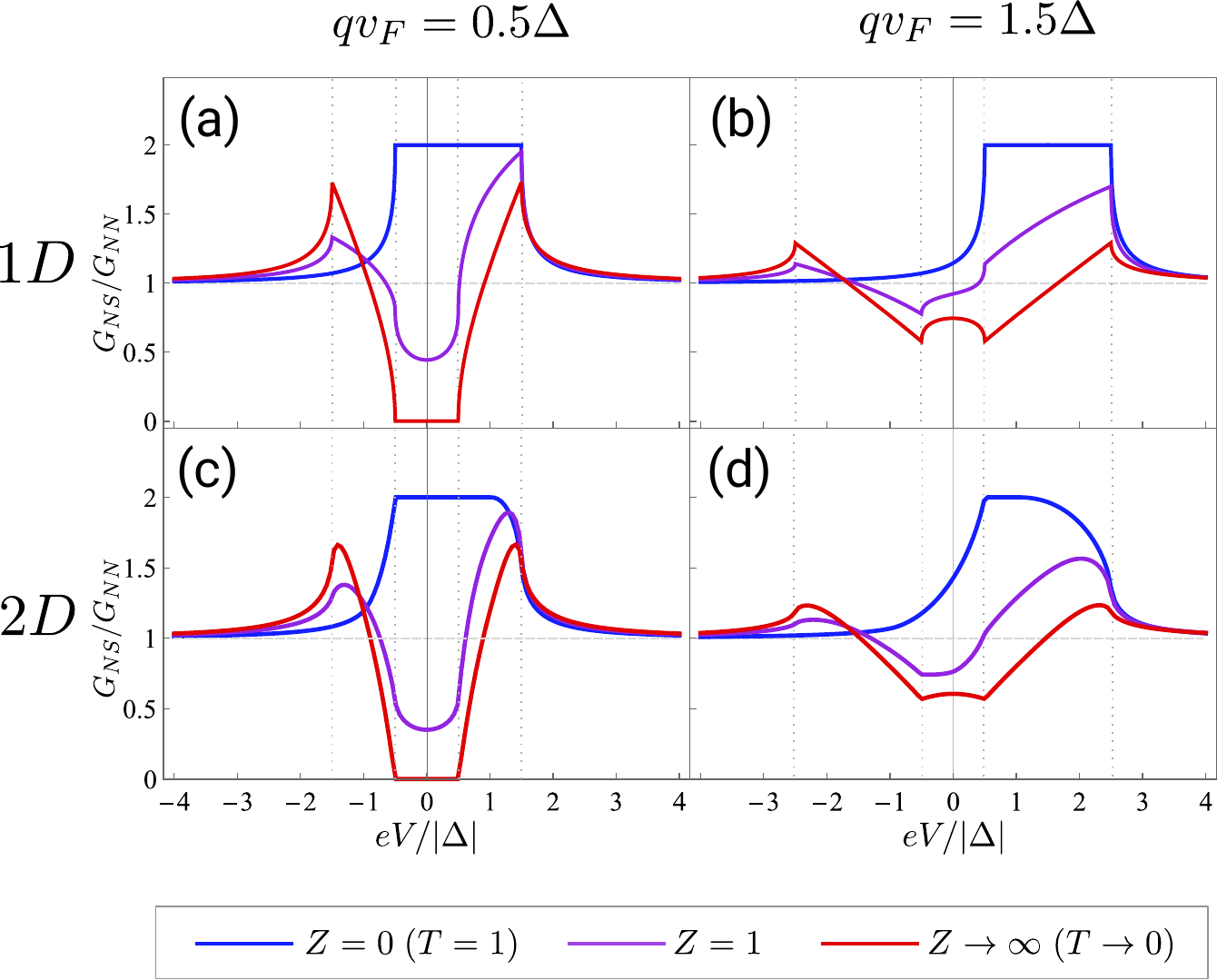}
	\caption{Differential conductance for a junction between a normal metal and nonreciprocal superconductor with finite Cooper pair momentum $q$ at different values of junction transparency computed for (a,b) one- and (c,d) two-dimensional geometries. Junction transparency is parameterized by the dimensional barrier strength $Z = \frac{m \lambda}{k_F}$, where $\lambda$ is the coefficient of the delta-function potential modeling a barrier at the junction. Blue, purple and red lines show $Z = 0$ ($T = 1$), $Z = 1$ and $Z = 20$, respectively. Intermediate barrier strength corresponds to intermediate transparencies $T_{1D}  = 0.5$  and $T_{2D} \approx 0.4$ for the plant junction, while  $Z = 20$ corresponds to the tunneling limit where $T_{1D, 2D} \approx 2 \times 10^{-3}$ (labeled as $Z \rightarrow \infty$, $T \rightarrow 0$ in the legend). } \label{fig:31}
\end{figure}

To calculate the differential conductance, we extend  the quasiclassical approach of Blonder, Tinkham, and Klapwijk~\cite{PhysRevB.25.4515}. The junction is captured by a spatially varying Hamiltonian obtained from \eqref{eq:BdG_matrix} by setting $\Delta(x) = \Delta \Theta(-x)$ (where $\Theta(x)$ is the Heavyside step function) and finite transparency is modeled adding a delta-function barrier $\lambda \delta(x) \tau_z$ at the boundary. For $x<0$, we have $\Delta = 0$ and $\bm q = 0$. 
The differential conductance $G_{NS}(\bm q, V) = \frac{\partial I }{\partial V}|$ can be found by evaluating quasiclassical expression for the current in the normal region on the far left from the junction (the details are provided in Appendix~\ref{sec:appendix_current}.   In 2D, the problem factors for each direction of the incoming quasiparticle, which we parametrize by the incidence angle $\theta$, and one finds:
\be \label{eq:cond} 
\begin{split}
\frac{G_{NS}(\bm q, v)}{G_{NN}(v)} = 
\frac{  \int \left ( 1- |r_{ee} (eV,\theta) |^2 + |r_{eh}(eV,\theta)|^2 \right ) \cos \theta d \theta} {\int \left ( 1- |r_{ee}^N (eV,\theta) |^2  \right ) \cos \theta d \theta},\ 
\end{split}
\ee
where the integration is only over particles incoming from one direction, i.e. over the domain $\left [ - \pi/2, \pi/2\right]$. $G_{NN}(eV)$ is the normal-state conductance,  $r_{ee}$ is the electron-electron reflection coefficient, $r_{eh}$ is the coefficient of reflection of a hole into an electron, $r_{ee}^N = \frac{Z}{\sqrt{Z^2 + \cos^2 \theta}}$ is the reflection coefficient of the same junction in the normal state,  and $Z = \frac{m \lambda}{k_F}$ is the dimensionless barrier strength. The expression for the differential conductance of a one-dimensional junction is obtained by removing integration and setting $\theta=0$. 

The differential conductance as a function of voltage is shown in Figs.~\ref{fig:3} and \ref{fig:31} for different values of the Cooper pair momentum (in the direction perpendicular to the junction) both for one- and two-dimensional junctions. Fig.~\ref{fig:3} corresponds to the fully transparent case, $T = 1$. For the one-dimensional junction, the conductance is simply shifted by $q v_F$ according to eq.~\eqref{eq:1D-jNS}. For the 2D geometry, because of the $\cos \theta$ factor, the contribution from quasiparticles whose wavevector is parallel to the current is weighted more than that of those that are at an angle. The region where the conductance is doubled due to total Andreev reflection  is narrowed to the range $-\Delta + q v_F < eV < \Delta$, where the dispersion for forward-moving electrons is fully gapped. 
Physically, the nonreciprocity in highly transparent junctions is possible because only one (positive- or negative-momentum region) of the quasiparticle dispersion is probed, which reveals its energy shift.

\begin{figure}[!t] 
	\includegraphics[width= 0.8\columnwidth]{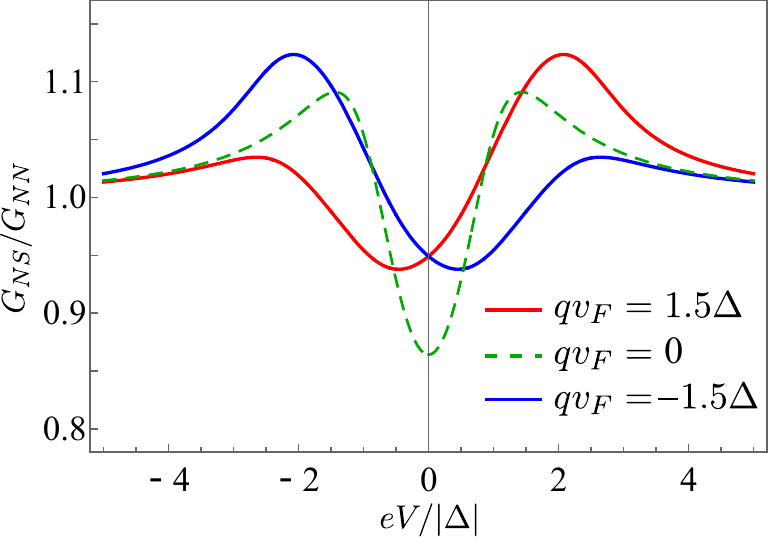}
	\caption{Conductance of a one-dimensional normal metal-nonreciprocal superconductor junction computed using BTK approach at transparency $T = 0.5$ with a  large broadening factor $\Gamma = 0.75 |\Delta|$. For concreteness, the nonreciprocity is quantified by the Cooper pair momentum in the superconductor. This is a simple model of a situation where an STM tip is located close to or to the side of an edge of a superconductor and operates in a partially transparent limit.  } \label{fig:stm_partial}
\end{figure}

For reduced junction transparency ($T < 1$), 
the interface barrier repeatedly reflects the Andreev-reflected quasiparticles back into the superconductor. As a consequence, the  quasiparticles incoming at the superconductor in total are both electrons and holes. As forward-moving electrons (holes) probe the quasiparticle dispersion at positive (negative) momenta, the spectrum of the superconductor around both positive and negative $k_F$ (which is shifted in the opposite directions in energy) is probed as a result. Thus, the N-S conductance becomes increasingly more symmetric as the interface transparency is reduced. 

Fig.~\ref{fig:31} shows the differential conductance for one- and two-dimensional junctions at finite junction transparency controlled by the dimensionless barrier strength $Z$.  In the one-dimensional case, the junction transparency is $T = \frac{1}{1 + Z^2}$, while in 2D, $T = \frac{1}{2} \int_{-\pi/2}^{\pi/2} \left ( 1 - \frac{Z^2}{Z^2 + \cos \theta^2}\right) \cos \theta d \theta$.
At moderate values of the barrier transparency, e.g. $Z = 1$ shown in Fig.~\ref{fig:31} (which corresponds to $T = 0.5$ in 1D and $T \approx 0.4$ in 2D case, respectively), the conductance is still significantly asymmetric but its features interpolate between the limits of transparent and tunneling junction.

In the limit $T \rightarrow 0$ the symmetry with respect to the voltage reversal becomes completely  restored. 
In ordinary superconductors, the conductance in the tunneling limit approaches the bulk quasiparticle density of states in the superdonductor, which is not the case when $\bm q$ is non-zero. We discuss this in the next section.


Finally, we remark that moderate barrier transparency can potentially be a reason for sometimes observed current-voltage asymmetry in scanning tunneling microscopy measurements. An example of such asymmetry was observed in UTe$_2$ whose microscopic origin was not understood~\cite{Jiao2020Mar}. As we show in Fig.~\ref{fig:stm_partial}, such transport characteristics can occur in the case of nonreciprocal superconductivity, for example, finite momentum pairing, arising under the circumstances discussed in this paper. This possible explanation relies on time-reversal symmetry breaking but does not invoke topological superconductivity or topological edge modes.

\section{Additional probes of nonreciprocal superconductivity}

\subsection{Tunneling conductance measurements} \label{sec:tunneling-stm}

\begin{figure}[!t] 
	\includegraphics[width= 1\columnwidth]{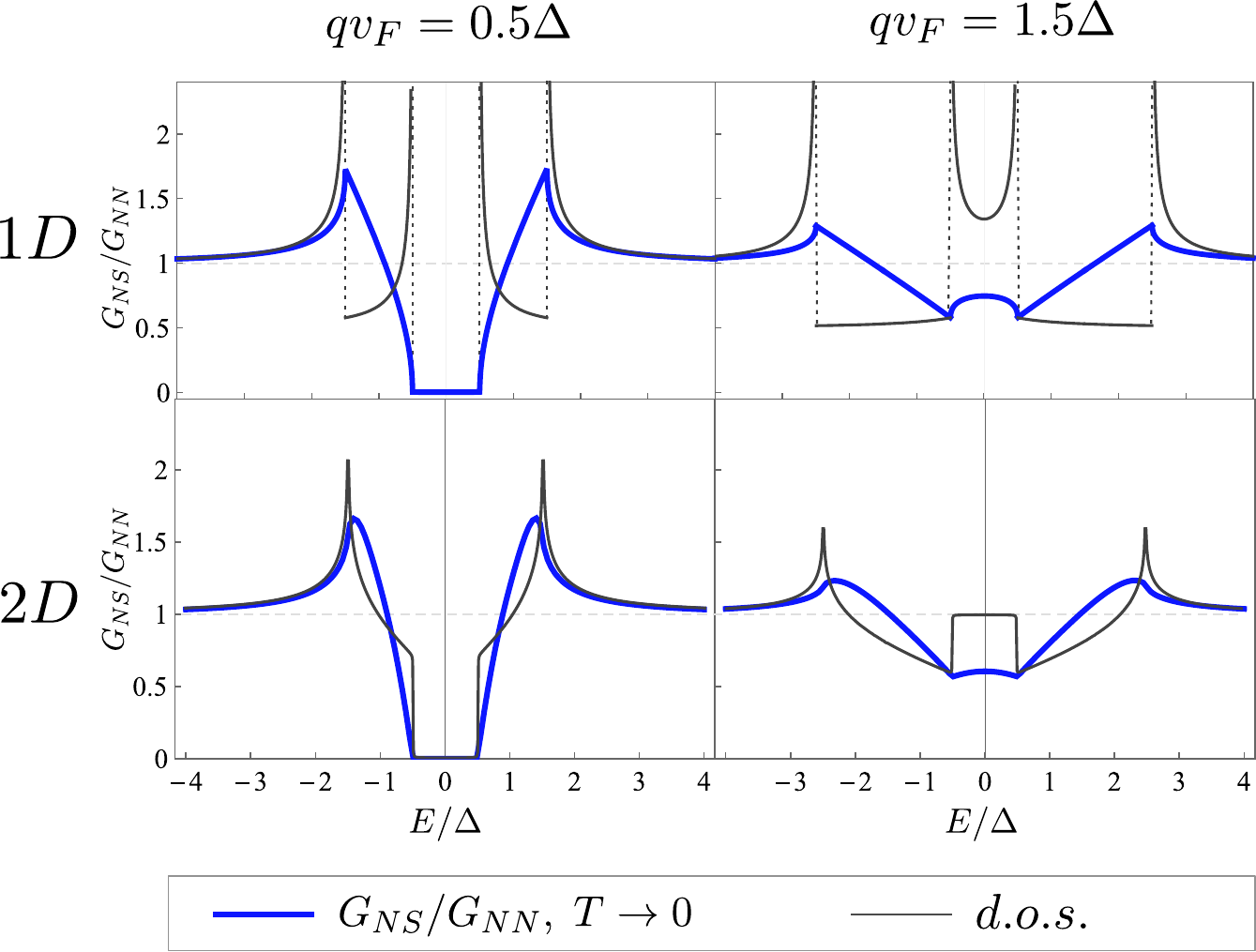}
	\caption{Tunneling-limit differential conductance at the edge of a nonreciprocal superconductor occupying half-space is shown by blue lines in (a,b) one and (c,d) two dimensions. The conductance was computed using the BTK approach described in sec.~\ref{sec:GNS} in the $T \rightarrow 0$ limit. This result coincides with the differential conductance at the edge of a nonreciprocal superconductor obtained using the transfer Hamiltonian method, eq.~\eqref{eq:GNS_boundary}. The latter approach describes local tunneling from an STM tip located at the edge of a superconductor. The bulk quasiparticle density of states (d.o.s)  at respective values of the Cooper pair momentum, which quantifies the degree of nonreciprocity, is plotted in black.  } \label{fig:32}
\end{figure}

Differential conductance measurements in scanning tunneling microscopy (STM) as well as normal-superconducting junctions in the tunneling regime are both excellent tools for studying spectral features of the superconductor (see~[\onlinecite{Fischer2007Mar}] and references therein).   STM experiments are usually designed such that a normal tip introduces only a small perturbation to the superconductor, and can be used to probe different regions locally. In contrast, because of its geometrical constraint, an in-plane tunneling N-S junction always probes a superconductor with an edge. 
The differential conductance measured this way can be used to determine the quasiparticle density of states, which is usually the same as the bulk quasiparticle density of states. Even when there is a difference, it usually occurs due to the edge states which lead to additional narrow peaks in energy and do not alter the global picture~\cite{Fischer2007Mar}.
Our results demonstrate that the nonreciprocal superconductors are a notable exception to this rule. There, the presence of an edge leads to a significant change in the quasiparticle wavefunctions and the quasiparticle density of states even far from the edge (i.e. at distances of a few superconducting coherence lengths). Because of this, the tunneling measurements in nonreciprocal superconductors can show vastly different results from the bulk quasiparticle density of states. In contrast, this effect doesn't take place in regular superconductors. The difference is especially drastic near the edges, as seen in Fig.~\ref{fig:32}.

The transport in normal-superconductor structures in the tunneling regime can be described using the transfer Hamiltonian approach~\cite{RevModPhys.36.200,PhysRevLett.6.57,PhysRevLett.9.147,PhysRevLett.8.316,PhysRevB.6.1747}, where the Hamiltonian is split into three parts
\be
H = H_S + H_N + H_T,
\ee
where $H_{N,S}$ describes isolated normal(superconducting) region, and $H_T$ is the transfer term coupling the regions by transporting a single electron:
\be
H_T = \sum_{\lambda \rho} t_{\lambda \eta} c^\dagger_{\lambda} b_{\eta} + h.c.
\ee
where $\lambda,\rho$ label single-particle modes coupled by tunneling, $c^\dagger_\lambda$($b^\dagger_\eta)$ creates an electron in state $\lambda$($\eta$) in the superconducting and normal regions, respectively, and $t_{\lambda \eta}$ is the tunneling matrix element. The concrete assumptions about $H_S$, $H_N$, and the tunneling matrix element depend on the geometry of the problem.

First, consider an ideal STM setup with point-like contact, in which case the tunneling occurs locally, and we assume that the tunneling modes are labeled by some coordinate $\bm  r$, giving $t_{\bm r\bm r'} = t \delta_{\bm  r \bm  r'}$. Let us assume that the tunneling is spin-independent, in which we can focus on one of the two spin copies.  Let $\gamma_{n,i}^\dagger$ be the quasiparticle creation operators for the eigenmodes of the superconducting Hamiltonian $H_S$ (assuming quasiparticle operators with well-defined electronic charge~\cite{PhysRevB.6.1747}), where $i = 1,2$ is an index enumerating the electron- and hole-like solutions that come in pairs. Let us expand the electron creation operator as 
\be \label{eq:c_through_gamma}
c^\dagger_{\bm  r} = \sum \varphi_{n,1}(\bm r) \gamma_{n,1}^{\dagger} + \varphi^*_{n,2}(\bm r) \gamma_{n,2},
\ee
where  $\varphi_{n,i}$ is the electron component of the single-particle wavefunction solving $H_S \varphi_{n,i} = E_n \varphi_{n,i} $, and $i$ labels electron-like and hole-like solutions.  For example, for an infinite superconductor the usual notation would be  $n \rightarrow \bm k$, $\varphi_{\bm k ,1}(\bm r) = e^{i \bm k \bm r} u_{\bm k}$, and  $\varphi_{\bm k ,2}(\bm r) = e^{-i \bm k \bm r} v_{-\bm k}$, where $u_{\bm k}$ and $v_{\bm k}$ are the coherence factors obtained by diagonalizing the BdG Hamiltonian $H_S$, see Appendix~\ref{app:BdG}. We note that for a finite-momentum superconductor, the pair of solutions occur at $\bm k$ and $-\bm k + 2\bm q$. Whenever the latter index occurs, we bookkeep it as $-\bm k$ for simplicity.

We use the Fermi golden rule approach to derive the tunneling conductance in this section (whose result coincides with the linear response theory summarized in Appendix~\ref{app:tunneling}). The current is determined by the tunneling rate from the normal metal to the superconductor, which can be found perturbatively:
\be
\Gamma_{i \rightarrow f}  = 2 \pi  |\langle f |H_T|i \rangle|^2 \delta(E_f - E_i - eV)
\ee
where $E_{f,i}$ are the energies of the initial and final states $\ket{i} = b^\dagger \ket{0_S} \ket{0_N}$ and $\ket{f} = \gamma^\dagger_{f} \ket{0_S} \ket{0_N}$, respectively, counted from the Fermi energy, and $\ket{0_{N,S}}$ are the ground states of the normal metal and superconducting regions. We label the final states by the quasiparticle mode that the tunneling occurs into.  Using eq.~\eqref{eq:c_through_gamma}, we find the matrix element to be $\langle f |H_T|i \rangle = t \varphi_{f}(\bm r)$. We can find the current as:
\be \label{eq:current_2} \nonumber
\begin{split}
I &\propto e |t|^2  \sum_{i,f} |\varphi_f(\bm r)|^2  \delta(E_f - E_i - eV) \left ( f_0(E_i) - f_0(E_f)\right),
\end{split}
\ee
where  $f_0$ is the Fermi-Dirac distribution. Assuming zero temperature and energy-independent density of states in the normal region, the differential conductance reads:
\be \label{eq:current_3}
\begin{split}
\frac{\partial I}{\partial V}  &\propto e |t|^2   \sum_{f: E_f= eV} |\varphi_f(\bm r)|^2  \nu_f(E),
\end{split}
\ee

Consider the example of nonreciprocal superconductivity associated with the finite-momentum pairing. In the case of an infinite superconductor, the eigenstates are enumerated by the wavevector $\bm k$ of the plane-wave solutions.  For each solution $\gamma^\dagger_{\bm k, i}$, we can define the momentum-resolved density of states $\nu_{\bm k}(E)  = -1/\pi \mathrm{Im}G^R_{\gamma}(E,\bm k)$.
\be
\begin{split}
\frac{\partial I}{\partial V} \propto \sum_{\bm k: E(\bm k) = eV} \left ( |u_{\bm k}|^2 \nu_{\bm k}(E) + |v_{\bm - \bm k}|^2\nu_{-\bm k}(E) \right)
\end{split}
\ee
We note that the densities of states $\nu_{\bm k}(E)$ and $\nu_{-\bm k}(E)$ are not equal because of the dispersion asymmetry. However, because of the summation over both negative and positive momenta, we use $|u_{\bm k}|^2 + |v_{\bm k}|^2 = 1$, and the expression becomes $\frac{\partial I}{\partial V} \propto \sum\left (\nu_{\bm k}(eV) + \nu_{-\bm k}(eV) \right) = \nu(eV)$. This recovers the expected result for scanning tunneling microscopy, namely the symmetric bulk quasiparticle density of states of the finite-momentum superconductor $\nu(eV)$ given in eq.~\eqref{eq:2Ddos}. 

The situation is different when the nonreciprocal superconductor occupies only half-space. Consider a momentum-$\bm q$ superconductor in one-dimensional geometry for simplicity. The quasiparticle eigenstates in this case are linear combinations of incoming and reflected waves. One of the solutions $\varphi_{ k,1}( r)$ comes from a combination of an electron-like quasiparticle state incoming at a barrier and reflected electron- and hole-like states. At distances from the edge less than coherence length $r < \xi$, the result can be simplified if we neglect the difference between momenta of electron-like and hole-like states (see Appendix~\ref{app:tunneling} for the general solution). In this approximation, we obtain:
\be \label{eq:boundary0}
\varphi_{1}( r) = u_{+k}(E) \frac{\left (u_{-k}(E)\right)^2 - \left (v_{-k}(E)\right)^2 }{u_{+k}(E)u_{-k}(E) - v_{+k}(E)v_{-k}(E)} \sin  k  r.
\ee
The other solution in the pair, $\varphi_{2}(\bm r)$, is formed by a hole-like quasiparticle coming at the barrier, and can be obtained from eq.~\eqref{eq:boundary0} by simply replacing $u \leftrightarrow v$ and $+ \leftrightarrow -$. In fact, for $k$ near the Fermi momentum, $u_{\pm k} (E) = u(E \mp q  v_F)$ and similarly for $v_{\pm k}$, and $u$ and $v$ are the coherence factors of a usual $s$-wave superconductor.

The densities of states corresponding to the contributions to the conductance from the two states are the same as those for the electron-like and hole-like quasiparticles coming at the barrier and are $\nu_{\mp k}(E)$. For asymmetric dispersion, these densities of states are shifted $\nu_{\mp k}(E) = \nu_0(E \pm q v_F)$, where $\nu_0(E)$ is the usual zero-momentum superconductor density of states. Finally, we obtain:
\be \label{eq:GNS_boundary}
\begin{split}
\frac{\partial I}{\partial V} \propto &\left [ |\varphi_{1}(\bm r)|^2 \nu_{0}(eV+ q v_F) \right .
+ \left . |\varphi_{2}(\bm r)|^2 \nu_{0}(eV - q v_F)\right ],
\end{split}
\ee
where we substitute $E = eV$ in the wavefunction weights. In fact, this is the same answer as the differential conductance of an N-S junction in the tunneling limit obtained from the BTK formalism. Unlike in the case of zero-momentum superconductors, this result is quite different from the bulk quasiparticle density of states, as shown in Fig.~\ref{fig:32}. 
The features of the resulting conductance, shown in Fig.~\ref{fig:32}, are dominated by the wavefunction weights \eqref{eq:boundary0} as opposed to the density of states.
The difference is especially drastic in the 1D geometry~\footnote{We also find a curious coincidence in that the BTK and transfer Hamiltonian-approach results for the 1D junction yield the same analytical expression as the bulk density of states of a finite-momentum three-dimensional superconductor. For the latter, see ref.~\cite{PhysRev.137.A783} and Appendix~\ref{app:tunneling}}.

The approximation above works relatively well for the N-S junction geometry as well as the conductance measured with an STM tip located within a coherence length distance $\xi$ from an edge of a nonreciprocal superconductor. Away from the edge, the difference between the wavevectors of the electron-like and hole-like states entering the eigenstates of $H_S$ cannot be neglected anymore. Nevertheless, the results approach the bulk quasiparticle density of states only at distances $r \gg \xi$. This situation is very different from the case of reciprocal superconductors.
The full result is provided in Appendix~\ref{app:tunneling}.

As a final remark, let us discuss a few other cases when STM measurements show nonreciprocal conductance. In the presence of an edge,  there is always an effect that comes from electron and hole weights in the states inside the superconductor that oscillate as $\sin \left ( k_F \pm  \frac{E}{v_F}\right )r$. This effect is larger at small distances from the edge and is only observed when measurements do not average over Fermi wavelength, which requires nearly-atomic resolution.

Nonreciprocal conductance is also frequently observed when tunneling occurs through magnetic impurities hosting localized Yu-Shiba-Rusinov states~\cite{Yazdani1997Mar,Heinrich2018Feb}. This is a common scenario in situations where the normal-state conductance between the tip and the sample is several orders of magnitudes below the conductance quantum.  
This asymmetry arises 
only in the presence of additional relaxation channels allowing a quasiparticle entering the superconductors to decay before it can Andreev reflect and exit as its conjugate \cite{Martin2014Sep}. 
Then, the conductance for incoming electrons (holes) is sensitive only to the electron (hole) part of the Yu-Shiba-Rusinov bound state wavefunction, which generically is asymmetric in energy. This situation is very distinct from the other sources of asymmetry discussed above.

\subsection{Tomography of the quasiparticle dispersion}%
\label{sec:collimated}
Normal metal-superconductor interfaces in planar geometries can be used to probe the quasiparticle dispersion in nonreciprocal superconductors with resolution in direction and energy, which we call tomography of the quasiparticle dispersion. In particular, this method can be used as an independent tool for studying the Bogoliubov Fermi surface.

\begin{figure}[!t] 	\includegraphics[width= 0.85\columnwidth]{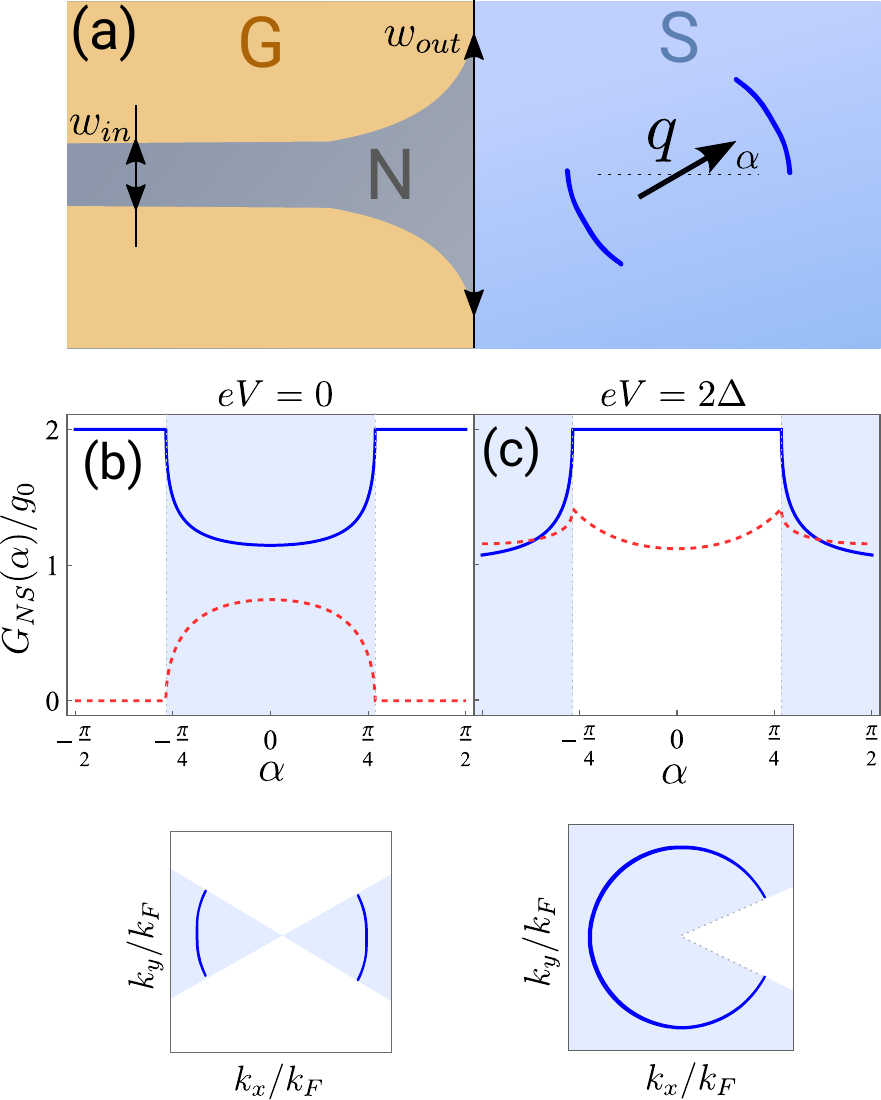}
	\caption{ Tomography of the asymmetric quasiparticle dispersion using collimation and rotating magnetic field. (a)  Adiabatic horn-shaped constriction is achieved by gating (G) that smoothly broadens from $w_{in}$ a $w_{out}$ at the junction, collimating the incoming electrons.   Differential conductance measurement reveals (b) the Bogoliubov Fermi surface at $q > \Delta/v_F$ (i.e. quasiparticle dispersion at the Fermi level) both in transparent (blue line) and in the tunneling (red dashed line) regimes as a function of the direction of the external in-plane magnetic field that controls the direction of the Cooper pair momentum $2\bm q$ that is perpendicular to it. In both cases, when the incoming quasiparticles are in the direction of the gaplessness of the quasiparticle dispersion, the differential conductance is close to that in the normal state. The range of angles $\alpha$ (measured between the direction of incidence and the Cooper pair momentum $\bm q$, as shown in panel (a))  corresponding to the gapless regions in the quasiparticle dispersion are shown in blue. The inset below shows the quasiparticle dispersion at $E = 0$, counted from the Fermi level. Panel (c) shows the differential conductance at finite voltage in transparent and tunneling regimes at finite voltage bias $e V = 2 \Delta$, which probes the quasiparticle dispersion at finite energy $E = 2 \Delta$ shown in the inset below.  The plots were obtained for $q= 1.5 \Delta/v_F$, $\Delta = 0.01 \mu$, and $g_0$ is the normalization by the number of modes times a quantum of conductance per mode.  } \label{fig:4}
\end{figure}

This is achieved by measuring the differential conductance in a structure where the electrons coming from the normal region are collimated to a narrow range of angles of incidence, which can be done by creating an adiabatic narrow constriction~\cite{PhysRevLett.125.107701,PhysRevB.101.165429}, or by setting up several constrictions separated in space from each other~\cite{PhysRevB.59.10176}. 
Consider an adiabatic constriction with slowly changing width $w(x)$ as shown Fig.~\ref{fig:4}. The constriction starts as a narrow few-mode channel $w_{in}$ ($n_0 = \frac{k_F w_{in}}{\pi}$) and gradually opens up to $w_{out} \gg w_{in}$ at the interface with the nonreciprocal superconductor.  For the $n$-th mode, we can solve a quasiclassical problem with a slowly varying potential  $E_n(x) = \frac{\pi^2  n^2}{2 m w^2 (x)}$, that varies with coordinate $x$ perpendicular to the junction, where $k_y w(x)$ serves as an adiabatic invariant. The modes with $n < n_0$ are reflectionless with exponential accuracy \cite{glazman1988reflectionless,PhysRevB.41.5341}, and the modes with $n> n_0$ are classically forbidden.  This leads to the narrow angular distribution of incoming quasiparticles $f(\theta) = \cos \theta \, T(\theta)$ with
\be
T(\theta) = \left\{\begin{matrix}
1, \ |\theta| \le  \frac{\pi  n_{max} }{k_F w_{out}} \approx  \frac{  w_{in} }{ w_{out}} \ll 1
\\ 
0 \ \ \ \ \   \text{otherwise}
\end{matrix}\right.
\ee
Consider the case where the nonreciprocal superconductivity comes from finite-momentum pairing for concreteness. When $q > \Delta/v_F$, the presence of the Bogoliubov Fermi surface shows as a feature in the conductance at $V = 0$. One can rotate quasiparticle dispersion and, consequently, the Fermi surface by rotating the direction of the in-plane magnetic field that induces finite-momentum pairing $\bm q$. The resulting differential conductance is shown in Fig.~\ref{fig:4}(b) as a function of angle $\alpha$ between the Cooper pair momentum $\bm q$ and the perpendicular to the junction, revealing the regions where the spectrum is fully gapped in the direction of the current.  Similarly, by varying the voltage in the normal lead, one can probe quasiparticle dispersion at different enegries, as shown for $e V = 2 \Delta$ in Fig.~\ref{fig:4}(c).

For probing intrinsic nonreciprocity that cannot be easily controlled by an external knob such as magnetic field, one could use several junctions with different orientations  on one device, or a multiple-constriction setup allowing to change the angle of the incoming collimated beam \cite{PhysRevB.59.10176}, which allows to probe several directions in the quasiparticle dispersion.

\section{Conclusions}

In this paper, we introduced the concept of nonreciprocal superconductors as a class of superconducting states that break both time-reversal and inversion symmetries. 
It is distinct from other classifications of unconventional superconductivity. 
In particular,  chiral superconductors~\cite{Kallin2016Apr}  break time-reversal symmetry but may or may not break inversion symmetry. Thus, nonreciprocal and chiral superconducting classes are not mutually exclusive but are nonetheless distinct.

Our work has identified universal signatures of nonreciprocal superconductivity in DC transport. 
One of them is nonreciprocal Andreev reflection occurring in transparent N-S junctions with nonreciprocal superconductors. This leads to asymmetric current-voltage characteristic which directly reveals the asymmetry of the quasiparticle dispersion of the nonreciprocal superconductor. In comparison to the usual superconducting diode effects, this nonreciprocal transport regime does not rely on reaching the critical current in the superconductor, thus, avoiding some of the typical complications such as heating.

The tunneling transport is another regime where nonreciprocal superconductivity exhibits distinct features. In the vicinity of an edge of a superconductor, the electronic part of the quasiparticle density of states in a finite-momentum superconductor undergoes a significant change that can be observed up to distances from the edge of the order of the coherence length $\xi$. This has crucial consequences for interpreting conductance measurements in scanning tunneling microscopy and tunneling junctions involving nonreciprocal superconductors.

Observation of the unusual transport behavior studied in this paper in the context of unconventional superconductors, such as 
UTe$_2$~\cite{Ran2019Aug,Aoki2019Mar,Aoki2022Apr}, iron pnictides~\cite{Fernandes2022Jan}, and multilayer graphene ~\cite{diez2021magnetic,lin2022zero,chiles2023nonreciprocal} could significantly improve our understanding of these systems.

%

\textbf{Acknowledgements.}
MD is grateful to Vlad Kurilovich and Pavel Kurilovich for helpful discussions. We thank Vidya Madhavan and Long Ju for useful discussions on experiments. 
This work is supported by a Simons Investigator Award from the Simons Foundation. 
MG acknowledges support from the German Research Foundation under the Walter Benjamin program (Grant Agreement No. 526129603).

\appendix
\section*{Appendices}

\section{The scattering problem}

\subsection{The BdG equation in the presence of the finite Cooper pair momentum } \label{app:BdG}

In the following will focus on one-dimensional geometry for simplicity, and will discuss the extension to the two-dimensional case later on.  The BdG equation for an $s$-wave superconductor with spin-degenerate bands and a finite Cooper pair momentum $2q$ is given in eq.~\eqref{eq:H_int} ~\cite{yuan2021supercurrent}. 
The BdG Hamiltonian bilinear in operators $(\psi_{k \uparrow},\psi_{k \downarrow},\psi^\dagger_{-k + 2 q \uparrow},\psi^\dagger_{-k +   2q \downarrow} )^T$ is
\begin{equation}
     H_{BdG} =\begin{pmatrix}
H(k) & - i \sigma_y \Delta \\
i \sigma_y \Delta^* & - H^*(-k + 2q) \\
\end{pmatrix}
\end{equation}
where in the main text we assumed absence of spin-orbit coupling leading to $H(k) = (\varepsilon(k) - \mu) \mathbb I$, where $\mathbb I$ is the identity matrix in the spin sector. More explicitly, we can write:
{
\begin{equation*} \label{eq_H_app}
     H_{BdG} =\left (\begin{smallmatrix}
\varepsilon(k)-\mu &  0& 0 &  \Delta\\
0 & \varepsilon(k)-\mu & -\Delta & 0 \\
0 & -\Delta^* & -\left ( \varepsilon(-k+2q)-\mu \right ) & 0\\
\Delta^* & 0 & 0 & -\left ( \varepsilon(-k+2 q)-\mu \right )  \\
\end{smallmatrix} \right)
\end{equation*}
}
In what follows, we will focus on a `spinless' case
by considering the $\Psi = (\psi_{k \uparrow},\psi^\dagger_{-k +   2q \downarrow} )^T$ part only. 
The linearized version of the BdG hamiltonian in this basis is
\begin{equation} \label{eq_app_BdG2}
     H_{BdG}^{\alpha} =
\alpha q v_F + \begin{pmatrix}
\alpha v_F(k-k_F - q)  &   \Delta\\ 
\Delta^*  & -\alpha v_F\left ( k - k_F - q\right ) 
\end{pmatrix}
\end{equation}
where $\alpha = \pm$ corresponds to right(left)-moving particles. 
Denoting $\xi(k) = \varepsilon(k) - \mu$, the  eigenstates of this Hamiltonian can be written as
\be
\begin{split}
E(k) &= \left ( \frac{\xi(k) - \xi(-k+2q) }{2}\right) \\
&\pm \sqrt{\left (  \frac{\xi(k) + \xi(-k+2q) }{2}\right )^2 + |\Delta|^2 }.
\end{split}
\ee

Linearizing dispersion near the Fermi level, one gets eq.~\eqref{eq:E_dop}. More explicitly, the energies of right-, and left-moving particles can be written as
\be
E( k) \approx   \alpha q   v_F \pm \sqrt{(v_F (k- k_F ) - q  v_F)^2 + |\Delta|^2 }
\ee
where without loss of generality we assumed $q v_F > 0$.
The eigenstates are 
\begin{equation} \label{app_sol_psi_2}
\psi_{e}^{\alpha} =  \sqrt{\frac{1}{2}} \begin{pmatrix}
u_0(E - \alpha q v_F)
 \\
v_0(E - \alpha q v_F)
\end{pmatrix}, \psi_{h}^{\alpha} = \sqrt{\frac{1}{2}} \begin{pmatrix}
v_0(E - \alpha q v_F)
\\
u_0(E - \alpha q v_F)\end{pmatrix}, 
\end{equation}
where $u_0$ and $v_0$ are the wavefunction amplitudes of the usual $s$-wave superconductor, which for the argument outside  the gap $|\xi| > |\Delta|$ are equal to:
\begin{equation}  \label{app_sol_u}
u_0 (\xi)=\sqrt{1 + \sqrt{1 - \left ( \frac{|\Delta|}{\xi } \right )^2}} 
\end{equation}

\begin{equation}  \label{app_sol_v}
\quad v_0 (\xi)= \mathrm{sgn}(\xi) \sqrt{1 - \sqrt{1 - \left ( \frac{|\Delta|}{\xi } \right )^2}}.
\end{equation}
and inside the gap, these are obtained from the condition on the evanescent wave in the superconductor, i.e. $\mathrm{Im}[k]> 0$. When $|\xi|< |\Delta|$, these are equal
\begin{equation}  
u_0 (\xi)=\sqrt{1 + i\text{sgn}(\xi )\sqrt{ \left ( \frac{|\Delta|}{\xi } \right )^2-1}} , \
\end{equation}
and $ v_0(\xi) = \text{sgn}(\xi ) u_0^*(\xi)$.
In the following, we will use the simplified notation $u_0^\alpha  = u_0 (E - \alpha q v_F)$ and $v_0^\alpha  = v_0 (E - \alpha q v_F)$ where it does not cause ambiguity.  

The generalization to two-dimensional geometry is straightforward.  In this case, we linearized the dispersion near direction $\bm k/k$, and for each wavevector, we have a separate problem where we define the Fermi velocity along the wavevector $\bm v_F = v_F \bm k / k $, as well as the directed Fermi wavevector $\bm k_F = k_F \bm k / k$. Assuming $k_x > 0$, the factors $q v_F$ turn into the scalar products $\bm q \bm v_F$.

\subsection{The scattering matrix}

\begin{figure}[b] 
	\includegraphics[width= 0.8\columnwidth]{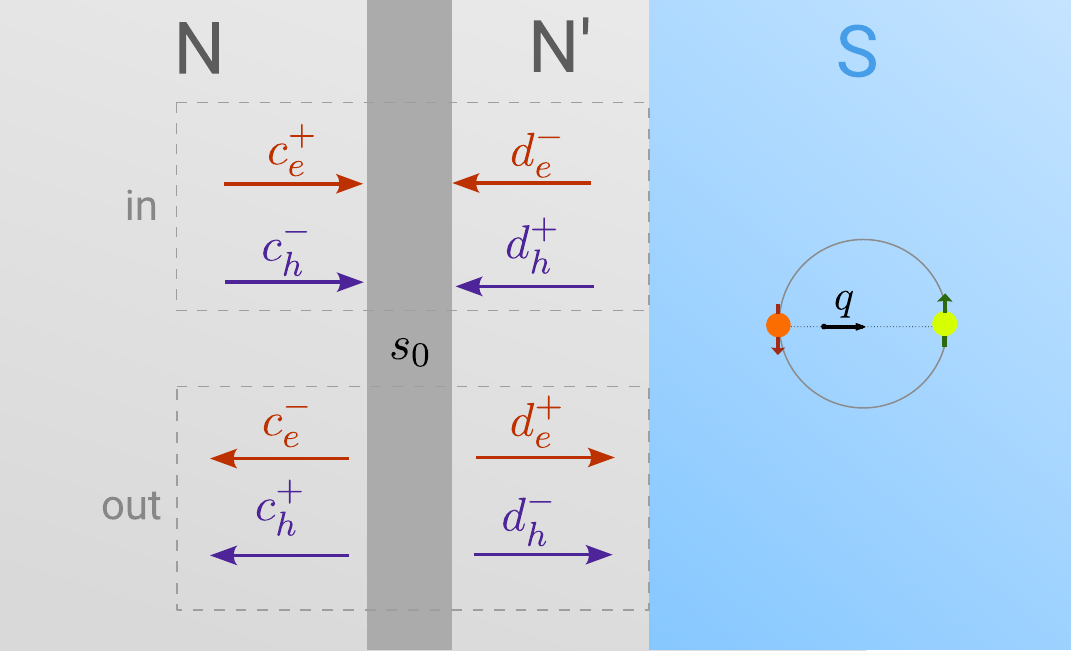}
	\caption{The $N$-$N'$-$S$ scattering setup, showing a barrier described by the scattering matrix $s_0$ and a transparent junction with a superconductor with finite Cooper pair momentum $2 \bm q$. For convenience of analysis, a finite distance $h$ is shown between the barrier and the transparent junction, but the limit of $h \rightarrow 0$ will be assumed throughout.}  
    \label{fig:app1_setup}
\end{figure}

    \begin{figure*}[t] 
	\includegraphics[width= 2\columnwidth]{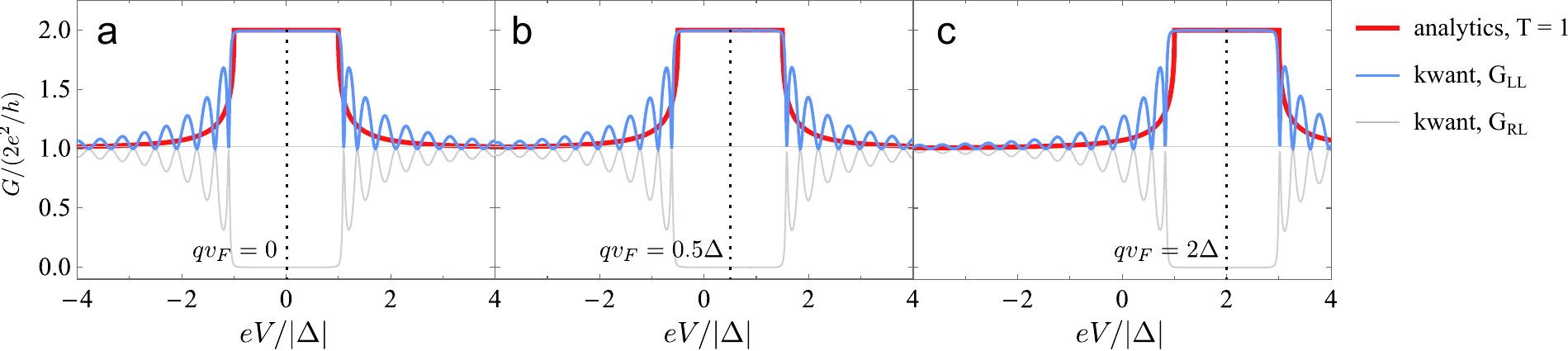}
	\caption{
    Analytically and numerically calculated differential conductance of a one-dimensional N-S junction with finite momentum superconductor. We assume a fully transparent barrier and show the conductance at several values of the finite Cooper pair momentum  $q$. The analytical results are shown in red, and the numerical calculation using kwant is shown in light blue and gray.  This calculation assumed two-lead N-S-N geometry.  $G_{LL}$ (light blue lines) corresponds to the conductance through the first lead (N-region), whereas $G_{LR}$ shown by the gray line is the conductance between the two normal leads. The oscillations are due to the finite length of the superconducting region.
    } \label{fig:app1_nonreciprocal}
\end{figure*}

Let us first solve the scattering problem~\cite{RevModPhys.69.731} in the geometry shown in Fig.~\ref{fig:app1_setup} that involves a normal lead with a scattering region and a junction with a finite-momentum superconductor. In the case of the junction in two dimensions, we fix the angle of incidence for the incoming electrons $\theta$, and for outgoing ones $- \theta$. Assuming elastic scattering only, we solve the scattering problem for each energy $E$ (counting from the Fermi level) and each angle $\theta$.  In the following, we will work in Andreev approximation, i.e. assuming that all momenta have magnitude $k_F$. 
The scattering states in both normal leads are shown in Fig.~\ref{fig:app1_setup} and are given by the amplitudes
\begin{equation}c^{\mathrm{in}}= \begin{pmatrix}
 c_e^+\\
 d_e^-\\
 c^-_h\\
d_h^+
\end{pmatrix}, \quad c^{\mathrm{out}}= \begin{pmatrix}
 c_e^-\\
 d_e^+\\
 c^+_h\\
d_h^-
\end{pmatrix}
\end{equation}
The scatterer $S_0$ in the normal region acts as $c^{\mathrm{out}} = S_0 c^{\mathrm{in}}$, where $S_0 = \mathrm{diag}(s_0(E), s_0^*(-E))$,
with the usual expression for the normal-state scattering matrix:
\begin{equation}
s_0(\xi)=\begin{pmatrix}
r_{11}(\xi) & t_{12}(\xi) \\
t_{21}(\xi) & r_{22}(\xi) \\
\end{pmatrix}.
\end{equation}
Next, use that the amplitudes $d_{e,h}^{\pm}$ are related by Andreev scattering at the $N'-S$ interface. We solve a separate problem at this interface and find $d_h^+ = r_A^e d_e^+$, $d_e^- = r_A^h d_h^-$ where the Andreev reflection amplitudes are 
\begin{align}
&r_A^e(E, \theta) = r_A(E- \bm q \bm v_F)\\
&r_A^h(E, \theta) = r_A(E+ \bm q \bm v_F)
\end{align}
where $r_A(\xi)$ is the conventional expression for the Andreev reflection amplitude in the absence of finite-momentum pairing, which is equal $e^{- i \arccos \frac{\xi}{|\Delta|}}$ when $|\xi|< |\Delta|$ and $e^{-  \mathrm{arcosh} \frac{|\xi|}{|\Delta|}}$ otherwise. The dependence on $\theta$ comes from the direction of $\bm v_F$; in the case of the Cooper pair momentum perpendicular to the junction, we have $\bm q \bm v_F = q v_F \cos \theta$.

Combining scattering between  $N$ and $N'$  states and at the $N'$-$S$ junction, we obtain the relation for the amplitudes in the normal region $N$ on the far left:
\begin{equation} \label{eq_s}
\begin{split}
&c_e^- = r_{ee} c_e^+ + r_{eh}  c_h^- \\
&c_h^+ = r_{he} c_e^++ r_{hh}  c_h^-
\end{split}
\end{equation}
where we have denoted the coefficients of the scattering matrix as $r_{ij}$ as opposed to the usual notation $s_{ij}$ for confomity with the BTK formalism. 
Suppressing the arguments for conciseness, we can write
\begin{align} 
    &r_{ee}= r_{11} + t_{12}  r_A^e r_{22}^* r_A^h   M t_{21}   \\
    &r_{eh} = t_{12} r_A^h  M^* t_{21}^*   \\
    &M = \left ( 1 - r_{22}^* r_A^e r_{22} r_A^h  \right)^{-1}, \label{eq:ree_reh_M}
\end{align}
Note that the scattering matrix in eq.~\eqref{eq_s} is unitary only when the scattering process at the interface with the superconductor is unitary. This is only true when quasiparticles do not enter the superconductor, which happens when the energy of the incoming particles is inside the gap of the superconductor. Assuming $|\Delta| > \bm q \bm v_F \ge 0$, this gives the condition  $-|\Delta| + \bm  q \bm  v_F< E< |\Delta| - \bm  q \bm 
 v_F$.

Finally, we specify the colsed-form equations for the scatteriing amplitudes in for the case of the delta-function barrier with effective strength $Z$:
\be
\begin{split}
    &r_{ee} = -i\widetilde{Z}^2(\theta)\left ( 1 -i \widetilde{Z}^2(\theta)\right )\frac{1 - r_A^h r_A^e}{1 + \widetilde{Z}^2(\theta) ( 1 - r_A^h r_A^e) } \\
    &r_{eh} = r_A^h(E, \theta)  \frac{1}{1 + \widetilde{Z}^2(\theta) ( 1 - r_A^h r_A^e) }
\end{split}
\ee
where we introduced $\widetilde{Z}(\theta) = \frac{Z}{\cos \theta}$. 

\subsection{Current through the  junction}
\label{sec:appendix_current}

Let us follow ref.~\cite{PhysRevB.25.4515} and assume zero temperature. Note that the discussion below, unless specified explicitly, is independent of the dimension of the junction. The incoming particles are distributed according to the Fermi-Dirac distribution, and to the far left,  voltage $V$ is applied. The voltage in the superconducting region is 0.  The distribution function for the incoming electrons ($c_e^+$) is $f_0(E  - eV)$ and for incoming holes ($c_h^-$) it is $1-f_0(-E  - eV)$. 

We denote the distribution function of electrons on the far left from the junction as $f_{\pm}(E, \theta)$, where $\pm$ stands for right(left)-moving electrons and $\theta \in [-\pi/2, \pi/2]$.  In the stationary regime, it is
\begin{equation}
  f_+(E, \theta) =  f_0(E  - eV)
\end{equation}
and
\begin{equation}
\begin{split}
 f_-(E, \theta)  &= |r_{ee}(E, \theta)|^2 f_0(E - eV) 
 \\
 &+  |r_{eh}(E, \theta)|^2 \left ( 1 - 
 f_0 (-E - e V )\right ).
 \end{split}
\end{equation}
Assuming the constant density of states near the Fermi energy $\nu(\mu)$, and ignoring the quasiparticle-imbalance contribution for simplicity, the current can be written as
\begin{equation}
    I(V) = 2 e v_F \nu (\mu) \int d E \int d \theta   \cos \theta \left ( f_+ - f_-\right ) 
\end{equation}
Finally, using $\frac{\partial f_0 ( E - e V)}{\partial V} = e  \delta ( E - e V)$, and the definition for the differential conductance $ G_{NS} = \frac{\partial I}{\partial V}$ we find
\begin{equation}
\begin{split}
   G_{NS}(eV) &= 2 e^2 v_F \nu(\mu)  \int d \theta \cos \theta  \left [ 1 - |r_{ee}(E = eV, \theta)|^2 \right .
   \\
   &+ \left. |r_{eh} (E = - eV, \theta)|^2 \right]
   \end{split}
\end{equation}
which is our final expression for the differential conductance. The one-dimensional case is obtained by removing the integration over the incidence angles and setting $\theta = 0$.

As an example of a result where the asymmetry of the differential conductance is seen explicitly, consider one-dimensional geometry and fully transparent junction $T = 1$. In this case $r_{ee} = 0$, $r_{eh} = |r_A(E +q v_F)|^2 $. Thus, in this case, the N-S conductance equals 
\begin{equation}
   G_{NS}^{T = 1 }  =  2 e^2 v_F \nu(\mu) \left [ 1  + |r_A (eV - q v_F)|^2 \right]
\end{equation}
Which is symmetric with respect to $ V = q v_F/e$ and not $V = 0$. 
The differential conductance for this case is shown in Fig.~\ref{fig:app1_nonreciprocal} as calculated both analytically and in a tight-binding model using kwant~\cite{Groth2014Jun}.
In the kwant calculations, we use a discretization with $\lambda_F = 12 a$ in terms of the lattice constant $a$, a superconducting gap $\Delta = 0.1 E_F$, and solve the scattering problem in a N-S-N geometry where S has a length of 250$a$ and the Cooper pair momentum is included explicitly as a phase gradient $\Delta e^{2iqx}$. 

The symmetry of the differential conductance with respect to $V \rightarrow -V$ is restored in the tunneling limit. The scattering matrix approach in the context of Fig.~\ref{fig:app1_setup} gives an intuitive explanation why the conductance becomes symmetric in the tunneling limit. As the normal reflection from the scattering region $s_0$ increases, the incoming particle in the $N'$ region experiences multiple reflections  between the scatterer and the superconductor on average before leaving the region (associated with  $M$ in eq.~\eqref{eq:ree_reh_M}). Upon each reflection from the superconductor, the quasiparticle type changes from electron to a hole (or vice versa), and upon reflection from the scatterer, the particle incoming at the superconductor belongs to the other side of the Fermi surface.
In the limit  $T \rightarrow 0$, the electron part of the current is formed almost equally due to incoming electrons and holes that subsequently reflect into electrons. The superconductor is thus effectively probed by the carriers from the both sides of the Fermi surface, which experience Andreev reflection with opposite shifts in energy. 

\begin{figure}[t] 
	\includegraphics[width= 1\columnwidth]{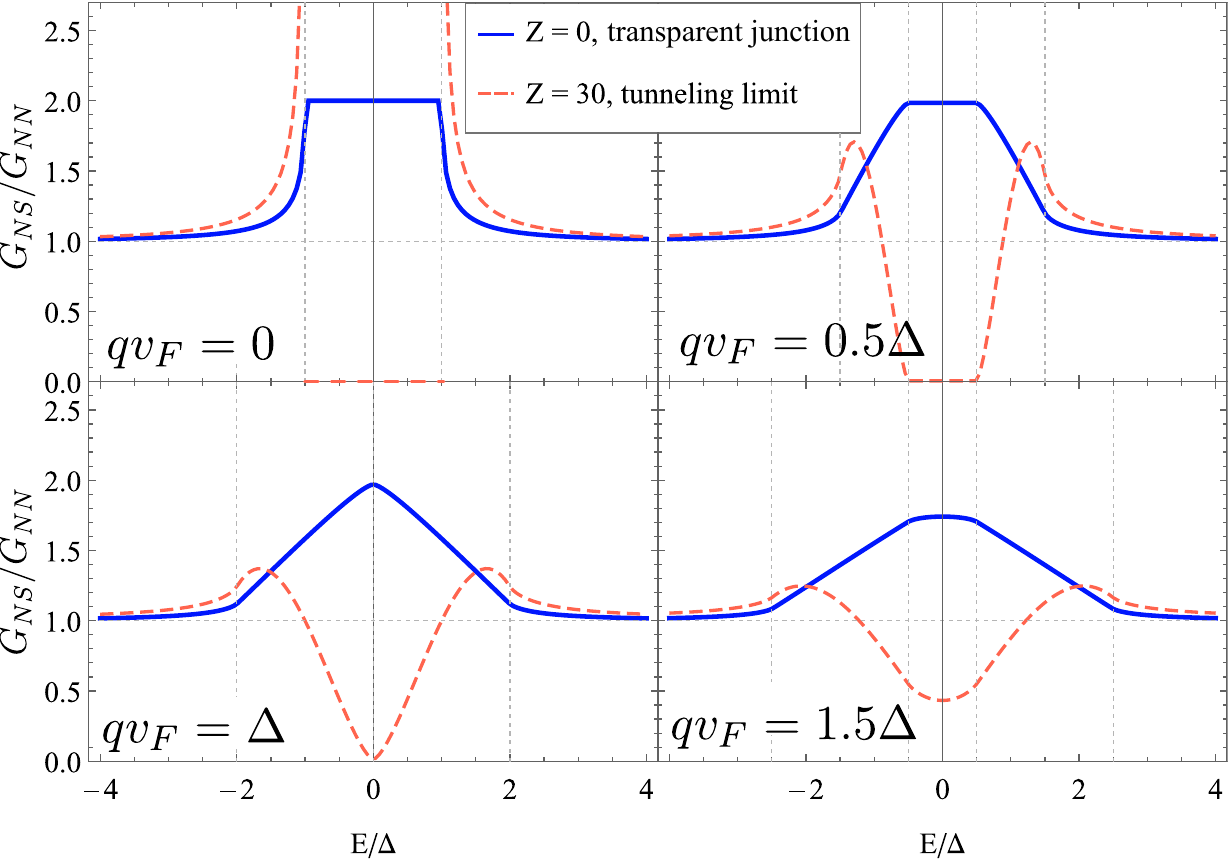}
	\caption{
    Differential conductance of an N-S junction in two dimensional geometry at different values of the Cooper pair momentum $q$ in the superconductor, directed parallel to the junction. 
    Blue lines show the case of the transparent junction, whereas red dashed lines show the tunneling limit $Z = 30$. 
    The vertical dashed lines show the energies $\pm \Delta \pm q v_F$. 
    } \label{fig:M6}
\end{figure}

The conductance also  becomes symmetric in the limit when the Cooper pair momentum  $\bm q$ becomes parallel to the junction. Only the component of the Cooper pair momentum along the direction of the current (i.e., perpendicular to the junction) leads to asymmetry. Fig.~\ref{fig:M6} shows the differential conductance for a junction in two dimensional geometry in the case when $\bm q$ is parallel to the junction. 

\section{Tunneling limit} \label{app:tunneling}

In this section, find the current between a normal lead and a superconductor for the transfer Hamiltonian approach formulated in sec.~\ref{sec:tunneling-stm} using linear response theory~\cite{levitov}. For simplicity, we describe the normal region as a single mode, $H_N = (E_N - \mu) b^\dagger b $. In order to be able to use the equilibrium Green's function technique, we perform gauge transformation which absorbs the effect of the voltage applied to the normal region  into the vector potential, i.e. $b \rightarrow b e^{i e V t}$, upon which the perturbation term becomes $H_T = T \left ( b^\dagger c(\bm r) e^{- i e V t}  + h.c.\right)$. The chemical potential is now equal in both leads.  One can write the current operator as $\hat I = \dot Q_N = - i [Q_N, H]=  i e T \left ( b^\dagger c(\bm r) e^{- i e V t}  + h.c.\right )$. Using Kubo formula, we can write $I (t) = \expval{\hat I (t)} = i \int_{-\infty}^t \expval{[H_T(t'), \hat I(t)]}dt'$. The dependence on the voltage is found by performing Fourier transform $$I(V) =  2 \Re\chi(\omega = e V),$$ where we introduced the susceptibility $\chi(\omega) = i \int_0^\infty dt  e^{i \omega t} \expval{[T c^\dagger (\bm r,t) b (t) , \hat I(0)]}$. We find the susceptibility by going from real $\omega$ to imaginary frequencies $\Omega_n$:
\be \nonumber 
\begin{split} 
&\chi(i \Omega_n) = \frac 1 2 T \int_{- \beta}^\beta \expval{T_\tau  c^\dagger(\bm r, \tau) b(\tau) \hat I(0)} e^{i \Omega_n \tau} d \tau\\
&= \frac 1 2 i e T^2 \int_{- \beta}^\beta \expval{T_\tau  c^\dagger(\bm r, \tau) c(\bm r,0)} \expval{T_\tau  b (\tau) b^\dagger(0)} e^{i \Omega_n \tau} d \tau
\\
&= \frac 1 2 i e T^2 \int_{- \beta}^\beta G_{e,S}(\tau, \bm r) G_{e,N} (-\tau) e^{i \Omega_n \tau} d \tau
\end{split}
\ee
where $\beta$ is the inverse temperature, and $G_{e, S/N}$ are the electronic Green's functions in equilibrium in the superconductor and the normal metal, respectively. We can express $G_{e,S}(\tau, \bm r)$ through quasiparticle Green's functions in the superconductor using \eqref{eq:c_through_gamma}, which we label as $G_\gamma(\tau, n, i)$ for the quasiparticle mode created by $\gamma^\dagger_{n,i}$.  We find 
\be \nonumber 
\begin{split} 
G_{e,S}(\tau, \bm r) = \sum_{n} &\left (|\varphi_{n,1} (\bm r)|^2 G_{\gamma}(\tau, n,1) \right. \\
&+ \left. |\varphi_{n,2} (\bm r)|^2 G_{\gamma}(-\tau, n,2) \right ) 
\end{split}
\ee
where the different arguments in respective Green's functions for modes $n,1$ and $n,2$ are because we separated electron-like and hole-like terms in the expression for $c^\dagger (\bm r, \tau)$. Finally, we use the Matsubara trick to change from integration over the imaginary time to the one over frequencies and we finally obtain:
\be
\begin{split}
I &\propto e |T|^2 \int dE \sum_{n,i: E_n = E} |\varphi_{n,i}(\bm r) |^2  \mathrm{Im}G^R_{\gamma}(E,n,i) 
\\
&\times\mathrm{Im} G^R_N(E + eV) \left ( f_0(E) - f_0(E + eV)\right),
\end{split}
\ee
where $G^R_{\gamma}(E, n,i)$ is the retarded quasiparticle Green's function at energy $E$ and mode $n,i$, and $f_0$ is the Fermi-Dirac distribution. Using $\nu_{n,i}(E)  = -1/\pi \mathrm{Im}G^R_{\gamma}(E,n,i)$, we obtain the same expression for the conductance as given in eq.~\eqref{eq:current_3}. We obtain that STM measurements describing tunneling between a  tip with a single localized mode and an infinite superconductor give the bulk quasiparticle density of states. In the presence of the edge in the superconductor, the quasiparticle density of states will be, interestingly, modified everywhere in space.

Let us also provide the full solution for the wavefunctions for the superconductor occupying half-space at $r > 0$. We use the linearized dispersion near the Fermi momentum, and the shorthand notation $u_{\pm } (E) = u(E \mp q  v_F)$ and similarly for $v_{\pm }$, where $u$ and $v$ are the coherence factors of a usual $s$-wave superconductor given in App.~\ref{app:BdG}. There are two scattering states, one corresponding 
\be \label{eq:boundary0_app}
\begin{split}
\psi_{ E,1}( r) &=  \psi_{e}^- e^{i k_{e,-} r} + \frac{u_+ (v_-^2 - u_-^2)}{u_+ u_- - v_+ v_-} \psi_{e}^+ e^{i k_{e,+} r} \\
&+ \frac{u_- v_+ - u_+ v_-}{u_+ u_- - v_+ v_-} \psi_{h}^- e^{i k_{h,-} r},
\end{split}
\ee
where $\psi_{e/h}^\pm$ are the electron- and hole-like plane waves given in eq.~\eqref{app_sol_psi_2}. The electron and hole momenta as the function of energy are
\be
\begin{split}
    k_{e/h,+} &= k_F + q \pm \sqrt{(E - q v_F)^2 - |\Delta|^2}, \\
    k_{e/h,-} &= -k_F + q \mp \sqrt{(E + q v_F)^2 - |\Delta|^2},
\end{split}
\ee
outside the gap at positive and negative momenta, respectively. In-gap, the momenta acquire small imaginary part and are
\be \nonumber
\begin{split}
    k_{e/h,+} &= k_F + q + i \sqrt{ |\Delta|^2 - (E - q v_F)^2},  \  |E - q v_F|< |\Delta|\\
    k_{e/h,-} &= -k_F + q + i \sqrt{ |\Delta|^2 -  (E + q v_F)^2}, \   |E + q v_F|< |\Delta|
\end{split}
\ee

The first coordinate of the wavefunction \eqref{eq:boundary0_app} is the electronic part $\varphi_{1}( r)$ entering the expression for the conductance. This expression is reduced to eq.~\eqref{eq:boundary0} when $r \ll \xi$.  Similarly, the scattering solution for the incoming hole-like wave gives
\be \label{eq:boundary1_app}
\begin{split}
\psi_{ E,2}( r) &=  \psi_{h}^+ e^{i k_{h,+} r} + \frac{u_+ (v_+^2 - u_+^2)}{u_+ u_- - v_+ v_-} \psi_{h}^- e^{i k_{h,-} r} \\
&+ \frac{u_+ v_- - u_- v_+}{u_+ u_- - v_+ v_-} \psi_{e}^+ e^{i k_{e,+} r},
\end{split}
\ee
whose first component gives $\varphi_{2}( r)$.

\begin{figure}[t] 
	\includegraphics[width= 0.6\columnwidth]{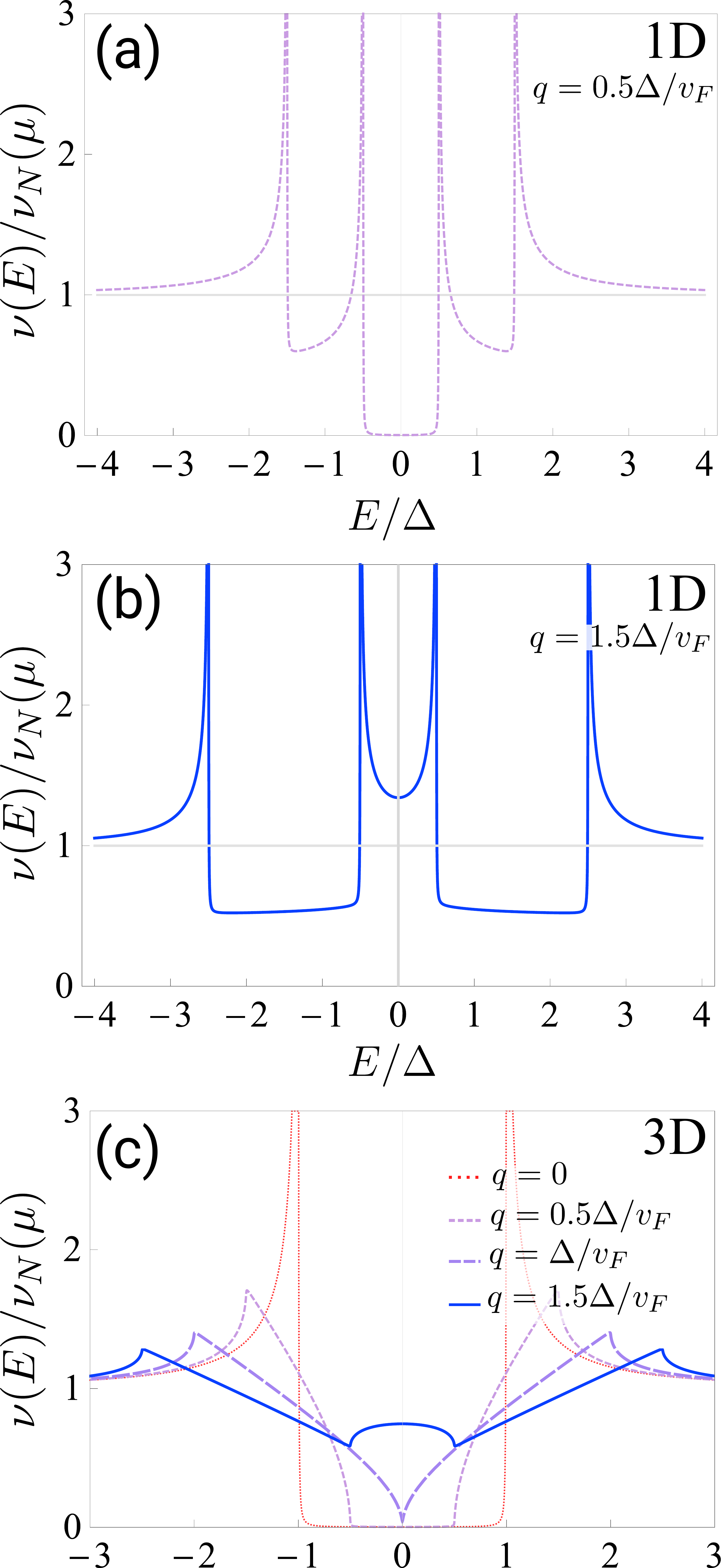}
	\caption{ The density of states for a finite-momentum superconductor in 1D (a,b) and 3D (c). The dashed lines indicate the positions of band extrema $\pm \left (|\Delta| \pm q v_F \right)$. Here, we included a broadening factor $\Gamma = 0.001 \Delta$.
	} \label{fig:app_dos}
\end{figure}

Finally, let us provide the bulk quasiparticle density of states  $\nu(E, \bm q)= - \frac{1}{\pi} \langle  \Im G_{\bm q}^R(E, \bm k) \rangle_{\bm k}$ for a finite-momentum supercondcutor for the reference. 
In one dimensions, it is:
\be
\begin{split}
    \nu_{1D}(E)&=\nu_{N, 1D}(\mu) \sum_{\alpha = \pm} \frac{ |E - \alpha q v_F \cos \theta|}{\sqrt{(E -  \alpha q v_F)^2 - \Delta^2}} \times \\
    &\times \Theta|(E -   \alpha q v_F | - \Delta),
\end{split}
\ee
while in two and three dimensions
\be 
\begin{split}
    \nu_{2D}(E, \bm q)&=\nu_{N, 2D}(\mu) \int \frac{d \Omega}{\Omega_{2D/3D}} \frac{ |E - q v_F \cos \theta|}{\sqrt{(E - q v_F \cos \theta)^2 - \Delta^2}}  \\
    &\times \Theta|(E -  q v_F \cos \theta| - \Delta),
\end{split}
\ee
where $d \Omega$ stands for spherical integration with respective measure  in 2 and 3 dimensions.

\section{Critical magnetic field for proximitized layers}
\label{sec:appendix_estimates}

In this section, we estimate  the in-lane magnetic field needed to induce Copper pair momentum on the order of gap closing one in proximitized 2D electron gas in the setup shown in Fig.~\ref{fig:schematic}. When the thickness of the parent superconductor is much larger than that of the proximitized layer, $d_{SC} \gg \min(w_{2DEG}, \xi)$ (where $w_{2DEG}$ is the width of the potential well that is used to form 2D electron gas layer and $\xi$ is the proximity-induced Cooper pair coherence length), we can estimate the orbital effect of the magnetic field by neglecting the contribution of the proximitized layer to the total supercurrent and assume a homogeneous superconductor. Then we can use London equations in order to find the distribution of the supercurrent in the superconducting film.  We us the London equation $\bm B =- \lambda _{\text{eff}} \nabla \times \bm j_s$      and Maxwell's equations leading to $\nabla^2 \bm B = \frac{1}{\lambda_{\text{eff}}^2} \bm B$. Assuming the superconductor of thickness $d$ occupying $-\frac{d}{2}\le z \le \frac{d}{2}$ and magnetic field in $y$-direction, one obtains $B_y = B_0 \cosh \frac{z}{2 \lambda_{\text{eff}}}/ \cosh \frac{d}{2 \lambda_{\text{eff}}}$, where $B_0$ is the magnitude of applied magnetic field. Using the gauge  $\partial_z A_x = B_y$ and $\phi = 0$, we obtain $(j_s)_x = -\frac{1}{\mu_0 \lambda_{\text{eff}}^2} A_x = -\frac{2}{\mu_0 \lambda_{\text{eff}}} B_0   \sinh \frac{z}{2 \lambda_{\text{eff}}}/ \cosh \frac{d}{2 \lambda_{\text{eff}}} $.  Thus, the supercurrent in the proximitized layer can be roughly estimated as $$j_s \approx \frac{2}{\mu_0 \lambda_{\text{eff}}} B_0 \tanh\frac{d}{2 \lambda_{\text{eff}}}.$$
Now, this supercurrent in the parent superconductor is equivalent to inducing proximity pairing potential $\Delta = \Delta_{prox}e^{i q x}$, where $q = \nabla \varphi = \frac{e m}{\pi \hbar} \lambda_{\text{eff}}^2 (j_s)_x$,  in the proximitized layer with $$q = \frac{2 e \lambda_{\text{eff}}}{\pi \hbar}B_0 \tanh \frac{d}{2\lambda_{\text{eff}}}.$$

The gap in proximitized layer closes when the condition $\hbar q v_F^{prox} = \Delta_{prox}$ is met. For a 100 nm-thick slab of Al, with $\lambda_{\text{eff}} \approx 60 $ nm and induced $\Delta_{prox} = 0.1 $ meV in InAs layer (hosting 2DEG), we obtain the critical field $B_c \approx $ 30 mT, at which the proximity-induced gap will close in the proximitized layer.

In case of epitaxial heterostructures, one needs to take into account the finite thickness of the proximitized layer and the dependence of the Cooper pair momentum on the interlayer spacing between 2DEG and superconductor \cite{banerjee2023phase}. There, with Al of 7 nm thickness and an insulating barrier between Al and 2DEG of 10 nm thickness, one finds $B_{c,\parallel} \sim 100$ mT.
 At the same time, the parent superconductor undergoes transition at the fields of the order of several Tesla \cite{vanWeerdenburg2023Mar}.

\vfill

\bibliography{ref}

\begin{thebibliography}{80}%
\makeatletter
\providecommand \@ifxundefined [1]{%
 \@ifx{#1\undefined}
}%
\providecommand \@ifnum [1]{%
 \ifnum #1\expandafter \@firstoftwo
 \else \expandafter \@secondoftwo
 \fi
}%
\providecommand \@ifx [1]{%
 \ifx #1\expandafter \@firstoftwo
 \else \expandafter \@secondoftwo
 \fi
}%
\providecommand \natexlab [1]{#1}%
\providecommand \enquote  [1]{``#1''}%
\providecommand \bibnamefont  [1]{#1}%
\providecommand \bibfnamefont [1]{#1}%
\providecommand \citenamefont [1]{#1}%
\providecommand \href@noop [0]{\@secondoftwo}%
\providecommand \href [0]{\begingroup \@sanitize@url \@href}%
\providecommand \@href[1]{\@@startlink{#1}\@@href}%
\providecommand \@@href[1]{\endgroup#1\@@endlink}%
\providecommand \@sanitize@url [0]{\catcode `\\12\catcode `\$12\catcode
  `\&12\catcode `\#12\catcode `\^12\catcode `\_12\catcode `\%12\relax}%
\providecommand \@@startlink[1]{}%
\providecommand \@@endlink[0]{}%
\providecommand \url  [0]{\begingroup\@sanitize@url \@url }%
\providecommand \@url [1]{\endgroup\@href {#1}{\urlprefix }}%
\providecommand \urlprefix  [0]{URL }%
\providecommand \Eprint [0]{\href }%
\providecommand \doibase [0]{https://doi.org/}%
\providecommand \selectlanguage [0]{\@gobble}%
\providecommand \bibinfo  [0]{\@secondoftwo}%
\providecommand \bibfield  [0]{\@secondoftwo}%
\providecommand \translation [1]{[#1]}%
\providecommand \BibitemOpen [0]{}%
\providecommand \bibitemStop [0]{}%
\providecommand \bibitemNoStop [0]{.\EOS\space}%
\providecommand \EOS [0]{\spacefactor3000\relax}%
\providecommand \BibitemShut  [1]{\csname bibitem#1\endcsname}%
\let\auto@bib@innerbib\@empty
\bibitem [{\citenamefont {Jiang}\ and\ \citenamefont
  {Hu}(2022)}]{jiang2022superconducting}%
  \BibitemOpen
  \bibfield  {author} {\bibinfo {author} {\bibfnamefont {K.}~\bibnamefont
  {Jiang}}\ and\ \bibinfo {author} {\bibfnamefont {J.}~\bibnamefont {Hu}},\
  }\bibfield  {title} {\bibinfo {title} {Superconducting diode effects},\
  }\href {https://www.nature.com/articles/s41467-021-23077-0} {\bibfield
  {journal} {\bibinfo  {journal} {Nature Physics}\ }\textbf {\bibinfo {volume}
  {18}},\ \bibinfo {pages} {1145} (\bibinfo {year} {2022})}\BibitemShut
  {NoStop}%
\bibitem [{\citenamefont {Nadeem}\ \emph {et~al.}(2023)\citenamefont {Nadeem},
  \citenamefont {Fuhrer},\ and\ \citenamefont {Wang}}]{Nadeem2023Oct}%
  \BibitemOpen
  \bibfield  {author} {\bibinfo {author} {\bibfnamefont {M.}~\bibnamefont
  {Nadeem}}, \bibinfo {author} {\bibfnamefont {M.~S.}\ \bibnamefont {Fuhrer}},\
  and\ \bibinfo {author} {\bibfnamefont {X.}~\bibnamefont {Wang}},\ }\bibfield
  {title} {\bibinfo {title} {{The superconducting diode effect}},\ }\href
  {https://doi.org/10.1038/s42254-023-00632-w} {\bibfield  {journal} {\bibinfo
  {journal} {Nat. Rev. Phys.}\ }\textbf {\bibinfo {volume} {5}},\ \bibinfo
  {pages} {558} (\bibinfo {year} {2023})}\BibitemShut {NoStop}%
\bibitem [{\citenamefont {Ando}\ \emph {et~al.}(2020)\citenamefont {Ando},
  \citenamefont {Miyasaka}, \citenamefont {Li}, \citenamefont {Ishizuka},
  \citenamefont {Arakawa}, \citenamefont {Shiota}, \citenamefont {Moriyama},
  \citenamefont {Yanase},\ and\ \citenamefont {Ono}}]{ando2020observation}%
  \BibitemOpen
  \bibfield  {author} {\bibinfo {author} {\bibfnamefont {F.}~\bibnamefont
  {Ando}}, \bibinfo {author} {\bibfnamefont {Y.}~\bibnamefont {Miyasaka}},
  \bibinfo {author} {\bibfnamefont {T.}~\bibnamefont {Li}}, \bibinfo {author}
  {\bibfnamefont {J.}~\bibnamefont {Ishizuka}}, \bibinfo {author}
  {\bibfnamefont {T.}~\bibnamefont {Arakawa}}, \bibinfo {author} {\bibfnamefont
  {Y.}~\bibnamefont {Shiota}}, \bibinfo {author} {\bibfnamefont
  {T.}~\bibnamefont {Moriyama}}, \bibinfo {author} {\bibfnamefont
  {Y.}~\bibnamefont {Yanase}},\ and\ \bibinfo {author} {\bibfnamefont
  {T.}~\bibnamefont {Ono}},\ }\bibfield  {title} {\bibinfo {title} {Observation
  of superconducting diode effect},\ }\href
  {https://doi.org/10.1038/s41586-020-2590-4} {\bibfield  {journal} {\bibinfo
  {journal} {Nature}\ }\textbf {\bibinfo {volume} {584}},\ \bibinfo {pages}
  {373} (\bibinfo {year} {2020})}\BibitemShut {NoStop}%
\bibitem [{\citenamefont {Yuan}\ and\ \citenamefont
  {Fu}(2022)}]{yuan2021supercurrent}%
  \BibitemOpen
  \bibfield  {author} {\bibinfo {author} {\bibfnamefont {N.~F.}\ \bibnamefont
  {Yuan}}\ and\ \bibinfo {author} {\bibfnamefont {L.}~\bibnamefont {Fu}},\
  }\bibfield  {title} {\bibinfo {title} {Supercurrent diode effect and
  finite-momentum superconductors},\ }\href
  {https://www.pnas.org/doi/10.1073/pnas.2119548119} {\bibfield  {journal}
  {\bibinfo  {journal} {Proceedings of the National Academy of Sciences}\
  }\textbf {\bibinfo {volume} {119}},\ \bibinfo {pages} {e2119548119} (\bibinfo
  {year} {2022})}\BibitemShut {NoStop}%
\bibitem [{\citenamefont {Daido}\ \emph {et~al.}(2022)\citenamefont {Daido},
  \citenamefont {Ikeda},\ and\ \citenamefont
  {Yanase}}]{PhysRevLett.128.037001}%
  \BibitemOpen
  \bibfield  {author} {\bibinfo {author} {\bibfnamefont {A.}~\bibnamefont
  {Daido}}, \bibinfo {author} {\bibfnamefont {Y.}~\bibnamefont {Ikeda}},\ and\
  \bibinfo {author} {\bibfnamefont {Y.}~\bibnamefont {Yanase}},\ }\bibfield
  {title} {\bibinfo {title} {Intrinsic superconducting diode effect},\ }\href
  {https://doi.org/10.1103/PhysRevLett.128.037001} {\bibfield  {journal}
  {\bibinfo  {journal} {Phys. Rev. Lett.}\ }\textbf {\bibinfo {volume} {128}},\
  \bibinfo {pages} {037001} (\bibinfo {year} {2022})}\BibitemShut {NoStop}%
\bibitem [{\citenamefont {Ili\ifmmode~\acute{c}\else \'{c}\fi{}}\ and\
  \citenamefont {Bergeret}(2022)}]{PhysRevLett.128.177001}%
  \BibitemOpen
  \bibfield  {author} {\bibinfo {author} {\bibfnamefont {S.}~\bibnamefont
  {Ili\ifmmode~\acute{c}\else \'{c}\fi{}}}\ and\ \bibinfo {author}
  {\bibfnamefont {F.~S.}\ \bibnamefont {Bergeret}},\ }\bibfield  {title}
  {\bibinfo {title} {{Theory of the Supercurrent Diode Effect in Rashba
  Superconductors with Arbitrary Disorder}},\ }\href
  {https://doi.org/10.1103/PhysRevLett.128.177001} {\bibfield  {journal}
  {\bibinfo  {journal} {Phys. Rev. Lett.}\ }\textbf {\bibinfo {volume} {128}},\
  \bibinfo {pages} {177001} (\bibinfo {year} {2022})}\BibitemShut {NoStop}%
\bibitem [{\citenamefont {Hou}\ \emph {et~al.}(2023)\citenamefont {Hou},
  \citenamefont {Nichele}, \citenamefont {Chi}, \citenamefont {Lodesani},
  \citenamefont {Wu}, \citenamefont {Ritter}, \citenamefont {Haxell},
  \citenamefont {Davydova}, \citenamefont
  {Ili{\ifmmode\acute{c}\else\'{c}\fi}}, \citenamefont {Glezakou-Elbert},
  \citenamefont {Varambally}, \citenamefont {Bergeret}, \citenamefont {Kamra},
  \citenamefont {Fu}, \citenamefont {Lee},\ and\ \citenamefont
  {Moodera}}]{hou2022ubiquitous}%
  \BibitemOpen
  \bibfield  {author} {\bibinfo {author} {\bibfnamefont {Y.}~\bibnamefont
  {Hou}}, \bibinfo {author} {\bibfnamefont {F.}~\bibnamefont {Nichele}},
  \bibinfo {author} {\bibfnamefont {H.}~\bibnamefont {Chi}}, \bibinfo {author}
  {\bibfnamefont {A.}~\bibnamefont {Lodesani}}, \bibinfo {author}
  {\bibfnamefont {Y.}~\bibnamefont {Wu}}, \bibinfo {author} {\bibfnamefont
  {M.~F.}\ \bibnamefont {Ritter}}, \bibinfo {author} {\bibfnamefont {D.~Z.}\
  \bibnamefont {Haxell}}, \bibinfo {author} {\bibfnamefont {M.}~\bibnamefont
  {Davydova}}, \bibinfo {author} {\bibfnamefont {S.}~\bibnamefont
  {Ili{\ifmmode\acute{c}\else\'{c}\fi}}}, \bibinfo {author} {\bibfnamefont
  {O.}~\bibnamefont {Glezakou-Elbert}}, \bibinfo {author} {\bibfnamefont
  {A.}~\bibnamefont {Varambally}}, \bibinfo {author} {\bibfnamefont {F.~S.}\
  \bibnamefont {Bergeret}}, \bibinfo {author} {\bibfnamefont {A.}~\bibnamefont
  {Kamra}}, \bibinfo {author} {\bibfnamefont {L.}~\bibnamefont {Fu}}, \bibinfo
  {author} {\bibfnamefont {P.~A.}\ \bibnamefont {Lee}},\ and\ \bibinfo {author}
  {\bibfnamefont {J.~S.}\ \bibnamefont {Moodera}},\ }\bibfield  {title}
  {\bibinfo {title} {{Ubiquitous Superconducting Diode Effect in Superconductor
  Thin Films}},\ }\href {https://doi.org/10.1103/PhysRevLett.131.027001}
  {\bibfield  {journal} {\bibinfo  {journal} {Phys. Rev. Lett.}\ }\textbf
  {\bibinfo {volume} {131}},\ \bibinfo {pages} {027001} (\bibinfo {year}
  {2023})}\BibitemShut {NoStop}%
\bibitem [{\citenamefont {Bauriedl}\ \emph {et~al.}(2022)\citenamefont
  {Bauriedl}, \citenamefont {B{\"a}uml}, \citenamefont {Fuchs}, \citenamefont
  {Baumgartner}, \citenamefont {Paulik}, \citenamefont {Bauer}, \citenamefont
  {Lin}, \citenamefont {Lupton}, \citenamefont {Taniguchi}, \citenamefont
  {Watanabe} \emph {et~al.}}]{bauriedl2022supercurrent}%
  \BibitemOpen
  \bibfield  {author} {\bibinfo {author} {\bibfnamefont {L.}~\bibnamefont
  {Bauriedl}}, \bibinfo {author} {\bibfnamefont {C.}~\bibnamefont {B{\"a}uml}},
  \bibinfo {author} {\bibfnamefont {L.}~\bibnamefont {Fuchs}}, \bibinfo
  {author} {\bibfnamefont {C.}~\bibnamefont {Baumgartner}}, \bibinfo {author}
  {\bibfnamefont {N.}~\bibnamefont {Paulik}}, \bibinfo {author} {\bibfnamefont
  {J.~M.}\ \bibnamefont {Bauer}}, \bibinfo {author} {\bibfnamefont {K.-Q.}\
  \bibnamefont {Lin}}, \bibinfo {author} {\bibfnamefont {J.~M.}\ \bibnamefont
  {Lupton}}, \bibinfo {author} {\bibfnamefont {T.}~\bibnamefont {Taniguchi}},
  \bibinfo {author} {\bibfnamefont {K.}~\bibnamefont {Watanabe}}, \emph
  {et~al.},\ }\bibfield  {title} {\bibinfo {title} {{Supercurrent diode effect
  and magnetochiral anisotropy in few-layer NbSe2}},\ }\href
  {https://www.nature.com/articles/s41467-022-31954-5} {\bibfield  {journal}
  {\bibinfo  {journal} {Nature communications}\ }\textbf {\bibinfo {volume}
  {13}},\ \bibinfo {pages} {4266} (\bibinfo {year} {2022})}\BibitemShut
  {NoStop}%
\bibitem [{\citenamefont {Kochan}\ \emph {et~al.}(2023)\citenamefont {Kochan},
  \citenamefont {Costa}, \citenamefont {Zhumagulov},\ and\ \citenamefont
  {Žutić}}]{kochan2023phenomenological}%
  \BibitemOpen
  \bibfield  {author} {\bibinfo {author} {\bibfnamefont {D.}~\bibnamefont
  {Kochan}}, \bibinfo {author} {\bibfnamefont {A.}~\bibnamefont {Costa}},
  \bibinfo {author} {\bibfnamefont {I.}~\bibnamefont {Zhumagulov}},\ and\
  \bibinfo {author} {\bibfnamefont {I.}~\bibnamefont {Žutić}},\ }\href@noop
  {} {\bibinfo {title} {{Phenomenological Theory of the Supercurrent Diode
  Effect: The Lifshitz Invariant}}} (\bibinfo {year} {2023}),\ \Eprint
  {https://arxiv.org/abs/2303.11975} {arXiv:2303.11975 [cond-mat.supr-con]}
  \BibitemShut {NoStop}%
\bibitem [{\citenamefont
  {Yuan}(2023)}]{yuan2023edelsteineffectsupercurrentdiode}%
  \BibitemOpen
  \bibfield  {author} {\bibinfo {author} {\bibfnamefont {N.~F.~Q.}\
  \bibnamefont {Yuan}},\ }\href@noop {} {\bibinfo {title} {Edelstein effect and
  supercurrent diode effect}} (\bibinfo {year} {2023}),\ \Eprint
  {https://arxiv.org/abs/2311.11087} {arXiv:2311.11087 [cond-mat.supr-con]}
  \BibitemShut {NoStop}%
\bibitem [{\citenamefont {Hasan}\ \emph {et~al.}(2024)\citenamefont {Hasan},
  \citenamefont {Shaffer}, \citenamefont {Khodas},\ and\ \citenamefont
  {Levchenko}}]{hasan2024supercurrentdiodeeffecthelical}%
  \BibitemOpen
  \bibfield  {author} {\bibinfo {author} {\bibfnamefont {J.}~\bibnamefont
  {Hasan}}, \bibinfo {author} {\bibfnamefont {D.}~\bibnamefont {Shaffer}},
  \bibinfo {author} {\bibfnamefont {M.}~\bibnamefont {Khodas}},\ and\ \bibinfo
  {author} {\bibfnamefont {A.}~\bibnamefont {Levchenko}},\ }\href@noop {}
  {\bibinfo {title} {Supercurrent diode effect in helical superconductors}}
  (\bibinfo {year} {2024}),\ \Eprint {https://arxiv.org/abs/2404.17072}
  {arXiv:2404.17072 [cond-mat.supr-con]} \BibitemShut {NoStop}%
\bibitem [{\citenamefont {Shaffer}\ \emph {et~al.}(2024)\citenamefont
  {Shaffer}, \citenamefont {Chichinadze},\ and\ \citenamefont
  {Levchenko}}]{shaffer2024superconductingdiodeeffectmultiphase}%
  \BibitemOpen
  \bibfield  {author} {\bibinfo {author} {\bibfnamefont {D.}~\bibnamefont
  {Shaffer}}, \bibinfo {author} {\bibfnamefont {D.~V.}\ \bibnamefont
  {Chichinadze}},\ and\ \bibinfo {author} {\bibfnamefont {A.}~\bibnamefont
  {Levchenko}},\ }\href@noop {} {\bibinfo {title} {Superconducting diode effect
  in multiphase superconductors}} (\bibinfo {year} {2024}),\ \Eprint
  {https://arxiv.org/abs/2406.14612} {arXiv:2406.14612 [cond-mat.supr-con]}
  \BibitemShut {NoStop}%
\bibitem [{\citenamefont {Fu}\ \emph {et~al.}(2024)\citenamefont {Fu},
  \citenamefont {Liu}, \citenamefont {Xu}, \citenamefont {Lee},\ and\
  \citenamefont {Ang}}]{PHFu2024May}%
  \BibitemOpen
  \bibfield  {author} {\bibinfo {author} {\bibfnamefont {P.-H.}\ \bibnamefont
  {Fu}}, \bibinfo {author} {\bibfnamefont {J.-F.}\ \bibnamefont {Liu}},
  \bibinfo {author} {\bibfnamefont {Y.}~\bibnamefont {Xu}}, \bibinfo {author}
  {\bibfnamefont {C.~H.}\ \bibnamefont {Lee}},\ and\ \bibinfo {author}
  {\bibfnamefont {Y.~S.}\ \bibnamefont {Ang}},\ }\bibfield  {title} {\bibinfo
  {title} {{Transverse Cooper-Pair Rectifier}},\ }\bibfield  {journal}
  {\bibinfo  {journal} {arXiv}\ }\href
  {https://doi.org/10.48550/arXiv.2405.04751} {10.48550/arXiv.2405.04751}
  (\bibinfo {year} {2024}),\ \Eprint {https://arxiv.org/abs/2405.04751}
  {2405.04751} \BibitemShut {NoStop}%
\bibitem [{\citenamefont {Hu}\ \emph {et~al.}(2007)\citenamefont {Hu},
  \citenamefont {Wu},\ and\ \citenamefont {Dai}}]{PhysRevLett.99.067004}%
  \BibitemOpen
  \bibfield  {author} {\bibinfo {author} {\bibfnamefont {J.}~\bibnamefont
  {Hu}}, \bibinfo {author} {\bibfnamefont {C.}~\bibnamefont {Wu}},\ and\
  \bibinfo {author} {\bibfnamefont {X.}~\bibnamefont {Dai}},\ }\bibfield
  {title} {\bibinfo {title} {{Proposed Design of a Josephson Diode}},\ }\href
  {https://doi.org/10.1103/PhysRevLett.99.067004} {\bibfield  {journal}
  {\bibinfo  {journal} {Phys. Rev. Lett.}\ }\textbf {\bibinfo {volume} {99}},\
  \bibinfo {pages} {067004} (\bibinfo {year} {2007})}\BibitemShut {NoStop}%
\bibitem [{\citenamefont {Baumgartner}\ \emph
  {et~al.}(2022{\natexlab{a}})\citenamefont {Baumgartner}, \citenamefont
  {Fuchs}, \citenamefont {Costa}, \citenamefont {Reinhardt}, \citenamefont
  {Gronin}, \citenamefont {Gardner}, \citenamefont {Lindemann}, \citenamefont
  {Manfra}, \citenamefont {Faria~Junior}, \citenamefont {Kochan}, \citenamefont
  {Fabian}, \citenamefont {Paradiso},\ and\ \citenamefont
  {Strunk}}]{Baumgartner2022Jan}%
  \BibitemOpen
  \bibfield  {author} {\bibinfo {author} {\bibfnamefont {C.}~\bibnamefont
  {Baumgartner}}, \bibinfo {author} {\bibfnamefont {L.}~\bibnamefont {Fuchs}},
  \bibinfo {author} {\bibfnamefont {A.}~\bibnamefont {Costa}}, \bibinfo
  {author} {\bibfnamefont {S.}~\bibnamefont {Reinhardt}}, \bibinfo {author}
  {\bibfnamefont {S.}~\bibnamefont {Gronin}}, \bibinfo {author} {\bibfnamefont
  {G.~C.}\ \bibnamefont {Gardner}}, \bibinfo {author} {\bibfnamefont
  {T.}~\bibnamefont {Lindemann}}, \bibinfo {author} {\bibfnamefont {M.~J.}\
  \bibnamefont {Manfra}}, \bibinfo {author} {\bibfnamefont {P.~E.}\
  \bibnamefont {Faria~Junior}}, \bibinfo {author} {\bibfnamefont
  {D.}~\bibnamefont {Kochan}}, \bibinfo {author} {\bibfnamefont
  {J.}~\bibnamefont {Fabian}}, \bibinfo {author} {\bibfnamefont
  {N.}~\bibnamefont {Paradiso}},\ and\ \bibinfo {author} {\bibfnamefont
  {C.}~\bibnamefont {Strunk}},\ }\bibfield  {title} {\bibinfo {title}
  {{Supercurrent rectification and magnetochiral effects in symmetric Josephson
  junctions}},\ }\href {https://doi.org/10.1038/s41565-021-01009-9} {\bibfield
  {journal} {\bibinfo  {journal} {Nat. Nanotechnol.}\ }\textbf {\bibinfo
  {volume} {17}},\ \bibinfo {pages} {39} (\bibinfo {year}
  {2022}{\natexlab{a}})}\BibitemShut {NoStop}%
\bibitem [{\citenamefont {Costa}\ \emph {et~al.}(2023)\citenamefont {Costa},
  \citenamefont {Baumgartner}, \citenamefont {Reinhardt}, \citenamefont
  {Berger}, \citenamefont {Gronin}, \citenamefont {Gardner}, \citenamefont
  {Lindemann}, \citenamefont {Manfra}, \citenamefont {Fabian}, \citenamefont
  {Kochan}, \citenamefont {Paradiso},\ and\ \citenamefont
  {Strunk}}]{Costa2023Nov}%
  \BibitemOpen
  \bibfield  {author} {\bibinfo {author} {\bibfnamefont {A.}~\bibnamefont
  {Costa}}, \bibinfo {author} {\bibfnamefont {C.}~\bibnamefont {Baumgartner}},
  \bibinfo {author} {\bibfnamefont {S.}~\bibnamefont {Reinhardt}}, \bibinfo
  {author} {\bibfnamefont {J.}~\bibnamefont {Berger}}, \bibinfo {author}
  {\bibfnamefont {S.}~\bibnamefont {Gronin}}, \bibinfo {author} {\bibfnamefont
  {G.~C.}\ \bibnamefont {Gardner}}, \bibinfo {author} {\bibfnamefont
  {T.}~\bibnamefont {Lindemann}}, \bibinfo {author} {\bibfnamefont {M.~J.}\
  \bibnamefont {Manfra}}, \bibinfo {author} {\bibfnamefont {J.}~\bibnamefont
  {Fabian}}, \bibinfo {author} {\bibfnamefont {D.}~\bibnamefont {Kochan}},
  \bibinfo {author} {\bibfnamefont {N.}~\bibnamefont {Paradiso}},\ and\
  \bibinfo {author} {\bibfnamefont {C.}~\bibnamefont {Strunk}},\ }\bibfield
  {title} {\bibinfo {title} {{Sign reversal of the Josephson inductance
  magnetochiral anisotropy and 0{\textendash}{$\pi$}-like transitions in
  supercurrent diodes}},\ }\href {https://doi.org/10.1038/s41565-023-01451-x}
  {\bibfield  {journal} {\bibinfo  {journal} {Nat. Nanotechnol.}\ }\textbf
  {\bibinfo {volume} {18}},\ \bibinfo {pages} {1266} (\bibinfo {year}
  {2023})}\BibitemShut {NoStop}%
\bibitem [{\citenamefont {Wu}\ \emph {et~al.}(2022)\citenamefont {Wu},
  \citenamefont {Wang}, \citenamefont {Xu}, \citenamefont {Sivakumar},
  \citenamefont {Pasco}, \citenamefont {Filippozzi}, \citenamefont {Parkin},
  \citenamefont {Zeng}, \citenamefont {McQueen},\ and\ \citenamefont
  {Ali}}]{wu2022field}%
  \BibitemOpen
  \bibfield  {author} {\bibinfo {author} {\bibfnamefont {H.}~\bibnamefont
  {Wu}}, \bibinfo {author} {\bibfnamefont {Y.}~\bibnamefont {Wang}}, \bibinfo
  {author} {\bibfnamefont {Y.}~\bibnamefont {Xu}}, \bibinfo {author}
  {\bibfnamefont {P.~K.}\ \bibnamefont {Sivakumar}}, \bibinfo {author}
  {\bibfnamefont {C.}~\bibnamefont {Pasco}}, \bibinfo {author} {\bibfnamefont
  {U.}~\bibnamefont {Filippozzi}}, \bibinfo {author} {\bibfnamefont {S.~S.}\
  \bibnamefont {Parkin}}, \bibinfo {author} {\bibfnamefont {Y.-J.}\
  \bibnamefont {Zeng}}, \bibinfo {author} {\bibfnamefont {T.}~\bibnamefont
  {McQueen}},\ and\ \bibinfo {author} {\bibfnamefont {M.~N.}\ \bibnamefont
  {Ali}},\ }\bibfield  {title} {\bibinfo {title} {{The field-free Josephson
  diode in a van der Waals heterostructure}},\ }\href
  {https://www.nature.com/articles/s41586-022-04504-8} {\bibfield  {journal}
  {\bibinfo  {journal} {Nature}\ }\textbf {\bibinfo {volume} {604}},\ \bibinfo
  {pages} {653} (\bibinfo {year} {2022})}\BibitemShut {NoStop}%
\bibitem [{\citenamefont {Davydova}\ \emph {et~al.}(2022)\citenamefont
  {Davydova}, \citenamefont {Prembabu},\ and\ \citenamefont
  {Fu}}]{davydova2022universal}%
  \BibitemOpen
  \bibfield  {author} {\bibinfo {author} {\bibfnamefont {M.}~\bibnamefont
  {Davydova}}, \bibinfo {author} {\bibfnamefont {S.}~\bibnamefont {Prembabu}},\
  and\ \bibinfo {author} {\bibfnamefont {L.}~\bibnamefont {Fu}},\ }\bibfield
  {title} {\bibinfo {title} {{Universal Josephson diode effect}},\ }\href
  {https://www.science.org/doi/10.1126/sciadv.abo0309} {\bibfield  {journal}
  {\bibinfo  {journal} {Science advances}\ }\textbf {\bibinfo {volume} {8}},\
  \bibinfo {pages} {eabo0309} (\bibinfo {year} {2022})}\BibitemShut {NoStop}%
\bibitem [{\citenamefont {Pal}\ \emph {et~al.}(2022)\citenamefont {Pal},
  \citenamefont {Chakraborty}, \citenamefont {Sivakumar}, \citenamefont
  {Davydova}, \citenamefont {Gopi}, \citenamefont {Pandeya}, \citenamefont
  {Krieger}, \citenamefont {Zhang}, \citenamefont {Date}, \citenamefont {Ju}
  \emph {et~al.}}]{pal2022josephson}%
  \BibitemOpen
  \bibfield  {author} {\bibinfo {author} {\bibfnamefont {B.}~\bibnamefont
  {Pal}}, \bibinfo {author} {\bibfnamefont {A.}~\bibnamefont {Chakraborty}},
  \bibinfo {author} {\bibfnamefont {P.~K.}\ \bibnamefont {Sivakumar}}, \bibinfo
  {author} {\bibfnamefont {M.}~\bibnamefont {Davydova}}, \bibinfo {author}
  {\bibfnamefont {A.~K.}\ \bibnamefont {Gopi}}, \bibinfo {author}
  {\bibfnamefont {A.~K.}\ \bibnamefont {Pandeya}}, \bibinfo {author}
  {\bibfnamefont {J.~A.}\ \bibnamefont {Krieger}}, \bibinfo {author}
  {\bibfnamefont {Y.}~\bibnamefont {Zhang}}, \bibinfo {author} {\bibfnamefont
  {M.}~\bibnamefont {Date}}, \bibinfo {author} {\bibfnamefont {S.}~\bibnamefont
  {Ju}}, \emph {et~al.},\ }\bibfield  {title} {\bibinfo {title} {{Josephson
  diode effect from Cooper pair momentum in a topological semimetal}},\ }\href
  {https://doi.org/10.1038/s41567-022-01699-5} {\bibfield  {journal} {\bibinfo
  {journal} {Nature physics}\ }\textbf {\bibinfo {volume} {18}},\ \bibinfo
  {pages} {1228} (\bibinfo {year} {2022})}\BibitemShut {NoStop}%
\bibitem [{\citenamefont {Zhang}\ \emph {et~al.}(2022)\citenamefont {Zhang},
  \citenamefont {Gu}, \citenamefont {Li}, \citenamefont {Hu},\ and\
  \citenamefont {Jiang}}]{zhang2022general}%
  \BibitemOpen
  \bibfield  {author} {\bibinfo {author} {\bibfnamefont {Y.}~\bibnamefont
  {Zhang}}, \bibinfo {author} {\bibfnamefont {Y.}~\bibnamefont {Gu}}, \bibinfo
  {author} {\bibfnamefont {P.}~\bibnamefont {Li}}, \bibinfo {author}
  {\bibfnamefont {J.}~\bibnamefont {Hu}},\ and\ \bibinfo {author}
  {\bibfnamefont {K.}~\bibnamefont {Jiang}},\ }\bibfield  {title} {\bibinfo
  {title} {{General theory of Josephson diodes}},\ }\href
  {https://doi.org/10.1103/PhysRevX.12.041013} {\bibfield  {journal} {\bibinfo
  {journal} {Physical Review X}\ }\textbf {\bibinfo {volume} {12}},\ \bibinfo
  {pages} {041013} (\bibinfo {year} {2022})}\BibitemShut {NoStop}%
\bibitem [{\citenamefont {Zazunov}\ \emph
  {et~al.}(2024{\natexlab{a}})\citenamefont {Zazunov}, \citenamefont {Rech},
  \citenamefont {Jonckheere}, \citenamefont {Gr\'emaud}, \citenamefont
  {Martin},\ and\ \citenamefont {Egger}}]{PhysRevB.109.024504}%
  \BibitemOpen
  \bibfield  {author} {\bibinfo {author} {\bibfnamefont {A.}~\bibnamefont
  {Zazunov}}, \bibinfo {author} {\bibfnamefont {J.}~\bibnamefont {Rech}},
  \bibinfo {author} {\bibfnamefont {T.}~\bibnamefont {Jonckheere}}, \bibinfo
  {author} {\bibfnamefont {B.}~\bibnamefont {Gr\'emaud}}, \bibinfo {author}
  {\bibfnamefont {T.}~\bibnamefont {Martin}},\ and\ \bibinfo {author}
  {\bibfnamefont {R.}~\bibnamefont {Egger}},\ }\bibfield  {title} {\bibinfo
  {title} {{Nonreciprocal charge transport and subharmonic structure in
  voltage-biased Josephson diodes}},\ }\href
  {https://doi.org/10.1103/PhysRevB.109.024504} {\bibfield  {journal} {\bibinfo
   {journal} {Phys. Rev. B}\ }\textbf {\bibinfo {volume} {109}},\ \bibinfo
  {pages} {024504} (\bibinfo {year} {2024}{\natexlab{a}})}\BibitemShut
  {NoStop}%
\bibitem [{\citenamefont {Zazunov}\ \emph
  {et~al.}(2024{\natexlab{b}})\citenamefont {Zazunov}, \citenamefont {Rech},
  \citenamefont {Jonckheere}, \citenamefont {Grémaud}, \citenamefont
  {Martin},\ and\ \citenamefont
  {Egger}}]{zazunov2024approachingidealrectificationsuperconducting}%
  \BibitemOpen
  \bibfield  {author} {\bibinfo {author} {\bibfnamefont {A.}~\bibnamefont
  {Zazunov}}, \bibinfo {author} {\bibfnamefont {J.}~\bibnamefont {Rech}},
  \bibinfo {author} {\bibfnamefont {T.}~\bibnamefont {Jonckheere}}, \bibinfo
  {author} {\bibfnamefont {B.}~\bibnamefont {Grémaud}}, \bibinfo {author}
  {\bibfnamefont {T.}~\bibnamefont {Martin}},\ and\ \bibinfo {author}
  {\bibfnamefont {R.}~\bibnamefont {Egger}},\ }\href@noop {} {\bibinfo {title}
  {{Approaching ideal rectification in superconducting diodes through multiple
  Andreev reflections}}} (\bibinfo {year} {2024}{\natexlab{b}}),\ \Eprint
  {https://arxiv.org/abs/2307.14698} {arXiv:2307.14698 [cond-mat.supr-con]}
  \BibitemShut {NoStop}%
\bibitem [{\citenamefont {Levitov}\ \emph {et~al.}(1985)\citenamefont
  {Levitov}, \citenamefont {Nazarov},\ and\ \citenamefont
  {Eliashberg}}]{PismaZhETF.41.365}%
  \BibitemOpen
  \bibfield  {author} {\bibinfo {author} {\bibfnamefont {L.~S.}\ \bibnamefont
  {Levitov}}, \bibinfo {author} {\bibfnamefont {Y.~V.}\ \bibnamefont
  {Nazarov}},\ and\ \bibinfo {author} {\bibfnamefont {G.~M.}\ \bibnamefont
  {Eliashberg}},\ }\bibfield  {title} {\bibinfo {title} {Magnetostatics of
  superconductors without an inversion center},\ }\href
  {http://jetpletters.ru/ps/0/article_22366.shtml} {\bibfield  {journal}
  {\bibinfo  {journal} {JETP Letters}\ }\textbf {\bibinfo {volume} {41}},\
  \bibinfo {pages} {365} (\bibinfo {year} {1985})}\BibitemShut {NoStop}%
\bibitem [{\citenamefont {Edelstein}(1996)}]{Edelstein1996Jan}%
  \BibitemOpen
  \bibfield  {author} {\bibinfo {author} {\bibfnamefont {V.~M.}\ \bibnamefont
  {Edelstein}},\ }\bibfield  {title} {\bibinfo {title} {{The Ginzburg - Landau
  equation for superconductors of polar symmetry}},\ }\href
  {https://doi.org/10.1088/0953-8984/8/3/012} {\bibfield  {journal} {\bibinfo
  {journal} {J. Phys.: Condens. Matter}\ }\textbf {\bibinfo {volume} {8}},\
  \bibinfo {pages} {339} (\bibinfo {year} {1996})}\BibitemShut {NoStop}%
\bibitem [{\citenamefont {He}\ \emph {et~al.}(2022)\citenamefont {He},
  \citenamefont {Tanaka},\ and\ \citenamefont {Nagaosa}}]{He_2022}%
  \BibitemOpen
  \bibfield  {author} {\bibinfo {author} {\bibfnamefont {J.~J.}\ \bibnamefont
  {He}}, \bibinfo {author} {\bibfnamefont {Y.}~\bibnamefont {Tanaka}},\ and\
  \bibinfo {author} {\bibfnamefont {N.}~\bibnamefont {Nagaosa}},\ }\bibfield
  {title} {\bibinfo {title} {A phenomenological theory of superconductor
  diodes},\ }\href {https://doi.org/10.1088/1367-2630/ac6766} {\bibfield
  {journal} {\bibinfo  {journal} {New Journal of Physics}\ }\textbf {\bibinfo
  {volume} {24}},\ \bibinfo {pages} {053014} (\bibinfo {year}
  {2022})}\BibitemShut {NoStop}%
\bibitem [{\citenamefont {Gaggioli}\ \emph {et~al.}(2024)\citenamefont
  {Gaggioli}, \citenamefont {Hou}, \citenamefont {Moodera},\ and\ \citenamefont
  {Kamra}}]{Gaggioli2024May}%
  \BibitemOpen
  \bibfield  {author} {\bibinfo {author} {\bibfnamefont {F.}~\bibnamefont
  {Gaggioli}}, \bibinfo {author} {\bibfnamefont {Y.}~\bibnamefont {Hou}},
  \bibinfo {author} {\bibfnamefont {J.~S.}\ \bibnamefont {Moodera}},\ and\
  \bibinfo {author} {\bibfnamefont {A.}~\bibnamefont {Kamra}},\ }\bibfield
  {title} {\bibinfo {title} {{Nonreciprocity of supercurrent along applied
  magnetic field}},\ }\bibfield  {journal} {\bibinfo  {journal} {arXiv}\ }\href
  {https://doi.org/10.48550/arXiv.2405.05306} {10.48550/arXiv.2405.05306}
  (\bibinfo {year} {2024}),\ \Eprint {https://arxiv.org/abs/2405.05306}
  {2405.05306} \BibitemShut {NoStop}%
\bibitem [{\citenamefont {Fulde}\ and\ \citenamefont
  {Ferrell}(1964)}]{PhysRev.135.A550}%
  \BibitemOpen
  \bibfield  {author} {\bibinfo {author} {\bibfnamefont {P.}~\bibnamefont
  {Fulde}}\ and\ \bibinfo {author} {\bibfnamefont {R.~A.}\ \bibnamefont
  {Ferrell}},\ }\bibfield  {title} {\bibinfo {title} {Superconductivity in a
  strong spin-exchange field},\ }\href
  {https://doi.org/10.1103/PhysRev.135.A550} {\bibfield  {journal} {\bibinfo
  {journal} {Phys. Rev.}\ }\textbf {\bibinfo {volume} {135}},\ \bibinfo {pages}
  {A550} (\bibinfo {year} {1964})}\BibitemShut {NoStop}%
\bibitem [{\citenamefont {Möckli}(2022)}]{M_ckli_2022}%
  \BibitemOpen
  \bibfield  {author} {\bibinfo {author} {\bibfnamefont {D.}~\bibnamefont
  {Möckli}},\ }\bibfield  {title} {\bibinfo {title} {Unconventional
  singlet-triplet superconductivity},\ }\href
  {https://doi.org/10.1088/1742-6596/2164/1/012009} {\bibfield  {journal}
  {\bibinfo  {journal} {Journal of Physics: Conference Series}\ }\textbf
  {\bibinfo {volume} {2164}},\ \bibinfo {pages} {012009} (\bibinfo {year}
  {2022})}\BibitemShut {NoStop}%
\bibitem [{\citenamefont {Hu}\ \emph {et~al.}(2021)\citenamefont {Hu},
  \citenamefont {Wang},\ and\ \citenamefont {Shang}}]{Hu_2021}%
  \BibitemOpen
  \bibfield  {author} {\bibinfo {author} {\bibfnamefont {L.-H.}\ \bibnamefont
  {Hu}}, \bibinfo {author} {\bibfnamefont {X.}~\bibnamefont {Wang}},\ and\
  \bibinfo {author} {\bibfnamefont {T.}~\bibnamefont {Shang}},\ }\bibfield
  {title} {\bibinfo {title} {Spontaneous magnetization in unitary
  superconductors with time reversal symmetry breaking},\ }\bibfield  {journal}
  {\bibinfo  {journal} {Physical Review B}\ }\textbf {\bibinfo {volume}
  {104}},\ \href {https://doi.org/10.1103/physrevb.104.054520}
  {10.1103/physrevb.104.054520} (\bibinfo {year} {2021})\BibitemShut {NoStop}%
\bibitem [{\citenamefont {Wang}\ \emph {et~al.}(2016)\citenamefont {Wang},
  \citenamefont {Cho}, \citenamefont {Hughes},\ and\ \citenamefont
  {Fradkin}}]{PhysRevB.93.134512}%
  \BibitemOpen
  \bibfield  {author} {\bibinfo {author} {\bibfnamefont {Y.}~\bibnamefont
  {Wang}}, \bibinfo {author} {\bibfnamefont {G.~Y.}\ \bibnamefont {Cho}},
  \bibinfo {author} {\bibfnamefont {T.~L.}\ \bibnamefont {Hughes}},\ and\
  \bibinfo {author} {\bibfnamefont {E.}~\bibnamefont {Fradkin}},\ }\bibfield
  {title} {\bibinfo {title} {Topological superconducting phases from inversion
  symmetry breaking order in spin-orbit-coupled systems},\ }\href
  {https://doi.org/10.1103/PhysRevB.93.134512} {\bibfield  {journal} {\bibinfo
  {journal} {Phys. Rev. B}\ }\textbf {\bibinfo {volume} {93}},\ \bibinfo
  {pages} {134512} (\bibinfo {year} {2016})}\BibitemShut {NoStop}%
\bibitem [{\citenamefont {Wang}\ and\ \citenamefont
  {Fu}(2017)}]{PhysRevLett.119.187003}%
  \BibitemOpen
  \bibfield  {author} {\bibinfo {author} {\bibfnamefont {Y.}~\bibnamefont
  {Wang}}\ and\ \bibinfo {author} {\bibfnamefont {L.}~\bibnamefont {Fu}},\
  }\bibfield  {title} {\bibinfo {title} {Topological phase transitions in
  multicomponent superconductors},\ }\href
  {https://doi.org/10.1103/PhysRevLett.119.187003} {\bibfield  {journal}
  {\bibinfo  {journal} {Phys. Rev. Lett.}\ }\textbf {\bibinfo {volume} {119}},\
  \bibinfo {pages} {187003} (\bibinfo {year} {2017})}\BibitemShut {NoStop}%
\bibitem [{\citenamefont {D{\'\i}ez-M{\'e}rida}\ \emph
  {et~al.}(2023)\citenamefont {D{\'\i}ez-M{\'e}rida}, \citenamefont
  {D{\'\i}ez-Carl{\'o}n}, \citenamefont {Yang}, \citenamefont {Xie},
  \citenamefont {Gao}, \citenamefont {Senior}, \citenamefont {Watanabe},
  \citenamefont {Taniguchi}, \citenamefont {Lu}, \citenamefont {Higginbotham}
  \emph {et~al.}}]{diez2021magnetic}%
  \BibitemOpen
  \bibfield  {author} {\bibinfo {author} {\bibfnamefont {J.}~\bibnamefont
  {D{\'\i}ez-M{\'e}rida}}, \bibinfo {author} {\bibfnamefont {A.}~\bibnamefont
  {D{\'\i}ez-Carl{\'o}n}}, \bibinfo {author} {\bibfnamefont {S.}~\bibnamefont
  {Yang}}, \bibinfo {author} {\bibfnamefont {Y.-M.}\ \bibnamefont {Xie}},
  \bibinfo {author} {\bibfnamefont {X.-J.}\ \bibnamefont {Gao}}, \bibinfo
  {author} {\bibfnamefont {J.}~\bibnamefont {Senior}}, \bibinfo {author}
  {\bibfnamefont {K.}~\bibnamefont {Watanabe}}, \bibinfo {author}
  {\bibfnamefont {T.}~\bibnamefont {Taniguchi}}, \bibinfo {author}
  {\bibfnamefont {X.}~\bibnamefont {Lu}}, \bibinfo {author} {\bibfnamefont
  {A.~P.}\ \bibnamefont {Higginbotham}}, \emph {et~al.},\ }\bibfield  {title}
  {\bibinfo {title} {{Symmetry-broken Josephson junctions and superconducting
  diodes in magic-angle twisted bilayer graphene}},\ }\href
  {https://doi.org/10.1038/s41467-023-38005-7} {\bibfield  {journal} {\bibinfo
  {journal} {Nature Communications}\ }\textbf {\bibinfo {volume} {14}},\
  \bibinfo {pages} {2396} (\bibinfo {year} {2023})}\BibitemShut {NoStop}%
\bibitem [{\citenamefont {Lin}\ \emph {et~al.}(2022)\citenamefont {Lin},
  \citenamefont {Siriviboon}, \citenamefont {Scammell}, \citenamefont {Liu},
  \citenamefont {Rhodes}, \citenamefont {Watanabe}, \citenamefont {Taniguchi},
  \citenamefont {Hone}, \citenamefont {Scheurer},\ and\ \citenamefont
  {Li}}]{lin2022zero}%
  \BibitemOpen
  \bibfield  {author} {\bibinfo {author} {\bibfnamefont {J.-X.}\ \bibnamefont
  {Lin}}, \bibinfo {author} {\bibfnamefont {P.}~\bibnamefont {Siriviboon}},
  \bibinfo {author} {\bibfnamefont {H.~D.}\ \bibnamefont {Scammell}}, \bibinfo
  {author} {\bibfnamefont {S.}~\bibnamefont {Liu}}, \bibinfo {author}
  {\bibfnamefont {D.}~\bibnamefont {Rhodes}}, \bibinfo {author} {\bibfnamefont
  {K.}~\bibnamefont {Watanabe}}, \bibinfo {author} {\bibfnamefont
  {T.}~\bibnamefont {Taniguchi}}, \bibinfo {author} {\bibfnamefont
  {J.}~\bibnamefont {Hone}}, \bibinfo {author} {\bibfnamefont {M.~S.}\
  \bibnamefont {Scheurer}},\ and\ \bibinfo {author} {\bibfnamefont
  {J.}~\bibnamefont {Li}},\ }\bibfield  {title} {\bibinfo {title} {Zero-field
  superconducting diode effect in small-twist-angle trilayer graphene},\ }\href
  {https://doi.org/10.1038/s41567-022-01700-1} {\bibfield  {journal} {\bibinfo
  {journal} {Nature Physics}\ }\textbf {\bibinfo {volume} {18}},\ \bibinfo
  {pages} {1221} (\bibinfo {year} {2022})}\BibitemShut {NoStop}%
\bibitem [{\citenamefont {Zhao}\ \emph {et~al.}(2021)\citenamefont {Zhao},
  \citenamefont {Poccia}, \citenamefont {Cui}, \citenamefont {Volkov},
  \citenamefont {Yoo}, \citenamefont {Engelke}, \citenamefont {Ronen},
  \citenamefont {Zhong}, \citenamefont {Gu}, \citenamefont {Plugge},
  \citenamefont {Tummuru}, \citenamefont {Franz}, \citenamefont {Pixley},\ and\
  \citenamefont {Kim}}]{zhao2021emergent}%
  \BibitemOpen
  \bibfield  {author} {\bibinfo {author} {\bibfnamefont {S.~Y.~F.}\
  \bibnamefont {Zhao}}, \bibinfo {author} {\bibfnamefont {N.}~\bibnamefont
  {Poccia}}, \bibinfo {author} {\bibfnamefont {X.}~\bibnamefont {Cui}},
  \bibinfo {author} {\bibfnamefont {P.~A.}\ \bibnamefont {Volkov}}, \bibinfo
  {author} {\bibfnamefont {H.}~\bibnamefont {Yoo}}, \bibinfo {author}
  {\bibfnamefont {R.}~\bibnamefont {Engelke}}, \bibinfo {author} {\bibfnamefont
  {Y.}~\bibnamefont {Ronen}}, \bibinfo {author} {\bibfnamefont
  {R.}~\bibnamefont {Zhong}}, \bibinfo {author} {\bibfnamefont
  {G.}~\bibnamefont {Gu}}, \bibinfo {author} {\bibfnamefont {S.}~\bibnamefont
  {Plugge}}, \bibinfo {author} {\bibfnamefont {T.}~\bibnamefont {Tummuru}},
  \bibinfo {author} {\bibfnamefont {M.}~\bibnamefont {Franz}}, \bibinfo
  {author} {\bibfnamefont {J.~H.}\ \bibnamefont {Pixley}},\ and\ \bibinfo
  {author} {\bibfnamefont {P.}~\bibnamefont {Kim}},\ }\href@noop {} {\bibinfo
  {title} {Emergent interfacial superconductivity between twisted cuprate
  superconductors}} (\bibinfo {year} {2021}),\ \Eprint
  {https://arxiv.org/abs/2108.13455} {arXiv:2108.13455 [cond-mat.supr-con]}
  \BibitemShut {NoStop}%
\bibitem [{\citenamefont {Zhu}\ \emph {et~al.}(2021)\citenamefont {Zhu},
  \citenamefont {Papaj}, \citenamefont {Nie}, \citenamefont {Xu}, \citenamefont
  {Gu}, \citenamefont {Yang}, \citenamefont {Guan}, \citenamefont {Wang},
  \citenamefont {Li}, \citenamefont {Liu} \emph {et~al.}}]{zhu2020discovery}%
  \BibitemOpen
  \bibfield  {author} {\bibinfo {author} {\bibfnamefont {Z.}~\bibnamefont
  {Zhu}}, \bibinfo {author} {\bibfnamefont {M.}~\bibnamefont {Papaj}}, \bibinfo
  {author} {\bibfnamefont {X.-A.}\ \bibnamefont {Nie}}, \bibinfo {author}
  {\bibfnamefont {H.-K.}\ \bibnamefont {Xu}}, \bibinfo {author} {\bibfnamefont
  {Y.-S.}\ \bibnamefont {Gu}}, \bibinfo {author} {\bibfnamefont
  {X.}~\bibnamefont {Yang}}, \bibinfo {author} {\bibfnamefont {D.}~\bibnamefont
  {Guan}}, \bibinfo {author} {\bibfnamefont {S.}~\bibnamefont {Wang}}, \bibinfo
  {author} {\bibfnamefont {Y.}~\bibnamefont {Li}}, \bibinfo {author}
  {\bibfnamefont {C.}~\bibnamefont {Liu}}, \emph {et~al.},\ }\bibfield  {title}
  {\bibinfo {title} {Discovery of segmented fermi surface induced by cooper
  pair momentum},\ }\href {https://doi.org/10.1126/science.abf1077} {\bibfield
  {journal} {\bibinfo  {journal} {Science}\ }\textbf {\bibinfo {volume}
  {374}},\ \bibinfo {pages} {1381} (\bibinfo {year} {2021})}\BibitemShut
  {NoStop}%
\bibitem [{\citenamefont {Chen}\ \emph {et~al.}(2018)\citenamefont {Chen},
  \citenamefont {Park}, \citenamefont {Gill}, \citenamefont {Xiao},
  \citenamefont {Reig-i Plessis}, \citenamefont {MacDougall}, \citenamefont
  {Gilbert},\ and\ \citenamefont {Mason}}]{chen2018finite}%
  \BibitemOpen
  \bibfield  {author} {\bibinfo {author} {\bibfnamefont {A.~Q.}\ \bibnamefont
  {Chen}}, \bibinfo {author} {\bibfnamefont {M.~J.}\ \bibnamefont {Park}},
  \bibinfo {author} {\bibfnamefont {S.~T.}\ \bibnamefont {Gill}}, \bibinfo
  {author} {\bibfnamefont {Y.}~\bibnamefont {Xiao}}, \bibinfo {author}
  {\bibfnamefont {D.}~\bibnamefont {Reig-i Plessis}}, \bibinfo {author}
  {\bibfnamefont {G.~J.}\ \bibnamefont {MacDougall}}, \bibinfo {author}
  {\bibfnamefont {M.~J.}\ \bibnamefont {Gilbert}},\ and\ \bibinfo {author}
  {\bibfnamefont {N.}~\bibnamefont {Mason}},\ }\bibfield  {title} {\bibinfo
  {title} {{Finite momentum Cooper pairing in three-dimensional topological
  insulator Josephson junctions}},\ }\href
  {https://www.nature.com/articles/s41467-018-05993-w} {\bibfield  {journal}
  {\bibinfo  {journal} {Nature communications}\ }\textbf {\bibinfo {volume}
  {9}},\ \bibinfo {pages} {3478} (\bibinfo {year} {2018})}\BibitemShut
  {NoStop}%
\bibitem [{\citenamefont {Rohlfing}\ \emph {et~al.}(2009)\citenamefont
  {Rohlfing}, \citenamefont {Tkachov}, \citenamefont {Otto}, \citenamefont
  {Richter}, \citenamefont {Weiss}, \citenamefont {Borghs},\ and\ \citenamefont
  {Strunk}}]{PhysRevB.80.220507}%
  \BibitemOpen
  \bibfield  {author} {\bibinfo {author} {\bibfnamefont {F.}~\bibnamefont
  {Rohlfing}}, \bibinfo {author} {\bibfnamefont {G.}~\bibnamefont {Tkachov}},
  \bibinfo {author} {\bibfnamefont {F.}~\bibnamefont {Otto}}, \bibinfo {author}
  {\bibfnamefont {K.}~\bibnamefont {Richter}}, \bibinfo {author} {\bibfnamefont
  {D.}~\bibnamefont {Weiss}}, \bibinfo {author} {\bibfnamefont
  {G.}~\bibnamefont {Borghs}},\ and\ \bibinfo {author} {\bibfnamefont
  {C.}~\bibnamefont {Strunk}},\ }\bibfield  {title} {\bibinfo {title} {{Doppler
  shift in Andreev reflection from a moving superconducting condensate in
  Nb/InAs Josephson junctions}},\ }\href
  {https://doi.org/10.1103/PhysRevB.80.220507} {\bibfield  {journal} {\bibinfo
  {journal} {Phys. Rev. B}\ }\textbf {\bibinfo {volume} {80}},\ \bibinfo
  {pages} {220507} (\bibinfo {year} {2009})}\BibitemShut {NoStop}%
\bibitem [{\citenamefont {Banerjee}\ \emph {et~al.}(2023)\citenamefont
  {Banerjee}, \citenamefont {Geier}, \citenamefont {Rahman}, \citenamefont
  {Thomas}, \citenamefont {Wang}, \citenamefont {Manfra}, \citenamefont
  {Flensberg},\ and\ \citenamefont {Marcus}}]{banerjee2023phase}%
  \BibitemOpen
  \bibfield  {author} {\bibinfo {author} {\bibfnamefont {A.}~\bibnamefont
  {Banerjee}}, \bibinfo {author} {\bibfnamefont {M.}~\bibnamefont {Geier}},
  \bibinfo {author} {\bibfnamefont {M.~A.}\ \bibnamefont {Rahman}}, \bibinfo
  {author} {\bibfnamefont {C.}~\bibnamefont {Thomas}}, \bibinfo {author}
  {\bibfnamefont {T.}~\bibnamefont {Wang}}, \bibinfo {author} {\bibfnamefont
  {M.~J.}\ \bibnamefont {Manfra}}, \bibinfo {author} {\bibfnamefont
  {K.}~\bibnamefont {Flensberg}},\ and\ \bibinfo {author} {\bibfnamefont
  {C.~M.}\ \bibnamefont {Marcus}},\ }\bibfield  {title} {\bibinfo {title}
  {{Phase Asymmetry of Andreev Spectra from Cooper-Pair Momentum}},\ }\href
  {https://doi.org/10.1103/PhysRevLett.131.196301} {\bibfield  {journal}
  {\bibinfo  {journal} {Phys. Rev. Lett.}\ }\textbf {\bibinfo {volume} {131}},\
  \bibinfo {pages} {196301} (\bibinfo {year} {2023})}\BibitemShut {NoStop}%
\bibitem [{\citenamefont {Hart}\ \emph {et~al.}(2017)\citenamefont {Hart},
  \citenamefont {Ren}, \citenamefont {Kosowsky}, \citenamefont {Ben-Shach},
  \citenamefont {Leubner}, \citenamefont {Br{\"u}ne}, \citenamefont {Buhmann},
  \citenamefont {Molenkamp}, \citenamefont {Halperin},\ and\ \citenamefont
  {Yacoby}}]{hart2017controlled}%
  \BibitemOpen
  \bibfield  {author} {\bibinfo {author} {\bibfnamefont {S.}~\bibnamefont
  {Hart}}, \bibinfo {author} {\bibfnamefont {H.}~\bibnamefont {Ren}}, \bibinfo
  {author} {\bibfnamefont {M.}~\bibnamefont {Kosowsky}}, \bibinfo {author}
  {\bibfnamefont {G.}~\bibnamefont {Ben-Shach}}, \bibinfo {author}
  {\bibfnamefont {P.}~\bibnamefont {Leubner}}, \bibinfo {author} {\bibfnamefont
  {C.}~\bibnamefont {Br{\"u}ne}}, \bibinfo {author} {\bibfnamefont
  {H.}~\bibnamefont {Buhmann}}, \bibinfo {author} {\bibfnamefont {L.~W.}\
  \bibnamefont {Molenkamp}}, \bibinfo {author} {\bibfnamefont {B.~I.}\
  \bibnamefont {Halperin}},\ and\ \bibinfo {author} {\bibfnamefont
  {A.}~\bibnamefont {Yacoby}},\ }\bibfield  {title} {\bibinfo {title}
  {Controlled finite momentum pairing and spatially varying order parameter in
  proximitized hgte quantum wells},\ }\href
  {https://www.nature.com/articles/nphys3877} {\bibfield  {journal} {\bibinfo
  {journal} {Nature Physics}\ }\textbf {\bibinfo {volume} {13}},\ \bibinfo
  {pages} {87} (\bibinfo {year} {2017})}\BibitemShut {NoStop}%
\bibitem [{\citenamefont {Baumgartner}\ \emph
  {et~al.}(2022{\natexlab{b}})\citenamefont {Baumgartner}, \citenamefont
  {Fuchs}, \citenamefont {Costa}, \citenamefont {Reinhardt}, \citenamefont
  {Gronin}, \citenamefont {Gardner}, \citenamefont {Lindemann}, \citenamefont
  {Manfra}, \citenamefont {Faria~Junior}, \citenamefont {Kochan} \emph
  {et~al.}}]{baumgartner2021josephson}%
  \BibitemOpen
  \bibfield  {author} {\bibinfo {author} {\bibfnamefont {C.}~\bibnamefont
  {Baumgartner}}, \bibinfo {author} {\bibfnamefont {L.}~\bibnamefont {Fuchs}},
  \bibinfo {author} {\bibfnamefont {A.}~\bibnamefont {Costa}}, \bibinfo
  {author} {\bibfnamefont {S.}~\bibnamefont {Reinhardt}}, \bibinfo {author}
  {\bibfnamefont {S.}~\bibnamefont {Gronin}}, \bibinfo {author} {\bibfnamefont
  {G.~C.}\ \bibnamefont {Gardner}}, \bibinfo {author} {\bibfnamefont
  {T.}~\bibnamefont {Lindemann}}, \bibinfo {author} {\bibfnamefont {M.~J.}\
  \bibnamefont {Manfra}}, \bibinfo {author} {\bibfnamefont {P.~E.}\
  \bibnamefont {Faria~Junior}}, \bibinfo {author} {\bibfnamefont
  {D.}~\bibnamefont {Kochan}}, \emph {et~al.},\ }\bibfield  {title} {\bibinfo
  {title} {{Supercurrent rectification and magnetochiral effects in symmetric
  Josephson junctions}},\ }\href
  {https://www.nature.com/articles/s41565-021-01009-9} {\bibfield  {journal}
  {\bibinfo  {journal} {Nature nanotechnology}\ }\textbf {\bibinfo {volume}
  {17}},\ \bibinfo {pages} {39} (\bibinfo {year}
  {2022}{\natexlab{b}})}\BibitemShut {NoStop}%
\bibitem [{\citenamefont {Yazdani}\ \emph {et~al.}(1997)\citenamefont
  {Yazdani}, \citenamefont {Jones}, \citenamefont {Lutz}, \citenamefont
  {Crommie},\ and\ \citenamefont {Eigler}}]{Yazdani1997Mar}%
  \BibitemOpen
  \bibfield  {author} {\bibinfo {author} {\bibfnamefont {A.}~\bibnamefont
  {Yazdani}}, \bibinfo {author} {\bibfnamefont {B.~A.}\ \bibnamefont {Jones}},
  \bibinfo {author} {\bibfnamefont {C.~P.}\ \bibnamefont {Lutz}}, \bibinfo
  {author} {\bibfnamefont {M.~F.}\ \bibnamefont {Crommie}},\ and\ \bibinfo
  {author} {\bibfnamefont {D.~M.}\ \bibnamefont {Eigler}},\ }\bibfield  {title}
  {\bibinfo {title} {{Probing the Local Effects of Magnetic Impurities on
  Superconductivity}},\ }\href {https://doi.org/10.1126/science.275.5307.1767}
  {\bibfield  {journal} {\bibinfo  {journal} {Science}\ }\textbf {\bibinfo
  {volume} {275}},\ \bibinfo {pages} {1767} (\bibinfo {year}
  {1997})}\BibitemShut {NoStop}%
\bibitem [{\citenamefont {Heinrich}\ \emph {et~al.}(2018)\citenamefont
  {Heinrich}, \citenamefont {Pascual},\ and\ \citenamefont
  {Franke}}]{Heinrich2018Feb}%
  \BibitemOpen
  \bibfield  {author} {\bibinfo {author} {\bibfnamefont {B.~W.}\ \bibnamefont
  {Heinrich}}, \bibinfo {author} {\bibfnamefont {J.~I.}\ \bibnamefont
  {Pascual}},\ and\ \bibinfo {author} {\bibfnamefont {K.~J.}\ \bibnamefont
  {Franke}},\ }\bibfield  {title} {\bibinfo {title} {{Single magnetic
  adsorbates on s-wave superconductors}},\ }\href
  {https://doi.org/10.1016/j.progsurf.2018.01.001} {\bibfield  {journal}
  {\bibinfo  {journal} {Prog. Surf. Sci.}\ }\textbf {\bibinfo {volume} {93}},\
  \bibinfo {pages} {1} (\bibinfo {year} {2018})}\BibitemShut {NoStop}%
\bibitem [{Note1()}]{Note1}%
  \BibitemOpen
  \bibinfo {note} {Generically, for a non-reciprocal metal, the Fermi
  velocities $v_{F+},\protect \, v_{F-}$ around the Fermi momenta $k_{F+}$ and
  $k_{F-}$ are distinct. Assuming that $\Delta , v_{F\pm } (\pm k_0 - k_{F\pm
  }) \ll E_F$, the quasiparticle dispersion, Eq.~\protect \eqref {eq:asymm1},
  for this case can be expressed as $$ E(k) = D(k)\pm \protect \sqrt {\left
  (\protect \frac {v_{F+}\delta k_{+}+v_{F-}\delta k_{-}}{2}\right )^{2}+\Delta
  ^{2}} $$ with $\delta k_\pm = \pm k-k_{F\pm }$ and $D(k) = \protect \frac
  {v_{F+}\delta k_{+}-v_{F-}\delta k_{-}}{2}$, which is valid around both
  $k_{F\pm }$. In this limit, the gap opening occurs around $k = \pm k_0$ with
  $k_{0}=\protect \frac
  {k_{F+}v_{F+}+k_{F-}v_{F-}}{v_{F+}-v_{F-}}$.}\BibitemShut {Stop}%
\bibitem [{\citenamefont {Chiles}\ \emph {et~al.}(2023)\citenamefont {Chiles},
  \citenamefont {Arnault}, \citenamefont {Chen}, \citenamefont {Larson},
  \citenamefont {Zhao}, \citenamefont {Watanabe}, \citenamefont {Taniguchi},
  \citenamefont {Amet},\ and\ \citenamefont
  {Finkelstein}}]{chiles2023nonreciprocal}%
  \BibitemOpen
  \bibfield  {author} {\bibinfo {author} {\bibfnamefont {J.}~\bibnamefont
  {Chiles}}, \bibinfo {author} {\bibfnamefont {E.~G.}\ \bibnamefont {Arnault}},
  \bibinfo {author} {\bibfnamefont {C.-C.}\ \bibnamefont {Chen}}, \bibinfo
  {author} {\bibfnamefont {T.~F.}\ \bibnamefont {Larson}}, \bibinfo {author}
  {\bibfnamefont {L.}~\bibnamefont {Zhao}}, \bibinfo {author} {\bibfnamefont
  {K.}~\bibnamefont {Watanabe}}, \bibinfo {author} {\bibfnamefont
  {T.}~\bibnamefont {Taniguchi}}, \bibinfo {author} {\bibfnamefont
  {F.}~\bibnamefont {Amet}},\ and\ \bibinfo {author} {\bibfnamefont
  {G.}~\bibnamefont {Finkelstein}},\ }\bibfield  {title} {\bibinfo {title}
  {{Nonreciprocal Supercurrents in a Field-Free Graphene Josephson Triode}},\
  }\href {https://pubs.acs.org/doi/full/10.1021/acs.nanolett.3c01276}
  {\bibfield  {journal} {\bibinfo  {journal} {Nano Letters}\ } (\bibinfo {year}
  {2023})}\BibitemShut {NoStop}%
\bibitem [{\citenamefont {Beenakker}(1997)}]{RevModPhys.69.731}%
  \BibitemOpen
  \bibfield  {author} {\bibinfo {author} {\bibfnamefont {C.~W.~J.}\
  \bibnamefont {Beenakker}},\ }\bibfield  {title} {\bibinfo {title}
  {Random-matrix theory of quantum transport},\ }\href
  {https://doi.org/10.1103/RevModPhys.69.731} {\bibfield  {journal} {\bibinfo
  {journal} {Rev. Mod. Phys.}\ }\textbf {\bibinfo {volume} {69}},\ \bibinfo
  {pages} {731} (\bibinfo {year} {1997})}\BibitemShut {NoStop}%
\bibitem [{Note2()}]{Note2}%
  \BibitemOpen
  \bibinfo {note} {In more detail, the symmetry under the voltage reversal of
  the differential conductance of an N-S interface can be derived from the
  elastic scattering theory by employing unitarity of the reflection matrix
  (i.e. the incoming quasiparticle current is entirely reflected into the same
  lead) and the particle-hole conjugation constraint inherited from Nambu
  formalism of the Bogoliubov-de Gennes description of the superconducting
  state.}\BibitemShut {Stop}%
\bibitem [{Note3()}]{Note3}%
  \BibitemOpen
  \bibinfo {note} {In addition, above the spectral gap of the superconductor,
  the incoming electrons and holes can enter the superconductor as Bogoliubov
  quasiparticles. Above this voltage, the reflection matrix is not unitary, and
  the local conductance is generically not symmetric in voltage.}\BibitemShut
  {Stop}%
\bibitem [{\citenamefont {Blonder}\ \emph {et~al.}(1982)\citenamefont
  {Blonder}, \citenamefont {Tinkham},\ and\ \citenamefont
  {Klapwijk}}]{PhysRevB.25.4515}%
  \BibitemOpen
  \bibfield  {author} {\bibinfo {author} {\bibfnamefont {G.~E.}\ \bibnamefont
  {Blonder}}, \bibinfo {author} {\bibfnamefont {M.}~\bibnamefont {Tinkham}},\
  and\ \bibinfo {author} {\bibfnamefont {T.~M.}\ \bibnamefont {Klapwijk}},\
  }\bibfield  {title} {\bibinfo {title} {Transition from metallic to tunneling
  regimes in superconducting microconstrictions: Excess current, charge
  imbalance, and supercurrent conversion},\ }\href
  {https://doi.org/10.1103/PhysRevB.25.4515} {\bibfield  {journal} {\bibinfo
  {journal} {Phys. Rev. B}\ }\textbf {\bibinfo {volume} {25}},\ \bibinfo
  {pages} {4515} (\bibinfo {year} {1982})}\BibitemShut {NoStop}%
\bibitem [{\citenamefont {Groth}\ \emph {et~al.}(2014)\citenamefont {Groth},
  \citenamefont {Wimmer}, \citenamefont {Akhmerov},\ and\ \citenamefont
  {Waintal}}]{Groth2014Jun}%
  \BibitemOpen
  \bibfield  {author} {\bibinfo {author} {\bibfnamefont {C.~W.}\ \bibnamefont
  {Groth}}, \bibinfo {author} {\bibfnamefont {M.}~\bibnamefont {Wimmer}},
  \bibinfo {author} {\bibfnamefont {A.~R.}\ \bibnamefont {Akhmerov}},\ and\
  \bibinfo {author} {\bibfnamefont {X.}~\bibnamefont {Waintal}},\ }\bibfield
  {title} {\bibinfo {title} {{Kwant: a software package for quantum
  transport}},\ }\href {https://doi.org/10.1088/1367-2630/16/6/063065}
  {\bibfield  {journal} {\bibinfo  {journal} {New J. Phys.}\ }\textbf {\bibinfo
  {volume} {16}},\ \bibinfo {pages} {063065} (\bibinfo {year}
  {2014})}\BibitemShut {NoStop}%
\bibitem [{\citenamefont {Partyka}\ \emph {et~al.}(2010)\citenamefont
  {Partyka}, \citenamefont {Sadzikowski},\ and\ \citenamefont
  {Tachibana}}]{Partyka_2010}%
  \BibitemOpen
  \bibfield  {author} {\bibinfo {author} {\bibfnamefont {T.~L.}\ \bibnamefont
  {Partyka}}, \bibinfo {author} {\bibfnamefont {M.}~\bibnamefont
  {Sadzikowski}},\ and\ \bibinfo {author} {\bibfnamefont {M.}~\bibnamefont
  {Tachibana}},\ }\bibfield  {title} {\bibinfo {title} {{Andreev reflection
  between a normal metal and the FFLO superconductor}},\ }\href
  {https://doi.org/10.1016/j.physc.2010.01.003} {\bibfield  {journal} {\bibinfo
   {journal} {Physica C: Superconductivity}\ }\textbf {\bibinfo {volume}
  {470}},\ \bibinfo {pages} {277–280} (\bibinfo {year} {2010})}\BibitemShut
  {NoStop}%
\bibitem [{\citenamefont {Kaczmarczyk}\ \emph {et~al.}(2011)\citenamefont
  {Kaczmarczyk}, \citenamefont {Sadzikowski},\ and\ \citenamefont
  {Spałek}}]{Kaczmarczyk_2011}%
  \BibitemOpen
  \bibfield  {author} {\bibinfo {author} {\bibfnamefont {J.}~\bibnamefont
  {Kaczmarczyk}}, \bibinfo {author} {\bibfnamefont {M.}~\bibnamefont
  {Sadzikowski}},\ and\ \bibinfo {author} {\bibfnamefont {J.}~\bibnamefont
  {Spałek}},\ }\bibfield  {title} {\bibinfo {title} {{Andreev reflection
  between a normal metal and the FFLO superconductor II: A self-consistent
  approach}},\ }\href {https://doi.org/10.1016/j.physc.2010.10.009} {\bibfield
  {journal} {\bibinfo  {journal} {Physica C: Superconductivity}\ }\textbf
  {\bibinfo {volume} {471}},\ \bibinfo {pages} {193–198} (\bibinfo {year}
  {2011})}\BibitemShut {NoStop}%
\bibitem [{\citenamefont {Burset}\ \emph {et~al.}(2015)\citenamefont {Burset},
  \citenamefont {Lu}, \citenamefont {Tkachov}, \citenamefont {Tanaka},
  \citenamefont {Hankiewicz},\ and\ \citenamefont
  {Trauzettel}}]{PhysRevB.92.205424}%
  \BibitemOpen
  \bibfield  {author} {\bibinfo {author} {\bibfnamefont {P.}~\bibnamefont
  {Burset}}, \bibinfo {author} {\bibfnamefont {B.}~\bibnamefont {Lu}}, \bibinfo
  {author} {\bibfnamefont {G.}~\bibnamefont {Tkachov}}, \bibinfo {author}
  {\bibfnamefont {Y.}~\bibnamefont {Tanaka}}, \bibinfo {author} {\bibfnamefont
  {E.~M.}\ \bibnamefont {Hankiewicz}},\ and\ \bibinfo {author} {\bibfnamefont
  {B.}~\bibnamefont {Trauzettel}},\ }\bibfield  {title} {\bibinfo {title}
  {Superconducting proximity effect in three-dimensional topological insulators
  in the presence of a magnetic field},\ }\href
  {https://doi.org/10.1103/PhysRevB.92.205424} {\bibfield  {journal} {\bibinfo
  {journal} {Phys. Rev. B}\ }\textbf {\bibinfo {volume} {92}},\ \bibinfo
  {pages} {205424} (\bibinfo {year} {2015})}\BibitemShut {NoStop}%
\bibitem [{\citenamefont {Yuan}\ and\ \citenamefont {Fu}(2018)}]{yuan2018}%
  \BibitemOpen
  \bibfield  {author} {\bibinfo {author} {\bibfnamefont {N.~F.~Q.}\
  \bibnamefont {Yuan}}\ and\ \bibinfo {author} {\bibfnamefont {L.}~\bibnamefont
  {Fu}},\ }\bibfield  {title} {\bibinfo {title} {Zeeman-induced gapless
  superconductivity with a partial fermi surface},\ }\href
  {https://doi.org/10.1103/PhysRevB.97.115139} {\bibfield  {journal} {\bibinfo
  {journal} {Phys. Rev. B}\ }\textbf {\bibinfo {volume} {97}},\ \bibinfo
  {pages} {115139} (\bibinfo {year} {2018})}\BibitemShut {NoStop}%
\bibitem [{\citenamefont {Papaj}\ and\ \citenamefont
  {Fu}(2021)}]{papaj2021creating}%
  \BibitemOpen
  \bibfield  {author} {\bibinfo {author} {\bibfnamefont {M.}~\bibnamefont
  {Papaj}}\ and\ \bibinfo {author} {\bibfnamefont {L.}~\bibnamefont {Fu}},\
  }\bibfield  {title} {\bibinfo {title} {Creating majorana modes from segmented
  fermi surface},\ }\href {https://doi.org/10.1038/s41467-020-20690-3}
  {\bibfield  {journal} {\bibinfo  {journal} {Nature communications}\ }\textbf
  {\bibinfo {volume} {12}},\ \bibinfo {pages} {1} (\bibinfo {year}
  {2021})}\BibitemShut {NoStop}%
\bibitem [{\citenamefont {Phan}\ \emph {et~al.}(2022)\citenamefont {Phan},
  \citenamefont {Senior}, \citenamefont {Ghazaryan}, \citenamefont
  {Hatefipour}, \citenamefont {Strickland}, \citenamefont {Shabani},
  \citenamefont {Serbyn},\ and\ \citenamefont
  {Higginbotham}}]{PhysRevLett.128.107701}%
  \BibitemOpen
  \bibfield  {author} {\bibinfo {author} {\bibfnamefont {D.}~\bibnamefont
  {Phan}}, \bibinfo {author} {\bibfnamefont {J.}~\bibnamefont {Senior}},
  \bibinfo {author} {\bibfnamefont {A.}~\bibnamefont {Ghazaryan}}, \bibinfo
  {author} {\bibfnamefont {M.}~\bibnamefont {Hatefipour}}, \bibinfo {author}
  {\bibfnamefont {W.~M.}\ \bibnamefont {Strickland}}, \bibinfo {author}
  {\bibfnamefont {J.}~\bibnamefont {Shabani}}, \bibinfo {author} {\bibfnamefont
  {M.}~\bibnamefont {Serbyn}},\ and\ \bibinfo {author} {\bibfnamefont {A.~P.}\
  \bibnamefont {Higginbotham}},\ }\bibfield  {title} {\bibinfo {title}
  {Detecting induced $p\ifmmode\pm\else\textpm\fi{}ip$ pairing at the al-inas
  interface with a quantum microwave circuit},\ }\href
  {https://doi.org/10.1103/PhysRevLett.128.107701} {\bibfield  {journal}
  {\bibinfo  {journal} {Phys. Rev. Lett.}\ }\textbf {\bibinfo {volume} {128}},\
  \bibinfo {pages} {107701} (\bibinfo {year} {2022})}\BibitemShut {NoStop}%
\bibitem [{\citenamefont {Zhang}\ \emph {et~al.}(2004)\citenamefont {Zhang},
  \citenamefont {Ting},\ and\ \citenamefont {Hu}}]{PhysRevB.70.172508}%
  \BibitemOpen
  \bibfield  {author} {\bibinfo {author} {\bibfnamefont {D.}~\bibnamefont
  {Zhang}}, \bibinfo {author} {\bibfnamefont {C.~S.}\ \bibnamefont {Ting}},\
  and\ \bibinfo {author} {\bibfnamefont {C.-R.}\ \bibnamefont {Hu}},\
  }\bibfield  {title} {\bibinfo {title} {Conductance characteristics between a
  normal metal and a clean superconductor carrying a supercurrent},\ }\href
  {https://doi.org/10.1103/PhysRevB.70.172508} {\bibfield  {journal} {\bibinfo
  {journal} {Phys. Rev. B}\ }\textbf {\bibinfo {volume} {70}},\ \bibinfo
  {pages} {172508} (\bibinfo {year} {2004})}\BibitemShut {NoStop}%
\bibitem [{\citenamefont {Lukic}\ and\ \citenamefont
  {Nicol}(2007)}]{PhysRevB.76.144508}%
  \BibitemOpen
  \bibfield  {author} {\bibinfo {author} {\bibfnamefont {V.}~\bibnamefont
  {Lukic}}\ and\ \bibinfo {author} {\bibfnamefont {E.~J.}\ \bibnamefont
  {Nicol}},\ }\bibfield  {title} {\bibinfo {title} {Conductance characteristics
  between a normal metal and a two-band superconductor carrying a
  supercurrent},\ }\href {https://doi.org/10.1103/PhysRevB.76.144508}
  {\bibfield  {journal} {\bibinfo  {journal} {Phys. Rev. B}\ }\textbf {\bibinfo
  {volume} {76}},\ \bibinfo {pages} {144508} (\bibinfo {year}
  {2007})}\BibitemShut {NoStop}%
\bibitem [{\citenamefont {Cui}\ \emph {et~al.}(2006)\citenamefont {Cui},
  \citenamefont {Hu}, \citenamefont {Wei},\ and\ \citenamefont
  {Yang}}]{PhysRevB.73.214514}%
  \BibitemOpen
  \bibfield  {author} {\bibinfo {author} {\bibfnamefont {Q.}~\bibnamefont
  {Cui}}, \bibinfo {author} {\bibfnamefont {C.-R.}\ \bibnamefont {Hu}},
  \bibinfo {author} {\bibfnamefont {J.~Y.~T.}\ \bibnamefont {Wei}},\ and\
  \bibinfo {author} {\bibfnamefont {K.}~\bibnamefont {Yang}},\ }\bibfield
  {title} {\bibinfo {title} {{Conductance characteristics between a normal
  metal and a two-dimensional Fulde-Ferrell-Larkin-Ovchinnikov superconductor:
  The Fulde-Ferrell state}},\ }\href
  {https://doi.org/10.1103/PhysRevB.73.214514} {\bibfield  {journal} {\bibinfo
  {journal} {Phys. Rev. B}\ }\textbf {\bibinfo {volume} {73}},\ \bibinfo
  {pages} {214514} (\bibinfo {year} {2006})}\BibitemShut {NoStop}%
\bibitem [{\citenamefont {Jiao}\ \emph {et~al.}(2020)\citenamefont {Jiao},
  \citenamefont {Howard}, \citenamefont {Ran}, \citenamefont {Wang},
  \citenamefont {Rodriguez}, \citenamefont {Sigrist}, \citenamefont {Wang},
  \citenamefont {Butch},\ and\ \citenamefont {Madhavan}}]{Jiao2020Mar}%
  \BibitemOpen
  \bibfield  {author} {\bibinfo {author} {\bibfnamefont {L.}~\bibnamefont
  {Jiao}}, \bibinfo {author} {\bibfnamefont {S.}~\bibnamefont {Howard}},
  \bibinfo {author} {\bibfnamefont {S.}~\bibnamefont {Ran}}, \bibinfo {author}
  {\bibfnamefont {Z.}~\bibnamefont {Wang}}, \bibinfo {author} {\bibfnamefont
  {J.~O.}\ \bibnamefont {Rodriguez}}, \bibinfo {author} {\bibfnamefont
  {M.}~\bibnamefont {Sigrist}}, \bibinfo {author} {\bibfnamefont
  {Z.}~\bibnamefont {Wang}}, \bibinfo {author} {\bibfnamefont {N.~P.}\
  \bibnamefont {Butch}},\ and\ \bibinfo {author} {\bibfnamefont
  {V.}~\bibnamefont {Madhavan}},\ }\bibfield  {title} {\bibinfo {title}
  {{Chiral superconductivity in heavy-fermion metal UTe2}},\ }\href
  {https://doi.org/10.1038/s41586-020-2122-2} {\bibfield  {journal} {\bibinfo
  {journal} {Nature}\ }\textbf {\bibinfo {volume} {579}},\ \bibinfo {pages}
  {523} (\bibinfo {year} {2020})}\BibitemShut {NoStop}%
\bibitem [{\citenamefont {Fischer}\ \emph {et~al.}(2007)\citenamefont
  {Fischer}, \citenamefont {Kugler}, \citenamefont {Maggio-Aprile},
  \citenamefont {Berthod},\ and\ \citenamefont {Renner}}]{Fischer2007Mar}%
  \BibitemOpen
  \bibfield  {author} {\bibinfo {author} {\bibfnamefont {{\O}.}~\bibnamefont
  {Fischer}}, \bibinfo {author} {\bibfnamefont {M.}~\bibnamefont {Kugler}},
  \bibinfo {author} {\bibfnamefont {I.}~\bibnamefont {Maggio-Aprile}}, \bibinfo
  {author} {\bibfnamefont {C.}~\bibnamefont {Berthod}},\ and\ \bibinfo {author}
  {\bibfnamefont {C.}~\bibnamefont {Renner}},\ }\bibfield  {title} {\bibinfo
  {title} {{Scanning tunneling spectroscopy of high-temperature
  superconductors}},\ }\href {https://doi.org/10.1103/RevModPhys.79.353}
  {\bibfield  {journal} {\bibinfo  {journal} {Rev. Mod. Phys.}\ }\textbf
  {\bibinfo {volume} {79}},\ \bibinfo {pages} {353} (\bibinfo {year}
  {2007})}\BibitemShut {NoStop}%
\bibitem [{\citenamefont {Schrieffer}(1964)}]{RevModPhys.36.200}%
  \BibitemOpen
  \bibfield  {author} {\bibinfo {author} {\bibfnamefont {J.~R.}\ \bibnamefont
  {Schrieffer}},\ }\bibfield  {title} {\bibinfo {title} {Theory of electron
  tunneling},\ }\href {https://doi.org/10.1103/RevModPhys.36.200} {\bibfield
  {journal} {\bibinfo  {journal} {Rev. Mod. Phys.}\ }\textbf {\bibinfo {volume}
  {36}},\ \bibinfo {pages} {200} (\bibinfo {year} {1964})}\BibitemShut
  {NoStop}%
\bibitem [{\citenamefont {Bardeen}(1961)}]{PhysRevLett.6.57}%
  \BibitemOpen
  \bibfield  {author} {\bibinfo {author} {\bibfnamefont {J.}~\bibnamefont
  {Bardeen}},\ }\bibfield  {title} {\bibinfo {title} {Tunnelling from a
  many-particle point of view},\ }\href
  {https://doi.org/10.1103/PhysRevLett.6.57} {\bibfield  {journal} {\bibinfo
  {journal} {Phys. Rev. Lett.}\ }\textbf {\bibinfo {volume} {6}},\ \bibinfo
  {pages} {57} (\bibinfo {year} {1961})}\BibitemShut {NoStop}%
\bibitem [{\citenamefont {Bardeen}(1962)}]{PhysRevLett.9.147}%
  \BibitemOpen
  \bibfield  {author} {\bibinfo {author} {\bibfnamefont {J.}~\bibnamefont
  {Bardeen}},\ }\bibfield  {title} {\bibinfo {title} {Tunneling into
  superconductors},\ }\href {https://doi.org/10.1103/PhysRevLett.9.147}
  {\bibfield  {journal} {\bibinfo  {journal} {Phys. Rev. Lett.}\ }\textbf
  {\bibinfo {volume} {9}},\ \bibinfo {pages} {147} (\bibinfo {year}
  {1962})}\BibitemShut {NoStop}%
\bibitem [{\citenamefont {Cohen}\ \emph {et~al.}(1962)\citenamefont {Cohen},
  \citenamefont {Falicov},\ and\ \citenamefont {Phillips}}]{PhysRevLett.8.316}%
  \BibitemOpen
  \bibfield  {author} {\bibinfo {author} {\bibfnamefont {M.~H.}\ \bibnamefont
  {Cohen}}, \bibinfo {author} {\bibfnamefont {L.~M.}\ \bibnamefont {Falicov}},\
  and\ \bibinfo {author} {\bibfnamefont {J.~C.}\ \bibnamefont {Phillips}},\
  }\bibfield  {title} {\bibinfo {title} {Superconductive tunneling},\ }\href
  {https://doi.org/10.1103/PhysRevLett.8.316} {\bibfield  {journal} {\bibinfo
  {journal} {Phys. Rev. Lett.}\ }\textbf {\bibinfo {volume} {8}},\ \bibinfo
  {pages} {316} (\bibinfo {year} {1962})}\BibitemShut {NoStop}%
\bibitem [{\citenamefont {Tinkham}(1972)}]{PhysRevB.6.1747}%
  \BibitemOpen
  \bibfield  {author} {\bibinfo {author} {\bibfnamefont {M.}~\bibnamefont
  {Tinkham}},\ }\bibfield  {title} {\bibinfo {title} {Tunneling generation,
  relaxation, and tunneling detection of hole-electron imbalance in
  superconductors},\ }\href {https://doi.org/10.1103/PhysRevB.6.1747}
  {\bibfield  {journal} {\bibinfo  {journal} {Phys. Rev. B}\ }\textbf {\bibinfo
  {volume} {6}},\ \bibinfo {pages} {1747} (\bibinfo {year} {1972})}\BibitemShut
  {NoStop}%
\bibitem [{Note4()}]{Note4}%
  \BibitemOpen
  \bibinfo {note} {We also find a curious coincidence in that the BTK and
  transfer Hamiltonian-approach results for the 1D junction yield the same
  analytical expression as the bulk density of states of a finite-momentum
  three-dimensional superconductor. For the latter, see ref.~\cite
  {PhysRev.137.A783} and Appendix~\ref {app:tunneling}}\BibitemShut {NoStop}%
\bibitem [{\citenamefont {Martin}\ and\ \citenamefont
  {Mozyrsky}(2014)}]{Martin2014Sep}%
  \BibitemOpen
  \bibfield  {author} {\bibinfo {author} {\bibfnamefont {I.}~\bibnamefont
  {Martin}}\ and\ \bibinfo {author} {\bibfnamefont {D.}~\bibnamefont
  {Mozyrsky}},\ }\bibfield  {title} {\bibinfo {title} {{Nonequilibrium theory
  of tunneling into a localized state in a superconductor}},\ }\href
  {https://doi.org/10.1103/PhysRevB.90.100508} {\bibfield  {journal} {\bibinfo
  {journal} {Phys. Rev. B}\ }\textbf {\bibinfo {volume} {90}},\ \bibinfo
  {pages} {100508} (\bibinfo {year} {2014})}\BibitemShut {NoStop}%
\bibitem [{\citenamefont {Freudenfeld}\ \emph {et~al.}(2020)\citenamefont
  {Freudenfeld}, \citenamefont {Geier}, \citenamefont {Umansky}, \citenamefont
  {Brouwer},\ and\ \citenamefont {Ludwig}}]{PhysRevLett.125.107701}%
  \BibitemOpen
  \bibfield  {author} {\bibinfo {author} {\bibfnamefont {J.}~\bibnamefont
  {Freudenfeld}}, \bibinfo {author} {\bibfnamefont {M.}~\bibnamefont {Geier}},
  \bibinfo {author} {\bibfnamefont {V.}~\bibnamefont {Umansky}}, \bibinfo
  {author} {\bibfnamefont {P.~W.}\ \bibnamefont {Brouwer}},\ and\ \bibinfo
  {author} {\bibfnamefont {S.}~\bibnamefont {Ludwig}},\ }\bibfield  {title}
  {\bibinfo {title} {Coherent electron optics with ballistically coupled
  quantum point contacts},\ }\href
  {https://doi.org/10.1103/PhysRevLett.125.107701} {\bibfield  {journal}
  {\bibinfo  {journal} {Phys. Rev. Lett.}\ }\textbf {\bibinfo {volume} {125}},\
  \bibinfo {pages} {107701} (\bibinfo {year} {2020})}\BibitemShut {NoStop}%
\bibitem [{\citenamefont {Geier}\ \emph {et~al.}(2020)\citenamefont {Geier},
  \citenamefont {Freudenfeld}, \citenamefont {Silva}, \citenamefont {Umansky},
  \citenamefont {Reuter}, \citenamefont {Wieck}, \citenamefont {Brouwer},\ and\
  \citenamefont {Ludwig}}]{PhysRevB.101.165429}%
  \BibitemOpen
  \bibfield  {author} {\bibinfo {author} {\bibfnamefont {M.}~\bibnamefont
  {Geier}}, \bibinfo {author} {\bibfnamefont {J.}~\bibnamefont {Freudenfeld}},
  \bibinfo {author} {\bibfnamefont {J.~T.}\ \bibnamefont {Silva}}, \bibinfo
  {author} {\bibfnamefont {V.}~\bibnamefont {Umansky}}, \bibinfo {author}
  {\bibfnamefont {D.}~\bibnamefont {Reuter}}, \bibinfo {author} {\bibfnamefont
  {A.~D.}\ \bibnamefont {Wieck}}, \bibinfo {author} {\bibfnamefont {P.~W.}\
  \bibnamefont {Brouwer}},\ and\ \bibinfo {author} {\bibfnamefont
  {S.}~\bibnamefont {Ludwig}},\ }\bibfield  {title} {\bibinfo {title}
  {Electrostatic potential shape of gate-defined quantum point contacts},\
  }\href {https://doi.org/10.1103/PhysRevB.101.165429} {\bibfield  {journal}
  {\bibinfo  {journal} {Phys. Rev. B}\ }\textbf {\bibinfo {volume} {101}},\
  \bibinfo {pages} {165429} (\bibinfo {year} {2020})}\BibitemShut {NoStop}%
\bibitem [{\citenamefont {Mortensen}\ \emph {et~al.}(1999)\citenamefont
  {Mortensen}, \citenamefont {Flensberg},\ and\ \citenamefont
  {Jauho}}]{PhysRevB.59.10176}%
  \BibitemOpen
  \bibfield  {author} {\bibinfo {author} {\bibfnamefont {N.~A.}\ \bibnamefont
  {Mortensen}}, \bibinfo {author} {\bibfnamefont {K.}~\bibnamefont
  {Flensberg}},\ and\ \bibinfo {author} {\bibfnamefont {A.-P.}\ \bibnamefont
  {Jauho}},\ }\bibfield  {title} {\bibinfo {title} {{Angle dependence of
  Andreev scattering at semiconductor--superconductor interfaces}},\ }\href
  {https://doi.org/10.1103/PhysRevB.59.10176} {\bibfield  {journal} {\bibinfo
  {journal} {Phys. Rev. B}\ }\textbf {\bibinfo {volume} {59}},\ \bibinfo
  {pages} {10176} (\bibinfo {year} {1999})}\BibitemShut {NoStop}%
\bibitem [{\citenamefont {Glazman}\ \emph {et~al.}(1988)\citenamefont
  {Glazman}, \citenamefont {Lesovik}, \citenamefont {Khmel'nitskiǐ},\ and\
  \citenamefont {Shekhter}}]{glazman1988reflectionless}%
  \BibitemOpen
  \bibfield  {author} {\bibinfo {author} {\bibfnamefont {L.}~\bibnamefont
  {Glazman}}, \bibinfo {author} {\bibfnamefont {G.}~\bibnamefont {Lesovik}},
  \bibinfo {author} {\bibfnamefont {D.}~\bibnamefont {Khmel'nitskiǐ}},\ and\
  \bibinfo {author} {\bibfnamefont {R.}~\bibnamefont {Shekhter}},\ }\bibfield
  {title} {\bibinfo {title} {Reflectionless quantum transport and fundamental
  ballistic-resistance steps in microscopic constrictions},\ }\href@noop {}
  {\bibfield  {journal} {\bibinfo  {journal} {Soviet JETP letters}\ }\textbf
  {\bibinfo {volume} {48}},\ \bibinfo {pages} {238} (\bibinfo {year}
  {1988})}\BibitemShut {NoStop}%
\bibitem [{\citenamefont {Yacoby}\ and\ \citenamefont
  {Imry}(1990)}]{PhysRevB.41.5341}%
  \BibitemOpen
  \bibfield  {author} {\bibinfo {author} {\bibfnamefont {A.}~\bibnamefont
  {Yacoby}}\ and\ \bibinfo {author} {\bibfnamefont {Y.}~\bibnamefont {Imry}},\
  }\bibfield  {title} {\bibinfo {title} {Quantization of the conductance of
  ballistic point contacts beyond the adiabatic approximation},\ }\href
  {https://doi.org/10.1103/PhysRevB.41.5341} {\bibfield  {journal} {\bibinfo
  {journal} {Phys. Rev. B}\ }\textbf {\bibinfo {volume} {41}},\ \bibinfo
  {pages} {5341} (\bibinfo {year} {1990})}\BibitemShut {NoStop}%
\bibitem [{\citenamefont {Kallin}\ and\ \citenamefont
  {Berlinsky}(2016)}]{Kallin2016Apr}%
  \BibitemOpen
  \bibfield  {author} {\bibinfo {author} {\bibfnamefont {C.}~\bibnamefont
  {Kallin}}\ and\ \bibinfo {author} {\bibfnamefont {J.}~\bibnamefont
  {Berlinsky}},\ }\bibfield  {title} {\bibinfo {title} {{Chiral
  superconductors}},\ }\href {https://doi.org/10.1088/0034-4885/79/5/054502}
  {\bibfield  {journal} {\bibinfo  {journal} {Rep. Prog. Phys.}\ }\textbf
  {\bibinfo {volume} {79}},\ \bibinfo {pages} {054502} (\bibinfo {year}
  {2016})}\BibitemShut {NoStop}%
\bibitem [{\citenamefont {Ran}\ \emph {et~al.}(2019)\citenamefont {Ran},
  \citenamefont {Eckberg}, \citenamefont {Ding}, \citenamefont {Furukawa},
  \citenamefont {Metz}, \citenamefont {Saha}, \citenamefont {Liu},
  \citenamefont {Zic}, \citenamefont {Kim}, \citenamefont {Paglione},\ and\
  \citenamefont {Butch}}]{Ran2019Aug}%
  \BibitemOpen
  \bibfield  {author} {\bibinfo {author} {\bibfnamefont {S.}~\bibnamefont
  {Ran}}, \bibinfo {author} {\bibfnamefont {C.}~\bibnamefont {Eckberg}},
  \bibinfo {author} {\bibfnamefont {Q.-P.}\ \bibnamefont {Ding}}, \bibinfo
  {author} {\bibfnamefont {Y.}~\bibnamefont {Furukawa}}, \bibinfo {author}
  {\bibfnamefont {T.}~\bibnamefont {Metz}}, \bibinfo {author} {\bibfnamefont
  {S.~R.}\ \bibnamefont {Saha}}, \bibinfo {author} {\bibfnamefont {I.-L.}\
  \bibnamefont {Liu}}, \bibinfo {author} {\bibfnamefont {M.}~\bibnamefont
  {Zic}}, \bibinfo {author} {\bibfnamefont {H.}~\bibnamefont {Kim}}, \bibinfo
  {author} {\bibfnamefont {J.}~\bibnamefont {Paglione}},\ and\ \bibinfo
  {author} {\bibfnamefont {N.~P.}\ \bibnamefont {Butch}},\ }\bibfield  {title}
  {\bibinfo {title} {{Nearly ferromagnetic spin-triplet superconductivity}},\
  }\href {https://doi.org/10.1126/science.aav8645} {\bibfield  {journal}
  {\bibinfo  {journal} {Science}\ }\textbf {\bibinfo {volume} {365}},\ \bibinfo
  {pages} {684} (\bibinfo {year} {2019})}\BibitemShut {NoStop}%
\bibitem [{\citenamefont {Aoki}\ \emph {et~al.}(2019)\citenamefont {Aoki},
  \citenamefont {Nakamura}, \citenamefont {Honda}, \citenamefont {Li},
  \citenamefont {Homma}, \citenamefont {Shimizu}, \citenamefont {Sato},
  \citenamefont {Knebel}, \citenamefont {Brison}, \citenamefont {Pourret},
  \citenamefont {Braithwaite}, \citenamefont {Lapertot}, \citenamefont {Niu},
  \citenamefont {Vali{\ifmmode\check{s}\else\v{s}\fi}ka}, \citenamefont
  {Harima},\ and\ \citenamefont {Flouquet}}]{Aoki2019Mar}%
  \BibitemOpen
  \bibfield  {author} {\bibinfo {author} {\bibfnamefont {D.}~\bibnamefont
  {Aoki}}, \bibinfo {author} {\bibfnamefont {A.}~\bibnamefont {Nakamura}},
  \bibinfo {author} {\bibfnamefont {F.}~\bibnamefont {Honda}}, \bibinfo
  {author} {\bibfnamefont {D.}~\bibnamefont {Li}}, \bibinfo {author}
  {\bibfnamefont {Y.}~\bibnamefont {Homma}}, \bibinfo {author} {\bibfnamefont
  {Y.}~\bibnamefont {Shimizu}}, \bibinfo {author} {\bibfnamefont {Y.~J.}\
  \bibnamefont {Sato}}, \bibinfo {author} {\bibfnamefont {G.}~\bibnamefont
  {Knebel}}, \bibinfo {author} {\bibfnamefont {J.-P.}\ \bibnamefont {Brison}},
  \bibinfo {author} {\bibfnamefont {A.}~\bibnamefont {Pourret}}, \bibinfo
  {author} {\bibfnamefont {D.}~\bibnamefont {Braithwaite}}, \bibinfo {author}
  {\bibfnamefont {G.}~\bibnamefont {Lapertot}}, \bibinfo {author}
  {\bibfnamefont {Q.}~\bibnamefont {Niu}}, \bibinfo {author} {\bibfnamefont
  {M.}~\bibnamefont {Vali{\ifmmode\check{s}\else\v{s}\fi}ka}}, \bibinfo
  {author} {\bibfnamefont {H.}~\bibnamefont {Harima}},\ and\ \bibinfo {author}
  {\bibfnamefont {J.}~\bibnamefont {Flouquet}},\ }\bibfield  {title} {\bibinfo
  {title} {{Unconventional Superconductivity in Heavy Fermion UTe2}},\ }\href
  {https://doi.org/10.7566/JPSJ.88.043702} {\bibfield  {journal} {\bibinfo
  {journal} {J. Phys. Soc. Jpn.}\ }\textbf {\bibinfo {volume} {88}},\ \bibinfo
  {pages} {043702} (\bibinfo {year} {2019})}\BibitemShut {NoStop}%
\bibitem [{\citenamefont {Aoki}\ \emph {et~al.}(2022)\citenamefont {Aoki},
  \citenamefont {Brison}, \citenamefont {Flouquet}, \citenamefont {Ishida},
  \citenamefont {Knebel}, \citenamefont {Tokunaga},\ and\ \citenamefont
  {Yanase}}]{Aoki2022Apr}%
  \BibitemOpen
  \bibfield  {author} {\bibinfo {author} {\bibfnamefont {D.}~\bibnamefont
  {Aoki}}, \bibinfo {author} {\bibfnamefont {J.-P.}\ \bibnamefont {Brison}},
  \bibinfo {author} {\bibfnamefont {J.}~\bibnamefont {Flouquet}}, \bibinfo
  {author} {\bibfnamefont {K.}~\bibnamefont {Ishida}}, \bibinfo {author}
  {\bibfnamefont {G.}~\bibnamefont {Knebel}}, \bibinfo {author} {\bibfnamefont
  {Y.}~\bibnamefont {Tokunaga}},\ and\ \bibinfo {author} {\bibfnamefont
  {Y.}~\bibnamefont {Yanase}},\ }\bibfield  {title} {\bibinfo {title}
  {{Unconventional superconductivity in UTe2}},\ }\href
  {https://doi.org/10.1088/1361-648X/ac5863} {\bibfield  {journal} {\bibinfo
  {journal} {J. Phys.: Condens. Matter}\ }\textbf {\bibinfo {volume} {34}},\
  \bibinfo {pages} {243002} (\bibinfo {year} {2022})}\BibitemShut {NoStop}%
\bibitem [{\citenamefont {Fernandes}\ \emph {et~al.}(2022)\citenamefont
  {Fernandes}, \citenamefont {Coldea}, \citenamefont {Ding}, \citenamefont
  {Fisher}, \citenamefont {Hirschfeld},\ and\ \citenamefont
  {Kotliar}}]{Fernandes2022Jan}%
  \BibitemOpen
  \bibfield  {author} {\bibinfo {author} {\bibfnamefont {R.~M.}\ \bibnamefont
  {Fernandes}}, \bibinfo {author} {\bibfnamefont {A.~I.}\ \bibnamefont
  {Coldea}}, \bibinfo {author} {\bibfnamefont {H.}~\bibnamefont {Ding}},
  \bibinfo {author} {\bibfnamefont {I.~R.}\ \bibnamefont {Fisher}}, \bibinfo
  {author} {\bibfnamefont {P.~J.}\ \bibnamefont {Hirschfeld}},\ and\ \bibinfo
  {author} {\bibfnamefont {G.}~\bibnamefont {Kotliar}},\ }\bibfield  {title}
  {\bibinfo {title} {{Iron pnictides and chalcogenides: a new paradigm for
  superconductivity}},\ }\href {https://doi.org/10.1038/s41586-021-04073-2}
  {\bibfield  {journal} {\bibinfo  {journal} {Nature}\ }\textbf {\bibinfo
  {volume} {601}},\ \bibinfo {pages} {35} (\bibinfo {year} {2022})}\BibitemShut
  {NoStop}%
\bibitem [{\citenamefont {Levitov}\ and\ \citenamefont
  {Shytov}(2004)}]{levitov}%
  \BibitemOpen
  \bibfield  {author} {\bibinfo {author} {\bibfnamefont {L.}~\bibnamefont
  {Levitov}}\ and\ \bibinfo {author} {\bibfnamefont {A.}~\bibnamefont
  {Shytov}},\ }\href {https://www.mit.edu/~levitov/book} {\emph {\bibinfo
  {title} {{Green's functions. Theory and practice}}}}\ (\bibinfo  {publisher}
  {FizMatLit-Nauka},\ \bibinfo {year} {2004})\BibitemShut {NoStop}%
\bibitem [{\citenamefont {van Weerdenburg}\ \emph {et~al.}(2023)\citenamefont
  {van Weerdenburg}, \citenamefont {Kamlapure}, \citenamefont {Fyhn},
  \citenamefont {Huang}, \citenamefont {van Mullekom}, \citenamefont
  {Steinbrecher}, \citenamefont {Krogstrup}, \citenamefont {Linder},\ and\
  \citenamefont {Khajetoorians}}]{vanWeerdenburg2023Mar}%
  \BibitemOpen
  \bibfield  {author} {\bibinfo {author} {\bibfnamefont {W.~M.~J.}\
  \bibnamefont {van Weerdenburg}}, \bibinfo {author} {\bibfnamefont
  {A.}~\bibnamefont {Kamlapure}}, \bibinfo {author} {\bibfnamefont {E.~H.}\
  \bibnamefont {Fyhn}}, \bibinfo {author} {\bibfnamefont {X.}~\bibnamefont
  {Huang}}, \bibinfo {author} {\bibfnamefont {N.~P.~E.}\ \bibnamefont {van
  Mullekom}}, \bibinfo {author} {\bibfnamefont {M.}~\bibnamefont
  {Steinbrecher}}, \bibinfo {author} {\bibfnamefont {P.}~\bibnamefont
  {Krogstrup}}, \bibinfo {author} {\bibfnamefont {J.}~\bibnamefont {Linder}},\
  and\ \bibinfo {author} {\bibfnamefont {A.~A.}\ \bibnamefont
  {Khajetoorians}},\ }\bibfield  {title} {\bibinfo {title} {{Extreme
  enhancement of superconductivity in epitaxial aluminum near the monolayer
  limit}},\ }\bibfield  {journal} {\bibinfo  {journal} {Sci. Adv.}\ }\textbf
  {\bibinfo {volume} {9}},\ \href {https://doi.org/10.1126/sciadv.adf5500}
  {10.1126/sciadv.adf5500} (\bibinfo {year} {2023})\BibitemShut {NoStop}%
\bibitem [{\citenamefont {Fulde}(1965)}]{PhysRev.137.A783}%
  \BibitemOpen
  \bibfield  {author} {\bibinfo {author} {\bibfnamefont {P.}~\bibnamefont
  {Fulde}},\ }\bibfield  {title} {\bibinfo {title} {Tunneling density of states
  for a superconductor carrying a current},\ }\href
  {https://doi.org/10.1103/PhysRev.137.A783} {\bibfield  {journal} {\bibinfo
  {journal} {Phys. Rev.}\ }\textbf {\bibinfo {volume} {137}},\ \bibinfo {pages}
  {A783} (\bibinfo {year} {1965})}\BibitemShut {NoStop}%
\end{thebibliography}%

\end{document}